\documentclass[11pt,reqno]{amsart}
\usepackage[utf8]{inputenc}
\usepackage{amsmath, amsthm, amssymb, mathrsfs, latexsym}
\usepackage{natbib}
\usepackage{bm} 
\usepackage{booktabs}

\usepackage{textcomp}
\usepackage{color}
\usepackage{graphicx}

\usepackage[margin=1in]{geometry}
\usepackage{setspace}
\numberwithin{equation}{section}
\numberwithin{figure}{section}
\numberwithin{table}{section}
\renewcommand{\cite}{\citet}

\newcommand{\E}{\mathbb{E}}
\newcommand{\Var}{\operatorname{Var}}
\newcommand{\RR}{\mathbb{R}}

\newcommand{\ba}{\begin{array}}
\newcommand{\ea}{\end{array}}
\newcommand{\bs}{\begin{align}\begin{split}\nonumber}
\newcommand{\bsnumber}{\begin{align}\begin{split}}
\newcommand{\es}{\end{split}\end{align}}

\newcommand{\be}{\begin{eqnarray}}
\newcommand{\ee}{\end{eqnarray}}

\renewcommand{\theequation}{\thesection.\arabic{equation}}

\newtheorem{theorem}{Theorem}
\newtheorem{lemma}{Lemma}

\newcommand{\VV}{\mathbb{V}}
\newcommand{\MM}{\mathbb{M}}

\newcommand{\inpr}{\overset{p}{\longrightarrow}}
\newcommand{\indist}{\overset{d}{\longrightarrow}}

\newcommand{\ud}{\textnormal{d}}
\newcommand{\prob}[1]{\textnormal{P} \left\{ #1 \right\}}
\newcommand{\ex}[1]{\textnormal{E} \left[ #1 \right]}
\newcommand{\exs}[1]{\textnormal{E}^* \left[ #1 \right]}
\DeclareMathOperator*{\argmin}{argmin}
\DeclareMathOperator*{\sign}{sgn}
\DeclareMathOperator*{\plim}{plim}
\newcommand{\zero}{\mathbf{0}}

\theoremstyle{definition}
\newtheorem{remark}{Remark}

\newtheorem{assumptionA}{}
\newtheorem{assumptionB}{}
\newtheorem{assumptionC}{}

\begin{document}

\vspace{10mm}

\begin{center}

\long\def\symbolfootnote[1]{\begingroup
\def\thefootnote{\fnsymbol{footnote}}\footnote[1]{This draft: \today. We would like to thank the Co-editor Xiaohong Chen, an Associate Editor, and two referees for their detailed and constructive comments which have improved the paper considerably. We are also grateful to Antonio Galvao and Matt Harding for helpful comments and suggestions as well as seminar participants at the University of Kentucky, the 2019 CFE/CMStatistics conference, and the 2021 New York Camp Econometrics meeting.}
\endgroup}

{\bf WILD BOOTSTRAP INFERENCE FOR PENALIZED QUANTILE REGRESSION FOR LONGITUDINAL DATA}\symbolfootnote[1] 

\long\def\symbolfootnote[2]{\begingroup 
\def\thefootnote{\fnsymbol{footnote}}\footnote[2]{Carlos Lamarche: Department of Economics, University of Kentucky, 223G Gatton College of Business \& Economics, Lexington, KY 40506. Email: clamarche@uky.edu. Thomas Parker: Department of Economics, University of Waterloo, 200 University Ave. West, Waterloo, ON, Canada N2L 3G1. Email: tmparker@uwaterloo.ca}
\endgroup}

\vspace{2.5mm}

{\small CARLOS LAMARCHE AND THOMAS PARKER}\symbolfootnote[2]

\end{center}

\vspace{5mm}

{\small  
\begin{quote}

{\bf Abstract}: The existing theory of penalized quantile regression for longitudinal data has focused primarily on point estimation. In this work, we investigate statistical inference. We propose a wild residual bootstrap procedure and show that it is asymptotically valid for approximating the distribution of the penalized estimator. The model puts no restrictions on individual effects, and the estimator achieves consistency by letting the shrinkage decay in importance asymptotically. The new method is easy to implement and simulation studies show that it has accurate small sample behavior in comparison with existing procedures. Finally, we illustrate the new approach using U.S. Census data to estimate a model that includes more than eighty thousand parameters.

\bigskip



\end{quote}}

\onehalfspacing

\section{Introduction}


We consider a longitudinal data model of conditional quantiles with individual
intercepts. Variations of this model have been extensively studied in the literature since at least
\citet{NS48}. Recent contributions to the literature using this model for
quantile regression have emphasized the drawbacks of estimating a large number
of individual intercepts ($N$) when the number of time periods ($T$) is small (see Galvao and Kato, 2018,\nocite{Galvao2018} for an excellent survey). Koenker (2004)\nocite{rK04}
proposed an estimator where $N$ individual parameters are regularized by a
Lasso-type penalty, shrinking them towards a common value. As in the case of the Gaussian random effect estimator, shrinkage can reduce the variability of the estimator of the slope parameter in the quantile regression model (Koenker, 2004). In models with short $T$, shrinkage can reduce the bias of the fixed effects estimator of the slope parameter as well (Harding and Lamarche, 2019\nocite{harding2019}).


Although the regularization procedure has advantages, the asymptotic
distribution of the estimator is difficult to approximate. It is known that
Lasso-type estimators have non-standard limiting distributions (Knight and Fu,
2000\nocite{kK00}), but in the case of quantile regression, there are new
challenges. Because individual intercepts are treated as parameters, the
increasing dimension of the parameter vector as the number of units increases
can be an issue. In the case of estimators without regularization, Kato,
Galvao, and Montes-Rojas (2012) and Galvao, Gu, and Volgushev
(2020)\nocite{AGalvao2020} found that $T$ must grow faster than $N$ for
consistency and asymptotic normality at rates that are, at best, similar to
standard non-linear panel data models \citep{Newey2004}. Second, the
covariance matrix of quantile regression estimators typically depends on
conditional densities and the penalized estimator of Koenker (2004) is no
exception. Inference based on the asymptotic distribution requires
non-parametric estimation of nuisance parameters, which can lead to important
size distortions (He, 2018\nocite{He2018}). 

Motivated by these limitations, cross-sectional pairs (or block)
bootstrap, which samples sets of covariate and response vectors over
individuals with replacement, appears to be a natural
alternative method for inference. However, we demonstrate that the cross-sectional pairs
bootstrap does not approximate well the limiting distribution of the penalized
estimator. We consider instead a wild residual bootstrap
procedure, which was previously employed by Feng, He, and Hu (2011)\nocite{xumingHe2011}, and Wang, Van Keilegom,
and Maidman (2018)\nocite{lanWang2018} in cross-sectional settings. We investigate the application of the procedure to
longitudinal data and show that the proposed wild bootstrap procedure is a
consistent estimator of the distribution of the penalized estimator. 

We begin by deriving consistency and asymptotic normality results for $\ell_1$ penalized
estimators of a longitudinal model in which individual effects can be
correlated with the regressors. Although our model might be considered to
be high-dimensional, the number of parameters is smaller than the number of
observations, as in the pioneering work by Koenker (2004), and thus our results are obtained without
assuming sparsity in terms of the individual intercepts. Consistency and asymptotic normality with $T$
growing faster than $N$ are achieved by letting the penalty parameter that
controls shrinkage diminish in importance asymptotically. Thus, relative to \citet{rK04},
the asymptotic bias of the estimator is zero in our case. The consistency and asymptotic normality results are new — they extend the heuristic
results in Koenker (2004) obtained for a model with individual effects as location shifts and are not
included in \citet{kK12} and \citet{AGalvao2020} because they did not consider
penalized estimation.

The main theoretical contribution is to show that the distribution of the wild
bootstrap estimator consistently estimates the asymptotic distribution and
covariance of the penalized estimator. The results include the special case of no penalization, and thus, these results also show the consistency of the wild bootstrap for the quantile regression estimator with fixed effects. The consistency
of the wild bootstrap is established using developments that are critically
different to those used in Wang, Van Keilegom, and Maidman
(2018)\nocite{lanWang2018}. We also consider bootstrap estimation of the asymptotic
covariance matrix of the slope parameter estimator, which is novel in the panel quantile
literature. As emphasized in \citet{GoncalvesWhite5}, \citet{Andreas2017}, and \citet{HahnLiao21}, the weak convergence of the bootstrap estimator does not necessarily imply convergence of the bootstrap second moment estimator. Therefore, we provide conditions and establish a result that supports using the second moment of the bootstrap distribution to estimate the asymptotic variance of the estimator.




Several penalized estimators for quantile regression models have been proposed
in the literature since \citet{rK04}. \citet{belloni2011} propose quantile regression estimators for
high-dimensional sparse models using cross-sectional data. \citet{WANG2013} considers a penalized least absolute deviation
estimator, and \citet{Wang2019} derives error bounds for the
penalized estimator under weak conditions. \citet{cL06}
investigates the selection of a regularization parameter, and \citet{SLee2018}
study estimation of a high-dimensional quantile regression model with a change
point, or threshold. \citet{harding2016,harding2019} investigate estimation of models with attrition and
correlated random effects. \citet{GU2019} propose a
method for estimation of models with unknown group membership. \citet{CHEN2009} establish the validity of a related weighted bootstrap procedure for the limiting distribution of a penalized sieve estimator and consider applications using quantile regression \citep[see also][]{CHEN2015}. The literature
on penalized estimation methods for linear panel data models has also grown in
the last decade \citep[see, e.g.,][among others.]{kock2013,KOCK2016,aBelloni2016, lSu2016, lSu2018, CANER2018143, kock_tang_2019}


This paper is organized as follows. The next section provides background and discusses the motivation of our study. It also introduces the proposed wild residual bootstrap approach. Section 3 presents theoretical results. Section 4 investigates the small sample performance of the method, showing that the estimator has satisfactory performance under different specifications and it performs better than the cross-sectional pairs bootstrap procedure. Section 5 presents extensions to the basic model. Section 6 illustrates the theory and provides practical guidelines from an application of the method. Considering data from the U.S. Census, we estimate a quantile function with more than eighty thousand parameters to study how wages of U.S. workers have been affected by the North American Free Trade Agreement. Finally, Section 7 concludes. One appendix contains proof of the main results, while a supplementary appendix contains additional technical results and proofs.

\section{Inference for penalized quantile regression}

In this section, we first introduce the model and the estimator, and then we discuss the validity of a cross-sectional pairs bootstrap method. Motivated by the limitations of existing procedures, we propose a new approach to estimate the asymptotic distribution of the estimator. 

\subsection{Background and Motivation}

We observe repeated measures $\{ ( y_{it},\bm{x}_{it}') \}_{t=1}^{T}$ for each
subject $1 \leq i \leq N$. The variable $y_{it} \in \RR$ denotes the response
for $i$ at time $t$ and $\bm{x}_{it}$ denotes a $p$-dimensional vector of
covariates. Although the number of repeated observations does not vary with $i$,
the analysis can be trivially extended to consider $T_i$ as long as $\max T_i /
\min T_i$ is bounded  for $1 \leq i \leq N$ (Gu and Volgushev, 2019).
The model considered in this paper is 
\begin{equation}
  Q_y(\tau | \bm{x}_{it})  = \bm{x}_{it}' \bm{\beta}_0(\tau) + \alpha_{i0}(\tau),
\end{equation}
where $\tau \in (0,1)$ and $ Q_y(\tau | \bm{x}_{it}) $ is the $\tau$-th quantile of the
conditional distribution of $y_{it}$ given $\bm{x}_{it}$.
It is assumed that the vector $\bm{x}_{it}$ does not contain an intercept.
The parameter of interest is $\bm{\beta}_0(\tau) \in \RR^p$ and
$\alpha_{i0}(\tau)$ is treated as a nuisance parameter.  Because we consider
just one value of $\tau$, we supress the dependence of the
parameters on $\tau$ in the sequel.

Let $\bm{\theta} = (\bm{\beta}',\bm{\alpha}')' \in \bm{\Theta} \subseteq
\RR^{p+N}$, where $\bm{\alpha} = (\alpha_{1},...,\alpha_{N})'$, and let
$\bm{\theta}_0 = (\bm{\beta}_0',\bm{\alpha}_0')'$. To estimate $\bm{\theta}_0$,
we consider the following estimator:
\begin{equation} \label{pqr}
  \hat{\bm{\theta}} = (\hat{\bm{\beta}}', \hat{\bm{\alpha}}')'
  = \argmin_{\bm{\theta} \in \bm{\Theta}} \sum_{i=1}^N \sum_{t=1}^T \rho_{\tau}
  (y_{it} - \bm{x}_{it}' \bm{\beta} - \alpha_i) + \lambda_T
  \sum_{i=1}^{N} | \alpha_i |,
\end{equation}
where $\rho _{\tau }(u)=$ $u(\tau -I(u<0))$ is the quantile regression loss
function.  The tuning parameter $\lambda_T \geq 0$ depends on $T$ and it can
also depend on data, as discussed below.


The penalty term in (2.2) helps improve the finite sample performance of the fixed effects estimator, which is defined for $\lambda_T=0$. Shrinkage of the individual effects can lead to reductions of the variance of the estimator. In models with incidental parameters, the penalty term reduces the noise in the estimation of individual intercepts, and consequently, it can also reduce the bias of the fixed effects estimator of $\bm{\beta}_0$. The online appendix presents simulation evidence to illustrate finite sample improvements when the time dimension is short, complementing the evidence presented in Koenker (2004) and Harding and Lamarche (2019). See Bester and Hansen (2009)\nocite{cH2009} for a related penalty approach to bias reduction in nonlinear models with fixed effects.

We establish conditions that result in 
a tractable asymptotic distribution for the estimator defined in \eqref{pqr}. However, we expect that
resampling methods offer a more accurate description of the
distribution of the estimator in finite samples. In practice, the cross-sectional pairs bootstrap, which samples over $i$ with replacement keeping the entire
block of time series observations for each $i$, has been used as a
method for inference, primarily in the fixed effects case when $\lambda_T = 0$.
However, the cross-sectional pairs bootstrap does not provide a good
approximation to the sampling distribution of the penalized estimator
\eqref{pqr}, as in the case of the pairs bootstrap procedure for the Lasso
estimator \citep{camponovo15}. 

\subsection{A cross-sectional pairs bootstrap procedure}\label{subsection:cs}
We now offer a heuristic illustration of some problems with using a cross-sectional pairs bootstrap and the penalized quantile regression estimator.  The cross-sectional pairs bootstrap can be used successfully to estimate the distribution of the quantile regression model with unpenalized fixed effects, but it will be shown below that the penalty causes problems for this approach to resampling.  We fix $N$ in this section to avoid the effect of a diverging number of parameters as the sample size increases (later, asymptotic approximations will be found assuming that $T$ grows faster than $N$).  This allows us to see problems with the cross-sectional pairs without the additional incidental parameters problem.  Define $\bm{\gamma} = (\bm{\delta}', \bm{\eta}')' \in \RR^{p + N}$, where $\bm{\delta} = \sqrt{NT} (\bm{\beta} - \bm{\beta}_0)$ and for $i = 1, \ldots N$, $\eta_i = \sqrt{T} (\alpha_i - \alpha_{i0})$.  Then let
\begin{equation}
  \VV_T(\bm{\gamma}) = \sum_{i=1}^N \sum_{t=1}^T \left\{ \rho_{\tau} \left( u_{it} - \frac{\bm{x}_{it}' \bm{\delta}}{\sqrt{NT}} - \frac{\eta_i}{\sqrt{T}} \right) - \rho_{\tau} (u_{it}) \right\} + \lambda_T \sum_{i=1}^{N} \left\{ \left| \alpha_{i0} + \frac{\eta_i}{\sqrt{T}} \right| - | \alpha_{i0} | \right\},  \label{pqr2}
\end{equation}
where $u_{it} = y_{it} - \bm{x}_{it}' \bm{\beta}_0 - \alpha_{i0}$. This objective function is equivalent to~\eqref{pqr}.  \citet{kK00} developed a method for dealing with the asymptotic behavior of this objective function, stated here as a lemma.

\begin{lemma}[\citet{kK00}] \label{L1}
Under Assumptions \ref{assume:data}-\ref{assume:Avar} below, if $N$ is fixed, $T
  \rightarrow \infty$ and $\lambda_T/\sqrt{T} \to \lambda_0 \geq 0$, the
  minimizer of \eqref{pqr2}, $\hat{\bm{\gamma}}$, converges weakly to the
  minimizer of $\VV: \RR^{p + N} \rightarrow \RR$ defined by
\begin{equation*}
  \VV(\bm{\gamma}) = - \bm{\gamma}' \bm{B} + \frac{1}{2} \bm{\gamma}' \bm{D}_1 \bm{\gamma} + \lambda_0 \sum_{i=1}^{N} \left( \eta_i \sign(\alpha_{i0}) I(\alpha_{i0} \neq 0) + |\eta_i| I(\alpha_{i0} = 0) \right),
\end{equation*}
  where $\bm{D}_1$ is positive definite and $\bm{B} \sim
  \mathcal{N}(\bm{\zero},\bm{D}_0)$.
\end{lemma}

To examine the validity of the cross-sectional pairs bootstrap, consider an analog loss function for resampled data.  Letting $\bm{y}_i$ and $\bm{X}_i$ denote the vector and matrix of response and covariate observations corresponding to unit $i$, a cross-sectional pairs bootstrap procedure resamples $N$ pairs $(\bm{y}_i, \bm{X}_i)$ for $1 \leq i \leq N$ with replacement. Let $n_i^*$ denote the number of times unit $i$ is redrawn from the original sample. Thus, the asymptotic distribution of $\hat{\bm{\gamma}}$ is approximated with $\tilde{\bm{\gamma}} = ( \sqrt{NT} (\tilde{\bm{\beta}} - \hat{\bm{\beta}})', \sqrt{T} ( \tilde{\bm{\alpha}} - \hat{\bm{\alpha}})' )'$ where
\begin{equation} \label{pairblock}
  \tilde{\bm{\theta}} = \left( \tilde{\bm{\beta}}', \tilde{\bm{\alpha}}' \right)' = \argmin_{ \bm{\theta} \in \bm{\Theta} } \sum_{i=1}^N  n_i^* \sum_{t=1}^T \rho_{\tau} \left( y_{it} - \bm{x}'_{it} \bm{\beta} - \alpha_i \right) + \lambda_T \sum_{i=1}^N  n_i^*  | \alpha_i |.
\end{equation}

Since $n_i^*$ is a multinomial weight with probability $1/N$, it is straightforward to calculate that the expected value of the objective function with respect to the bootstrap weights (i.e., conditional on the observations) is minimized at $\hat{\bm{\theta}} = (\hat{\bm{\beta}}, \hat{\bm{\alpha}})$. However, a finite sample problem is associated with the presence of the penalty in the objective function. To see this, let $\alpha_i^\ast = n_i^* |\alpha_i|$ and $\mathcal{A}=\{i : \alpha_i^* \neq 0\}$ denote the ``active" set corresponding to the penalty term in \eqref{pairblock}. In each bootstrap repetition, the cardinality of $\mathcal{A} < N$, leading to solutions $\tilde{\bm{\theta}}$ that can be potentially very different than the minimizer $\hat{\bm{\theta}}$. This may be especially so when $\alpha_i$ is correlated with $\bm{x}_{it}$. 

To see other problems with the cross-sectional bootstrap, we can find the weak limit of the bootstrap objective function \eqref{pairblock} similarly to Lemma~\ref{L1}. When we recenter~\eqref{pairblock} employing $\hat{\bm{\theta}}$, using the $i$ chosen by resampling, we find a naive bootstrap analog of the original objective function~\eqref{pqr2}, denoting $\hat{u}_{it} = y_{it} - \hat{\bm{\beta}}' \bm{x}_{it} - \hat{\alpha}_i$:
\begin{equation} 
  \tilde{\VV}_{T}(\bm{\gamma}) = \sum_{i=1}^N n_i^* \sum_{t=1}^T \left\{ \rho_{\tau} \left( \hat{u}_{it} - \frac{\bm{\delta}'\bm{x}_{it}}{\sqrt{NT}}  - \frac{\eta_i}{\sqrt{T}} \right) - \rho_{\tau}( \hat{u}_{it} ) \right\} + \lambda_T \sum_{i=1}^N n_i^* \left\{ \left| \hat{\alpha}_i + \frac{\eta_i}{\sqrt{T}} \right| - | \hat{\alpha}_i | \right\}.
\end{equation}

As $T \rightarrow \infty$, assuming $\hat{\eta}_i = \sqrt{T}(\hat{\alpha}_i - \alpha_{i0}) \indist A_i$ for $i = 1, \ldots N$ as $T \rightarrow \infty$, $\tilde{\VV}_T$ converges weakly to
\begin{equation*}
  \tilde{\VV}(\bm{\gamma}) = - \bm{\gamma}' \tilde{\bm{B}} + \frac{1}{2} \bm{\gamma}' \tilde{\bm{D}}_1 \bm{\gamma} + \lambda_0 \sum_{i=1}^N n_i^* \big( \eta_i \sign(\alpha_{i0}) I(\alpha_{i0} \neq 0) + \left( |\eta_i + A_i| - |A_i| \right) I(\alpha_{i0} = 0) \big).
\end{equation*}

However, there are two key differences with the resulting expression. The first problem with this limiting objective function is that $\tilde{\bm{B}} \neq \bm{B}$ and  $\tilde{\bm{D}_1} \neq \bm{D}_1$ from Lemma~\ref{L1}, due to the fact that recentering uses $\hat{\bm{\theta}}$, which is asymptotically biased if $\lambda_0>0$. Second, there is additional randomness arising from variable selection and resampling. (In the online appendix, we illustrate these issues with fixed $N$ and $T$). In the next section, we propose a wild residual bootstrap that does not suffer from these shortcomings. Then we expect that the distribution of the wild bootstrap estimator $\bm{\gamma}^\ast$ provides a better approximation to the distribution of $\hat{\bm{\gamma}}$ in Lemma \ref{L1}.

\subsection{Wild bootstrap procedures} \label{subsec:wild}
Let $\hat{u}_{it} = y_{it} - \bm{x}_{it}' \hat{\bm{\beta}} - \hat{\alpha}_i$ be the $\tau$-th quantile residual. Let $u_{it}^\ast = w_{it} | \hat{u}_{it} |$ denote bootstrap residuals, where $w_{it}$ is drawn randomly from a pre-determined distribution $G_W$ that satisfies the following conditions:

\begin{assumptionA} \label{C1}
The $\tau$-th quantile of $G_W$ is equal to zero, i.e. $G_W(0) = \tau$.
\end{assumptionA}
\begin{assumptionA} \label{C2}
The support of $G_W$ is bounded and contained in the interval $(-\infty,-c_1] \cup [c_2, \infty)$, where $c_1 > 0$ and $c_2 > 0$.  
\end{assumptionA}
\begin{assumptionA} \label{C3}
The weight distribution $G_W$ satisfies $- \int_{-\infty}^0 w^{-1} dG_W(w) = \int_0^{+\infty} w^{-1} dG_W(w) = \frac{1}{2}$.
\end{assumptionA}

Several weight distributions have been proposed in the
quantile regression literature that satisfy these conditions. \citet{xumingHe2011} propose, for $1/8 \leq \tau \leq 7/8$, the continuous weight density
$g_W(w) = - w I(-2 \tau - 1/4 \leq w \leq -2 \tau + 1/4) + w I(2 (1-\tau) - 1/4
\leq w \leq 2 (1-\tau) + 1/4)$. Another distribution that satisfies
\ref{C1}-\ref{C3} is the two-point distribution at $w = 2 (1-\tau)$ with
probability $\tau$ and at $w = - 2 \tau$ with probability $(1-\tau)$. We adopt
this distribution in the numerical examples. See Appendix 3 in
\citet{lanWang2018} for additional examples of the weight
distribution.

Using the bootstrap sample of residuals and the penalized quantile estimator as defined in equation \eqref{pqr}, we can form $y_{it}^\ast = \bm{x}_{it}' \hat{\bm{\beta}} + \hat{\alpha}_i + u_{it}^\ast$ to obtain the bootstrap estimator:
\begin{equation} \label{penboot}
  \bm{\theta}^\ast = (\bm{\beta}^{\ast'}, \bm{\alpha}^{\ast'})' = \argmin_{\bm{\theta} \in \bm{\Theta}} \sum_{i=1}^N \sum_{t=1}^T \rho_{\tau} (y_{it}^\ast - \bm{x}_{it}' \bm{\beta} - \alpha_i) + \lambda_T \sum_{i=1}^{N} | \alpha_{i} | .
\end{equation}

Given a bootstrap sample $\{\bm{\beta}_b^\ast\}_{b=1}^B$, we can obtain confidence intervals that are asymptotically valid, as demonstrated in Theorem \ref{thm:boot} below. Let $G_{j}^*(\alpha/2)$ and $G_{j}^*(1-\alpha/2)$ be the $(\alpha/2)$-th quantile and $(1-\alpha/2)$-th quantile of the bootstrap distribution of $\sqrt{NT} ( \beta_{j}^\ast - \hat{\beta}_{j})$ for $j = 1,2,\hdots,p$. We obtain asymptotically valid $100 (1-\alpha)\%$ confidence intervals for $\beta_{j}$ by $[ \hat{\beta}_{j} - (NT)^{-1/2} G_{j}^*(1 - \alpha/2), \hat{\beta}_{j} - (NT)^{-1/2} G_{j}^*(\alpha/2)]$.  Alternatively, Theorem~\ref{thm:var} shows that we may also estimate the covariance matrix of $\sqrt{NT}(\hat{\bm{\beta}} - \bm{\beta}_0)$ using the estimated covariance matrix from the bootstrap sample, which can be used to estimate the variance without requiring density estimation and to construct bootstrap-$t$ statistics for inference.

We may also consider a threshold estimator for $1 \leq i \leq N$, $\alpha_i^{**} = \hat{\alpha}_i I( | \hat{\alpha}_i | \geq a_{T})$, where $a_T$ is a constant that satisfies $a_T \to 0$ as $T \to \infty$. Define $v_{it}^\ast = w_{it} | \hat{v}_{it} |$, where $\hat{v}_{it} = y_{it} - \bm{x}_{it}' \hat{\bm{\beta}} - \alpha_i^{**}$. The response variable is generated as $y_{it}^{**} = \bm{x}_{it}' \hat{\bm{\beta}} + \alpha_i^{**} + v_{it}^\ast$, and the threshold estimator is defined as 
\begin{equation} \label{penbootthr}
  \bm{\theta}^{**} = \argmin_{\bm{\theta} \in \bm{\Theta}} \sum_{i=1}^N \sum_{t=1}^T \rho_{\tau} (y_{it}^{**} - \bm{x}_{it}' \bm{\beta} - \alpha_i) + \lambda_T \sum_{i=1}^{N} | \alpha_{i} | .
\end{equation}
As in the case of the estimator defined in \eqref{penboot}, we estimate the
distribution of $\hat{\bm{\theta}}$ based on the estimator $\bm{\theta}^{**}$. Given the similarities between estimators \eqref{penboot} and \eqref{penbootthr}, we derive below consistency and asymptotic normality results for \eqref{penboot} only. The performance of the bootstrap with this estimator is examined in the online appendix.

\subsection{Tuning parameter selection}

The tuning parameter $\lambda_T$ controls the degree of shrinkage of the
individual effect $\alpha_i$ towards zero and the penalty helps to control the bias
and variance of $\hat{\bm{\beta}}$. We restrict the tuning parameter to
$\lambda_T \in \mathcal{L} \subset [0, \lambda_U]$, where $\lambda_U$ is an
upper bound. As shown in Lemma~\ref{lem:upperbnd} in the supplementary
appendix, $\lambda_U = \max\{\tau, 1 - \tau\} T$ is a natural choice because if
$\lambda_T$ is set larger than this value, all the individual effects will be
set equal to zero. If the number of observed time periods $T_i$ vary over $i$,
then one would need to replace the $T$ in these bounds with $\max_i T_i$. This
estimator accommodates the choice of $\lambda_T = 0$, which means that the
results below continue to hold for the corresponding unpenalized estimator. 

The selection $\lambda_T$ in related settings has been investigated in several papers \citep[see, e.g.,][]{cL06,ERLee2014,lanWang2018}. We follow \citet{lanWang2018} and employ cross-validation for tuning parameter selection. To the best of our knowledge, theory has not yet been developed for the stochastic order of $\lambda_T$ when chosen using cross-validation, but in extensive simulations we have found that it tends to grow much more slowly than $T$, as required in Theorems~\ref{thm:consistent} and~\ref{thm:AN} below. 



\section{Asymptotic theory} 

This section investigates the large sample properties of the proposed estimator. We consider the following assumptions: 

\begin{assumptionB} \label{assume:data}
  Suppose that $\{ (y_{it}, \bm{x}_{it}): t \geq 1 \}$ are independent across $i$ and independent and identically distributed (i.i.d.) within each unit $i$.
\end{assumptionB} 

\begin{assumptionB}\label{assume:ID}
  For each $\phi > 0$,
\begin{equation*}
  \inf_{i\geq1} \inf_{\| \bm{\theta}_i \|_1 = \phi} \ex{ \int_0^{ (\alpha_i - \alpha_{i0}) + \bm{x}_{it}' (\bm{\beta} - \bm{\beta}_0)} \left( F_i(s | \bm{x}_{it}) - \tau \right) \ud s } = \epsilon_\phi > 0,
\end{equation*}
  where $F_i := F_{u_{it}|\bm{x}_{it}}$ is the distribution function of $u_{it} = y_{it} - \alpha_{i0} - \bm{x}_{it}' \bm{\beta}_0$ conditional on $\bm{x}_{it}$.
\end{assumptionB}

\begin{assumptionB} \label{assume:xsupport}
  The covariate vector $\bm{x}_{it}$ satisfies $\sup_{i,t} \| \bm{x}_{it} \| < M < \infty$ a.s.
\end{assumptionB}

These conditions are standard in the literature on quantile regression with individual effects.  Conditions \ref{assume:data} and \ref{assume:ID} are the same as Assumptions (A1) and (A3) in Kato, Galvao, and Montes-Rojas (2012). Condition \ref{assume:data} is relaxed in Kato et al. (2012) and in Section 5 below to allow for time dependence. Condition \ref{assume:ID} is an identification condition and it is sufficient for consistency. Slightly weaker than the assumption that $F_i$ has a continuous density given $\bm{x}_{it}$, it allows an expansion that guarantees the convexity of the limiting objective function, and therefore, the uniqueness of $(\bm{\beta}_0',\alpha_{i0})$ for all $1 \leq i \leq N$. Assumption \ref{assume:xsupport} is a simple way to assume appropriate moment conditions on the covariates and it is similar to (B1) in Kato, Galvao, and Montes-Rojas (2012) and (A1) in \citet{GU2019}. The condition can be relaxed as in Kato, Galvao and Montes-Rojas (2012). Condition B3 can be replaced with the moment condition $\sup_{i \geq 1} \ex{ \| \bm{x}_{i1} \|^{2s} } < \infty$ for some $s \geq 1$. The implication of this weaker condition is that $N / T^s \to 0$ instead of $\log(N)/T \to \infty$ to achieve consistency, as demonstrated in Theorem \ref{thm:consistent}.

The consistency of the estimator $\hat{\bm{\theta}}$ is needed to establish the main result stated in Theorem \ref{thm:boot}.  
\begin{theorem} \label{thm:consistent}
  Under Assumptions \ref{assume:data}-\ref{assume:xsupport}, if $\log(N)/T \to 0$ and $\lambda_T = o_p(T)$ as $N, T \to \infty$, then the estimator $\hat{\bm{\theta}}$ defined in equation \eqref{pqr} is a consistent estimator of $\bm{\theta}_0$.
\end{theorem} 

\begin{remark}
Theorem \ref{thm:consistent} is of independent interest as it has not been established the consistency of the penalized estimator under arbitrary dependence between regressors and individual effects. The result depends on the condition that $\lambda_T$, the parameter governing penalization of the individual effects, grows slowly as $T$ increases.
\end{remark}

We now focus our attention on weak convergence and we present a series of results to facilitate the estimation of standard errors and confidence intervals. To show asymptotic normality of the estimator, it is necessary to strengthen the conditions required for consistency slightly with the following conditions routinely adopted in the panel quantile regression literature (see, e.g., assumptions (B2) and (B3) in Kato, Galvao, and Montes-Rojas, 2012, and assumption (A2) in Gu and Volgushev, 2019). 

\begin{assumptionB}\label{assume:ID_weakconv}
  The conditional density function $f_i := f_{u_{it}|\bm{x}_{it}}$ corresponding to $F_i$ is uniformly bounded and has a bounded first derivative:
  \begin{equation*}
    \overline{f} := \sup_i \sup_{u \in \RR, \bm{x} \in \RR^p} | f_i(u | \bm{x}) | < \infty
  \end{equation*}
  and
  \begin{equation*}
    \overline{f'} := \sup_i \sup_{u \in \RR, \bm{x} \in \RR^p} | f'_i(u | \bm{x}) | < \infty.
  \end{equation*}
  Assume that in an open neighborhood $\mathcal{U}$ of $0$, $f_i$ is bounded away from zero for all realizations of $\bm{x}_{it}$:
  \begin{equation*}
    \underline{f} := \inf_i \inf_{u \in \mathcal{U}, \bm{x} \in \RR^p} | f_i(u | \bm{x}) | < \infty.
  \end{equation*}
\end{assumptionB}

\begin{assumptionB} \label{assume:Avar}
  Let $\varphi_i := \ex{f_i(0 | \bm{x}_{i1})}$, $\bm{E}_i := \ex{f_i(0 | \bm{x}_{i1}) \bm{x}_{i1}}$ and $\bm{J}_i := \ex{f_i(0 | \bm{x}_{i1}) \bm{x}_{i1} \bm{x}_{i1}'}$.  Let
  \begin{equation*}
    \bm{D}_N = \frac{1}{N} \sum_{i=1}^N  \left( \bm{J}_i - \varphi_i^{-1} \bm{E}_i \bm{E}_i' \right).
  \end{equation*}
  Suppose that $\bm{D}_N$ is positive definite for all $N$ and there is a positive definite matrix $\bm{D}$ such that $\bm{D} = \lim_{N \rightarrow \infty} \bm{D}_N$.  Also assume that
  \begin{equation*}
    \bm{V} = \tau(1-\tau) \times \lim_{N \to \infty} \frac{1}{N} \sum_{i=1}^N \ex{ \left( \bm{x}_{i1} - \varphi_i^{-1} \bm{E}_i \right) \left( \bm{x}_{i1} - \varphi_i^{-1} \bm{E}_i \right)' }
  \end{equation*}
  is positive definite.
\end{assumptionB}

Then we have the following result:
\begin{theorem} \label{thm:AN}
  Under Assumptions \ref{assume:data}-\ref{assume:Avar}, if $N^2 (\log N)^3 / T \to 0$ and $\lambda_T = o_p(T^{1/2} (\log N)^{1/2})$ as $N, T \rightarrow \infty$, then
  \begin{equation*}
    \sqrt{NT} (\hat{\bm{\beta}} - \bm{\beta}_0) \indist \mathcal{N}(\bm{0}, \bm{\Omega}),
  \end{equation*}
  where $\bm{\Omega} = \bm{D}^{-1} \bm{V} \bm{D}^{-1}$.
\end{theorem}

\begin{remark}
As with Condition G in Theorem 3.2 in \citet{GU2019}, Theorem \ref{thm:AN} provides a selection rule for candidate values of the tuning parameters that are justified by theory. The limiting distribution for this estimator matches that of the conventional fixed effects estimator derived in \citet{kK12} because the tuning parameter $\lambda_T$ diverges at a slow rate.
\end{remark}

\begin{remark}
Because the goal of the shrinkage estimator here is not variable selection but regularization of the estimated $\hat{\alpha}_i$, the rate of growth of $\lambda_T$ is different than what would usually be used in high-dimensional models (\citet[p. 86]{belloni2011}, \citet[eq. 2.3]{SLee2018}, and \citet[Theorem 3.2]{Wang2019}). This difference in stochastic order is because the individual effects $\{\alpha_{i0}\}_i$ are not assumed sparse and this condition on $\lambda_T$ is needed for consistency in models with regressors correlated with individual latent effects.
Moreover, perhaps not surprisingly, the rates derived for linear models \citep[see, e.g.,][]{kock2013,KOCK2016} are also different to the rate required for establishing the asymptotic normality of the quantile estimator.
\end{remark}

The wild residual bootstrap procedure is consistent as an estimator of the asymptotic distribution of $\hat{\bm{\beta}}$, as the next theorem shows.
\begin{theorem} \label{thm:boot}
Under Assumptions \ref{C1}-\ref{C3} and the conditions of Theorem~\ref{thm:AN},
  \begin{equation*}
    \sup_{b \in \RR^p} \left| \prob{\sqrt{NT}( \bm{\beta}^* - \hat{\bm{\beta}}) \leq b | \bm{S}} - \prob{\sqrt{NT}( \hat{\bm{\beta}} - \bm{\beta}_0) \leq b} \right| \inpr 0
  \end{equation*}
  where $\bm{S}$ denotes the observed sample and $\bm{\beta}^\ast$ denotes the slope estimator defined by~\eqref{penboot}.
\end{theorem}

\begin{remark}
By setting $\lambda_T = 0$, Theorem~\ref{thm:boot} also implies consistency of the wild residual bootstrap for the unpenalized estimator with individual effects and i.i.d. errors.
\end{remark}

\begin{remark} \label{rk:bootlmda}
The results allow for a data-dependent $\lambda_T$ but they do not allow selecting the tuning parameter at each bootstrap repetition. While theoretical developments are out of the scope of this paper, we investigated if this idea leads to improvements in the finite sample performance of the estimator. We did not find significant changes relative to the results presented in Section 4, although the computational cost of the procedure is higher. 
\end{remark}

Theorem~\ref{thm:boot} only shows consistency of the bootstrap distribution estimator.  Theorem~\ref{thm:var} ahead shows that the bootstrap covariance matrix, defined as
\begin{equation*}
  \bm{\Omega}^* = \exs{ NT \left( \bm{\beta}^* - \hat{\bm{\beta}} \right) \left( \bm{\beta}^* - \hat{\bm{\beta}} \right)' },
\end{equation*}
may be used to estimate the covariance of $\sqrt{NT}(\hat{\bm{\beta}} - \bm{\beta}_0)$.  In practice, one simply uses the sample covariance of all the bootstrap repetitions, increasing the number of repetitions to bring the sample average as close as desired to the bootstrap expectation.  Variance estimation using the bootstrap was formally investigated for quantile regression with clustered data in \citet{Andreas2017}, but the model in this paper is complicated by the diverging number of individual effects as $N \rightarrow \infty$ and the penalty term in~\eqref{penboot}.

\begin{theorem} \label{thm:var}
  Under Assumptions \ref{C1}-\ref{C3} and the conditions of Theorem~\ref{thm:AN}, if $\bm{\theta}_i$ for $1 \leq i \leq N$ lie in a compact set and
  \begin{equation*}
    \sup_{N,T} \ex{| \sqrt{N} \lambda_T / \sqrt{T} |^q} < \infty
  \end{equation*}
  for $q > 2$, then
  \begin{equation*}
    \| \bm{\Omega}^* - \bm{\Omega} \| \stackrel{p^\ast}{\longrightarrow} 0.
  \end{equation*}
\end{theorem}
The assumptions that are required for Theorem~\ref{thm:var} are slightly stronger than those used in Theorem~\ref{thm:boot}.  The requirement on $\lambda_T$ is due to its presence in asymptotic expansions leading to the Bahadur representation of $\bm{\beta}^\ast$ and is similar to the moment requirement made on the covariates in \citet{Andreas2017}.  The compactness assumption must be made to ensure that expansions used in the asymptotic approximation are uniformly bounded.

\section{Simulation Study}
In this section, we report the results of several simulation experiments designed to evaluate the performance of the method in finite samples. We consider a data generating process similar to the ones considered in Koenker (2004) and
Kato, Galvao and Montes-Rojas (2012). The dependent variable is $y_{it} = \alpha_i + x_{it} + (1 + \zeta x_{it}) u_{it}$, where $x_{it} = 0.5 \alpha_i + z_i + \epsilon_{it}$, and $z_i$
and $\epsilon_{it}$ are i.i.d. random variables distributed as $\chi^2$ with 3 degrees of freedom ($\chi_3^2$). The corresponding quantile regression function is
$Q_y (\tau | x_{it}) = \alpha_{0i} + \beta_0 x_{it}$, where $\alpha_{0i} = \alpha_i + F_u(\tau)^{-1}$, $\beta_0 = 1 + \zeta F_u(\tau)^{-1}$, and $F_u(\cdot)$ denotes the distribution of the error term, $u_{it}$.

We generate data from several variations of the basic model. In one variant of the model, $\alpha_i$ is an i.i.d. Gaussian random variable.
In another, we generate $\alpha_i = i/N$ for $1 \leq i \leq N$ as in Galvao, Gu, and Volgushev (2020). We use $\zeta \in \{0,0.5\}$, and thus, $\beta_0 = 1$ in the location shift version of the model and $\beta_0 = 1 + 0.5 F_u(\tau)^{-1}$ in the location-scale shift case. Lastly, we consider three
different distributions for the error term. We assume that $u_{it}$ is
distributed as $\mathcal{N}(0,1)$, a $t$ distribution with 3 degrees of
freedom ($t_3$), or $\chi_3^2$. 

\begin{singlespace}
\begin{table}
\begin{center}\footnotesize
\begin{tabular}{c c c c c c c c c c c c c c} \hline
 &  & \multicolumn{6}{c}{Quantile 0.5} & \multicolumn{6}{c}{Quantile 0.75} \\
&  &	\multicolumn{3}{c}{Method:} &	\multicolumn{3}{c}{Method:} &	\multicolumn{3}{c}{Method:}	&	\multicolumn{3}{c}{Method:} \\ \cmidrule(lr){3-5} \cmidrule(lr){6-8} \cmidrule(lr){9-11} \cmidrule(lr){12-14}
$N$	&	$T$  &	CS    &	\multicolumn{2}{c}{WB}	        &	CS &	\multicolumn{2}{c}{WB}  &	CS &	\multicolumn{2}{c}{WB} &	CS &	\multicolumn{2}{c}{WB} \\ \cmidrule(lr){4-5} \cmidrule(lr){7-8} \cmidrule(lr){10-11} \cmidrule(lr){13-14}
    &	     &	  PQR    &	PQR &  FE  & PQR	&	PQR &  FE &	PQR &	PQR &  FE &	PQR &	PQR &  FE \\  \hline 
\multicolumn{2}{c}{} & \multicolumn{12}{c}{Location shift model ($\zeta=0$) and $u \sim \mathcal{N}(0,1)$}  \\ 
 &  & \multicolumn{3}{c}{$\alpha_i \sim \mathcal{N}(0,1)$} & \multicolumn{3}{c}{$\alpha_i = i/N$}  & \multicolumn{3}{c}{$\alpha_i \sim \mathcal{N}(0,1)$} & \multicolumn{3}{c}{$\alpha_i = i/N$} \\ \hline
100	&	5	  &	0.697	&	0.902	&	0.905	&	0.675	&	0.909	&	0.913	&	0.702	&	0.854	&	0.854	&	0.640	&	0.848	&	0.859	\\
100	&	10	&	0.720	&	0.908	&	0.868	&	0.683	&	0.915	&	0.876	&	0.737	&	0.887	&	0.900	&	0.683	&	0.885	&	0.903	\\
200	&	5		&	0.677	&	0.923	&	0.925	&	0.670	&	0.916	&	0.920	&	0.668	&	0.862	&	0.861	&	0.641	&	0.873	&	0.877	\\
200	&	10	&	0.650	&	0.925	&	0.857	&	0.700	&	0.927	&	0.870	&	0.662	&	0.898	&	0.909	&	0.691	&	0.897	&	0.915	\\ \hline
25	&	50	&	0.886	&	0.908	&	0.904	&	0.779	&	0.887	&	0.882	&	0.857	&	0.881	&	0.880	&	0.758	&	0.885	&	0.888	\\
25	&	100	&	0.903	&	0.910	&	0.911	&	0.811	&	0.902	&	0.905	&	0.908	&	0.898	&	0.901	&	0.816	&	0.892	&	0.898	\\
50	&	50	&	0.833	&	0.905	&	0.903	&	0.769	&	0.884	&	0.880	&	0.831	&	0.893	&	0.888	&	0.768	&	0.889	&	0.892	\\
50	&	100	&	0.847	&	0.900	&	0.895	&	0.831	&	0.903	&	0.900	&	0.854	&	0.902	&	0.897	&	0.827	&	0.898	&	0.902	\\ \hline
\multicolumn{2}{c}{} & \multicolumn{12}{c}{Location shift model ($\zeta=0$) and $u \sim t_3$}  \\ 
 &  & \multicolumn{3}{c}{$\alpha_i \sim \mathcal{N}(0,1)$} & \multicolumn{3}{c}{$\alpha_i = i/N$}  & \multicolumn{3}{c}{$\alpha_i \sim \mathcal{N}(0,1)$} & \multicolumn{3}{c}{$\alpha_i = i/N$} \\ \hline
100	&	5	  &	0.710	&	0.906	&	0.911	&	0.674	&	0.902	&	0.912	&	0.701	&	0.828	&	0.833	&	0.623	&	0.819	&	0.840	\\
100	&	10	&	0.713	&	0.923	&	0.880	&	0.660	&	0.919	&	0.881	&	0.734	&	0.881	&	0.893	&	0.680	&	0.864	&	0.880	\\
200	&	5	  &	0.681	&	0.932	&	0.936	&	0.645	&	0.907	&	0.927	&	0.669	&	0.841	&	0.852	&	0.573	&	0.816	&	0.852	\\
200	&	10	&	0.650	&	0.922	&	0.834	&	0.661	&	0.931	&	0.845	&	0.641	&	0.881	&	0.892	&	0.678	&	0.859	&	0.889	\\ \hline
25	&	50	&	0.887	&	0.921	&	0.916	&	0.784	&	0.906	&	0.906	&	0.852	&	0.887	&	0.886	&	0.738	&	0.881	&	0.881	\\
25	&	100	&	0.905	&	0.901	&	0.901	&	0.819	&	0.892	&	0.891	&	0.870	&	0.883	&	0.891	&	0.801	&	0.891	&	0.900	\\
50	&	50	&	0.850	&	0.898	&	0.895	&	0.759	&	0.884	&	0.884	&	0.839	&	0.886	&	0.889	&	0.761	&	0.900	&	0.895	\\
50	&	100	&	0.848	&	0.886	&	0.887	&	0.816	&	0.899	&	0.892	&	0.837	&	0.885	&	0.890	&	0.770	&	0.863	&	0.875	\\ \hline
\multicolumn{2}{c}{} & \multicolumn{12}{c}{Location shift model ($\zeta=0$) and $u \sim \chi_3^2$} \\
 &  & \multicolumn{3}{c}{$\alpha_i \sim \mathcal{N}(0,1)$} & \multicolumn{3}{c}{$\alpha_i = i/N$}  & \multicolumn{3}{c}{$\alpha_i \sim \mathcal{N}(0,1)$} & \multicolumn{3}{c}{$\alpha_i = i/N$} \\ \hline
100	&	5	  &	0.730	&	0.906	&	0.912	&	0.633	&	0.897	&	0.907	&	0.690	&	0.728	&	0.745	&	0.589	&	0.716	&	0.716	\\
100	&	10	&	0.673	&	0.894	&	0.871	&	0.647	&	0.870	&	0.855	&	0.718	&	0.828	&	0.844	&	0.707	&	0.827	&	0.838	\\
200	&	5	  &	0.708	&	0.920	&	0.927	&	0.657	&	0.915	&	0.910	&	0.652	&	0.745	&	0.753	&	0.602	&	0.763	&	0.763	\\
200	&	10	&	0.686	&	0.901	&	0.860	&	0.642	&	0.880	&	0.838	&	0.703	&	0.833	&	0.853	&	0.703	&	0.818	&	0.831	\\  \hline
25	&	50	&	0.791	&	0.882	&	0.883	&	0.716	&	0.891	&	0.896	&	0.749	&	0.839	&	0.839	&	0.707	&	0.861	&	0.862	\\
25	&	100	&	0.845	&	0.887	&	0.890	&	0.749	&	0.873	&	0.874	&	0.781	&	0.871	&	0.879	&	0.683	&	0.856	&	0.860	\\
50	&	50	&	0.768	&	0.869	&	0.872	&	0.726	&	0.898	&	0.897	&	0.758	&	0.862	&	0.862	&	0.727	&	0.876	&	0.877	\\
50	&	100	&	0.798	&	0.879	&	0.880	&	0.735	&	0.891	&	0.892	&	0.770	&	0.861	&	0.867	&	0.710	&	0.864	&	0.877	\\  \hline
\end{tabular}
\vspace{3mm}
\end{center}
\caption{\emph{Empirical coverage probabilities of the bootstrap confidence interval for a nominal 90\% level. A location shift model is considered. CS denotes cross-sectional pairs bootstrap, WB denotes wild bootstrap, PQR denotes the penalized estimator, and FE is the unpenalized fixed effects estimator.}}
\label{mc.table1}
\end{table}
\end{singlespace}

\begin{singlespace}
\begin{table}
\begin{center}\footnotesize
\begin{tabular}{c c c c c c c c c c c c c c} \hline
 &  & \multicolumn{6}{c}{Quantile 0.5} & \multicolumn{6}{c}{Quantile 0.75} \\
&  &	\multicolumn{3}{c}{Method:} &	\multicolumn{3}{c}{Method:} &	\multicolumn{3}{c}{Method:}	&	\multicolumn{3}{c}{Method:} \\ \cmidrule(lr){3-5} \cmidrule(lr){6-8} \cmidrule(lr){9-11} \cmidrule(lr){12-14}
$N$	&	$T$  &	CS    &	\multicolumn{2}{c}{WB}	        &	CS &	\multicolumn{2}{c}{WB}  &	CS &	\multicolumn{2}{c}{WB} &	CS &	\multicolumn{2}{c}{WB} \\ \cmidrule(lr){4-5} \cmidrule(lr){7-8} \cmidrule(lr){10-11} \cmidrule(lr){13-14}
    &	     &	  PQR    &	PQR &  FE  & PQR	&	PQR &  FE &	PQR &	PQR &  FE &	PQR &	PQR &  FE \\  \hline 
\multicolumn{2}{c}{} & \multicolumn{12}{c}{Location-scale shift model ($\zeta=0.5$) and $u \sim \mathcal{N}(0,1)$}  \\ 
&  & \multicolumn{3}{c}{$\alpha_i \sim \mathcal{N}(0,1)$} & \multicolumn{3}{c}{$\alpha_i = i/N$}  & \multicolumn{3}{c}{$\alpha_i \sim \mathcal{N}(0,1)$} & \multicolumn{3}{c}{$\alpha_i = i/N$} \\ \hline
100	&	5	  &	0.688	&	0.861	&	0.881	&	0.699	&	0.884	&	0.893	&	0.592	&	0.798	&	0.824	&	0.690	&	0.819	&	0.827	\\
100	&	10	&	0.619	&	0.899	&	0.865	&	0.657	&	0.896	&	0.868	&	0.611	&	0.864	&	0.867	&	0.652	&	0.877	&	0.887	\\
200	&	5	  &	0.660	&	0.871	&	0.892	&	0.674	&	0.868	&	0.892	&	0.540	&	0.825	&	0.823	&	0.605	&	0.850	&	0.848	\\
200	&	10	&	0.650	&	0.913	&	0.852	&	0.651	&	0.904	&	0.860	&	0.567	&	0.876	&	0.880	&	0.627	&	0.868	&	0.884	\\ \hline
25	&	50	&	0.698	&	0.898	&	0.893	&	0.676	&	0.900	&	0.899	&	0.696	&	0.881	&	0.882	&	0.672	&	0.866	&	0.865	\\
25	&	100	&	0.762	&	0.906	&	0.906	&	0.678	&	0.894	&	0.896	&	0.749	&	0.894	&	0.897	&	0.678	&	0.892	&	0.893	\\
50	&	50	&	0.685	&	0.896	&	0.892	&	0.678	&	0.888	&	0.883	&	0.679	&	0.888	&	0.889	&	0.669	&	0.886	&	0.888	\\
50	&	100	&	0.720	&	0.900	&	0.902	&	0.674	&	0.905	&	0.903	&	0.697	&	0.879	&	0.881	&	0.677	&	0.901	&	0.902	\\ \hline
\multicolumn{2}{c}{} & \multicolumn{12}{c}{Location-scale shift model ($\zeta=0.5$) and $u \sim t_3$}  \\
 &  & \multicolumn{3}{c}{$\alpha_i \sim \mathcal{N}(0,1)$} & \multicolumn{3}{c}{$\alpha_i = i/N$}  & \multicolumn{3}{c}{$\alpha_i \sim \mathcal{N}(0,1)$} & \multicolumn{3}{c}{$\alpha_i = i/N$} \\ \hline
100	&	5	  &	0.723	&	0.857	&	0.886	&	0.758	&	0.866	&	0.890	&	0.620	&	0.751	&	0.769	&	0.721	&	0.769	&	0.781	\\
100	&	10	&	0.664	&	0.902	&	0.882	&	0.676	&	0.895	&	0.873	&	0.638	&	0.862	&	0.874	&	0.672	&	0.855	&	0.857	\\
200	&	5	  &	0.754	&	0.875	&	0.910	&	0.722	&	0.873	&	0.909	&	0.513	&	0.767	&	0.775	&	0.707	&	0.780	&	0.795	\\
200	&	10	&	0.650	&	0.899	&	0.857	&	0.649	&	0.910	&	0.843	&	0.554	&	0.816	&	0.835	&	0.634	&	0.822	&	0.824	\\   \hline
25	&	50	&	0.742	&	0.910	&	0.910	&	0.682	&	0.914	&	0.916	&	0.694	&	0.874	&	0.875	&	0.672	&	0.869	&	0.873	\\
25	&	100	&	0.750	&	0.885	&	0.884	&	0.678	&	0.896	&	0.897	&	0.711	&	0.874	&	0.876	&	0.688	&	0.885	&	0.888	\\
50	&	50	&	0.683	&	0.891	&	0.888	&	0.675	&	0.880	&	0.877	&	0.690	&	0.881	&	0.879	&	0.678	&	0.885	&	0.891	\\
50	&	100	&	0.714	&	0.886	&	0.885	&	0.661	&	0.883	&	0.883	&	0.691	&	0.879	&	0.889	&	0.668	&	0.864	&	0.872	\\   \hline
\multicolumn{2}{c}{} & \multicolumn{12}{c}{Location-scale model ($\zeta=0.5$) and $u \sim \chi_3^2$} \\ 
 &  & \multicolumn{3}{c}{$\alpha_i \sim \mathcal{N}(0,1)$} & \multicolumn{3}{c}{$\alpha_i = i/N$}  & \multicolumn{3}{c}{$\alpha_i \sim \mathcal{N}(0,1)$} & \multicolumn{3}{c}{$\alpha_i = i/N$} \\ \hline
100	&	5	  &	0.735	&	0.840	&	0.862	&	0.693	&	0.818	&	0.842	&	0.746	&	0.719	&	0.676	&	0.784	&	0.727	&	0.701	\\
100	&	10	&	0.666	&	0.868	&	0.859	&	0.645	&	0.849	&	0.837	&	0.641	&	0.783	&	0.790	&	0.679	&	0.774	&	0.762	\\
200	&	5	  &	0.714	&	0.831	&	0.856	&	0.685	&	0.839	&	0.842	&	0.722	&	0.750	&	0.669	&	0.770	&	0.731	&	0.612	\\
200	&	10	&	0.661	&	0.886	&	0.850	&	0.658	&	0.865	&	0.835	&	0.637	&	0.777	&	0.771	&	0.642	&	0.751	&	0.758	\\ \hline
25	&	50	&	0.662	&	0.850	&	0.859	&	0.663	&	0.880	&	0.884	&	0.613	&	0.847	&	0.843	&	0.653	&	0.848	&	0.846	\\
25	&	100	&	0.674	&	0.888	&	0.886	&	0.676	&	0.883	&	0.886	&	0.662	&	0.877	&	0.879	&	0.651	&	0.864	&	0.872	\\
50	&	50	&	0.633	&	0.863	&	0.869	&	0.685	&	0.896	&	0.899	&	0.661	&	0.852	&	0.856	&	0.672	&	0.866	&	0.868	\\
50	&	100	&	0.646	&	0.861	&	0.874	&	0.685	&	0.887	&	0.887	&	0.670	&	0.863	&	0.873	&	0.668	&	0.856	&	0.860	\\ \hline
\end{tabular}
\vspace{3mm}
\end{center}
\caption{\emph{Empirical coverage probabilities of the bootstrap confidence interval for a nominal 90\% level. A location-scale shift model is considered. CS denotes cross-sectional pairs bootstrap, WB denotes wild bootstrap, PQR denotes the penalized estimator, and FE is the unpenalized fixed effects estimator.}}
\label{mc.table2}
\end{table}
\end{singlespace}

\begin{singlespace}
\begin{table}
\begin{center}\footnotesize
\begin{tabular}{c c c c c c c c c c c c c c} \hline
 &  & \multicolumn{6}{c}{Quantile 0.5} & \multicolumn{6}{c}{Quantile 0.75} \\
&  &	\multicolumn{3}{c}{Method:} &	\multicolumn{3}{c}{Method:} &	\multicolumn{3}{c}{Method:}	&	\multicolumn{3}{c}{Method:} \\ \cmidrule(lr){3-5} \cmidrule(lr){6-8} \cmidrule(lr){9-11} \cmidrule(lr){12-14}
$N$	&	$T$  &	CS    &	\multicolumn{2}{c}{WB}	        &	CS &	\multicolumn{2}{c}{WB}  &	CS &	\multicolumn{2}{c}{WB} &	CS &	\multicolumn{2}{c}{WB} \\ \cmidrule(lr){4-5} \cmidrule(lr){7-8} \cmidrule(lr){10-11} \cmidrule(lr){13-14}
    &	     &	  PQR    &	PQR &  FE  & PQR	&	PQR &  FE &	PQR &	PQR &  FE &	PQR &	PQR &  FE \\  \hline 
\multicolumn{2}{c}{} & \multicolumn{12}{c}{Location shift model ($\zeta=0$) and $u \sim \mathcal{N}(0,1)$}  \\ 
 &  & \multicolumn{3}{c}{$\alpha_i \sim \mathcal{N}(0,1)$} & \multicolumn{3}{c}{$\alpha_i = i/N$}  & \multicolumn{3}{c}{$\alpha_i \sim \mathcal{N}(0,1)$} & \multicolumn{3}{c}{$\alpha_i = i/N$} \\ \hline
100	&	5	&	0.951	&	0.908	&	0.910	&	0.862	&	0.922	&	0.919	&	0.930	&	0.896	&	0.894	&	0.845	&	0.899	&	0.899	\\
100	&	10	&	0.971	&	0.917	&	0.862	&	0.864	&	0.913	&	0.870	&	0.969	&	0.909	&	0.913	&	0.861	&	0.910	&	0.922	\\
200	&	5	&	0.953	&	0.925	&	0.927	&	0.849	&	0.919	&	0.922	&	0.918	&	0.902	&	0.902	&	0.831	&	0.912	&	0.913	\\
200	&	10	&	0.974	&	0.933	&	0.853	&	0.868	&	0.925	&	0.868	&	0.975	&	0.910	&	0.920	&	0.876	&	0.904	&	0.924	\\  \hline
25	&	50	&	0.998	&	0.911	&	0.909	&	0.916	&	0.894	&	0.894	&	1.000	&	0.889	&	0.892	&	0.912	&	0.886	&	0.886	\\
25	&	100	&	1.000	&	0.910	&	0.910	&	0.972	&	0.908	&	0.907	&	1.000	&	0.901	&	0.904	&	0.944	&	0.902	&	0.904	\\
50	&	50	&	1.000	&	0.907	&	0.905	&	0.922	&	0.888	&	0.885	&	1.000	&	0.895	&	0.897	&	0.918	&	0.907	&	0.908	\\
50	&	100	&	1.000	&	0.908	&	0.904	&	0.973	&	0.904	&	0.903	&	1.000	&	0.900	&	0.900	&	0.965	&	0.913	&	0.907	\\ \hline
\multicolumn{2}{c}{} & \multicolumn{12}{c}{Location shift model ($\zeta=0$) and $u \sim t_3$}  \\ 
 &  & \multicolumn{3}{c}{$\alpha_i \sim \mathcal{N}(0,1)$} & \multicolumn{3}{c}{$\alpha_i = i/N$}  & \multicolumn{3}{c}{$\alpha_i \sim \mathcal{N}(0,1)$} & \multicolumn{3}{c}{$\alpha_i = i/N$} \\ \hline
100	&	5	&	0.945	&	0.922	&	0.918	&	0.860	&	0.920	&	0.916	&	0.917	&	0.887	&	0.885	&	0.814	&	0.865	&	0.901	\\
100	&	10	&	0.969	&	0.923	&	0.873	&	0.856	&	0.923	&	0.878	&	0.964	&	0.908	&	0.916	&	0.857	&	0.894	&	0.907	\\
200	&	5	&	0.946	&	0.940	&	0.939	&	0.826	&	0.913	&	0.931	&	0.911	&	0.898	&	0.901	&	0.774	&	0.841	&	0.903	\\
200	&	10	&	0.967	&	0.920	&	0.822	&	0.850	&	0.933	&	0.838	&	0.960	&	0.901	&	0.915	&	0.848	&	0.880	&	0.897	\\ \hline
25	&	50	&	0.997	&	0.926	&	0.921	&	0.924	&	0.912	&	0.906	&	0.989	&	0.890	&	0.889	&	0.901	&	0.890	&	0.889	\\
25	&	100	&	1.000	&	0.895	&	0.893	&	0.950	&	0.893	&	0.892	&	0.999	&	0.879	&	0.880	&	0.935	&	0.902	&	0.902	\\
50	&	50	&	0.998	&	0.904	&	0.899	&	0.900	&	0.887	&	0.881	&	0.999	&	0.895	&	0.897	&	0.910	&	0.906	&	0.909	\\
50	&	100	&	1.000	&	0.896	&	0.891	&	0.955	&	0.902	&	0.898	&	0.998	&	0.890	&	0.890	&	0.917	&	0.875	&	0.874	\\ \hline
\multicolumn{2}{c}{} & \multicolumn{12}{c}{Location shift model ($\zeta=0$) and $u \sim \chi_3^2$} \\
 &  & \multicolumn{3}{c}{$\alpha_i \sim \mathcal{N}(0,1)$} & \multicolumn{3}{c}{$\alpha_i = i/N$}  & \multicolumn{3}{c}{$\alpha_i \sim \mathcal{N}(0,1)$} & \multicolumn{3}{c}{$\alpha_i = i/N$} \\ \hline
100	&	5	&	0.897	&	0.917	&	0.918	&	0.832	&	0.892	&	0.916	&	0.857	&	0.805	&	0.824	&	0.791	&	0.750	&	0.800	\\
100	&	10	&	0.905	&	0.898	&	0.873	&	0.830	&	0.879	&	0.844	&	0.886	&	0.835	&	0.852	&	0.852	&	0.836	&	0.848	\\
200	&	5	&	0.897	&	0.914	&	0.922	&	0.824	&	0.893	&	0.917	&	0.828	&	0.784	&	0.820	&	0.798	&	0.784	&	0.820	\\
200	&	10	&	0.912	&	0.904	&	0.845	&	0.823	&	0.877	&	0.831	&	0.884	&	0.834	&	0.858	&	0.849	&	0.831	&	0.846	\\ \hline
25	&	50	&	0.968	&	0.881	&	0.883	&	0.864	&	0.894	&	0.896	&	0.904	&	0.851	&	0.852	&	0.846	&	0.868	&	0.870	\\
25	&	100	&	0.994	&	0.889	&	0.890	&	0.870	&	0.878	&	0.879	&	0.956	&	0.878	&	0.885	&	0.831	&	0.856	&	0.860	\\
50	&	50	&	0.965	&	0.878	&	0.874	&	0.871	&	0.895	&	0.894	&	0.916	&	0.865	&	0.864	&	0.859	&	0.876	&	0.878	\\
50	&	100	&	0.992	&	0.882	&	0.882	&	0.890	&	0.894	&	0.894	&	0.965	&	0.870	&	0.870	&	0.853	&	0.868	&	0.868	\\  \hline
\end{tabular}
\vspace{3mm}
\end{center}
\caption{\emph{Empirical coverage probabilities of the asymptotic Gaussian confidence interval for a nominal 90\% level. A location shift model is considered. CS denotes cross-sectional pairs bootstrap, WB denotes wild bootstrap, PQR denotes the penalized estimator, and FE is the unpenalized fixed effects estimator.}}
\label{mc.table3}
\end{table}
\end{singlespace}

\begin{singlespace}
\begin{table}
\begin{center}\footnotesize
\begin{tabular}{c c c c c c c c c c c c c c} \hline
 &  & \multicolumn{6}{c}{Quantile 0.5} & \multicolumn{6}{c}{Quantile 0.75} \\
&  &	\multicolumn{3}{c}{Method:} &	\multicolumn{3}{c}{Method:} &	\multicolumn{3}{c}{Method:}	&	\multicolumn{3}{c}{Method:} \\ \cmidrule(lr){3-5} \cmidrule(lr){6-8} \cmidrule(lr){9-11} \cmidrule(lr){12-14}
$N$	&	$T$  &	CS    &	\multicolumn{2}{c}{WB}	        &	CS &	\multicolumn{2}{c}{WB}  &	CS &	\multicolumn{2}{c}{WB} &	CS &	\multicolumn{2}{c}{WB} \\ \cmidrule(lr){4-5} \cmidrule(lr){7-8} \cmidrule(lr){10-11} \cmidrule(lr){13-14}
    &	     &	  PQR    &	PQR &  FE  & PQR	&	PQR &  FE &	PQR &	PQR &  FE &	PQR &	PQR &  FE \\  \hline 
\multicolumn{2}{c}{} & \multicolumn{12}{c}{Location-scale shift model ($\zeta=0.5$) and $u \sim \mathcal{N}(0,1)$}  \\ 
&  & \multicolumn{3}{c}{$\alpha_i \sim \mathcal{N}(0,1)$} & \multicolumn{3}{c}{$\alpha_i = i/N$}  & \multicolumn{3}{c}{$\alpha_i \sim \mathcal{N}(0,1)$} & \multicolumn{3}{c}{$\alpha_i = i/N$} \\ \hline
100	&	5	&	0.851	&	0.893	&	0.886	&	0.863	&	0.909	&	0.905	&	0.812	&	0.841	&	0.837	&	0.830	&	0.876	&	0.843	\\
100	&	10	&	0.833	&	0.912	&	0.865	&	0.821	&	0.910	&	0.860	&	0.819	&	0.873	&	0.871	&	0.825	&	0.887	&	0.877	\\
200	&	5	&	0.838	&	0.899	&	0.896	&	0.834	&	0.906	&	0.895	&	0.784	&	0.838	&	0.831	&	0.810	&	0.884	&	0.837	\\
200	&	10	&	0.833	&	0.917	&	0.845	&	0.817	&	0.913	&	0.847	&	0.812	&	0.873	&	0.870	&	0.801	&	0.884	&	0.879	\\  \hline
25	&	50	&	0.897	&	0.903	&	0.903	&	0.837	&	0.909	&	0.905	&	0.883	&	0.894	&	0.893	&	0.813	&	0.873	&	0.874	\\
25	&	100	&	0.938	&	0.912	&	0.909	&	0.827	&	0.902	&	0.901	&	0.922	&	0.901	&	0.900	&	0.824	&	0.898	&	0.898	\\
50	&	50	&	0.887	&	0.901	&	0.897	&	0.801	&	0.896	&	0.886	&	0.882	&	0.890	&	0.891	&	0.821	&	0.888	&	0.890	\\
50	&	100	&	0.943	&	0.909	&	0.906	&	0.824	&	0.904	&	0.900	&	0.919	&	0.891	&	0.888	&	0.838	&	0.907	&	0.907	\\ \hline
\multicolumn{2}{c}{} & \multicolumn{12}{c}{Location-scale shift model ($\zeta=0.5$) and $u \sim t_3$}  \\
 &  & \multicolumn{3}{c}{$\alpha_i \sim \mathcal{N}(0,1)$} & \multicolumn{3}{c}{$\alpha_i = i/N$}  & \multicolumn{3}{c}{$\alpha_i \sim \mathcal{N}(0,1)$} & \multicolumn{3}{c}{$\alpha_i = i/N$} \\ \hline
100	&	5	&	0.864	&	0.915	&	0.898	&	0.882	&	0.918	&	0.894	&	0.816	&	0.816	&	0.789	&	0.857	&	0.871	&	0.793	\\
100	&	10	&	0.845	&	0.912	&	0.864	&	0.845	&	0.910	&	0.870	&	0.847	&	0.897	&	0.873	&	0.832	&	0.874	&	0.854	\\
200	&	5	&	0.884	&	0.919	&	0.921	&	0.855	&	0.910	&	0.916	&	0.778	&	0.803	&	0.756	&	0.836	&	0.860	&	0.792	\\
200	&	10	&	0.830	&	0.913	&	0.842	&	0.821	&	0.919	&	0.827	&	0.786	&	0.842	&	0.817	&	0.793	&	0.849	&	0.817	\\  \hline
25	&	50	&	0.894	&	0.919	&	0.912	&	0.851	&	0.916	&	0.914	&	0.856	&	0.877	&	0.878	&	0.814	&	0.883	&	0.881	\\
25	&	100	&	0.902	&	0.887	&	0.883	&	0.826	&	0.903	&	0.903	&	0.874	&	0.886	&	0.886	&	0.831	&	0.890	&	0.893	\\
50	&	50	&	0.882	&	0.894	&	0.889	&	0.791	&	0.885	&	0.873	&	0.856	&	0.882	&	0.884	&	0.820	&	0.899	&	0.897	\\
50	&	100	&	0.908	&	0.886	&	0.882	&	0.815	&	0.889	&	0.885	&	0.893	&	0.892	&	0.892	&	0.805	&	0.879	&	0.878	\\   \hline
\multicolumn{2}{c}{} & \multicolumn{12}{c}{Location-scale model ($\zeta=0.5$) and $u \sim \chi_3^2$} \\ 
 &  & \multicolumn{3}{c}{$\alpha_i \sim \mathcal{N}(0,1)$} & \multicolumn{3}{c}{$\alpha_i = i/N$}  & \multicolumn{3}{c}{$\alpha_i \sim \mathcal{N}(0,1)$} & \multicolumn{3}{c}{$\alpha_i = i/N$} \\ \hline
100	&	5	&	0.857	&	0.887	&	0.868	&	0.829	&	0.857	&	0.847	&	0.881	&	0.818	&	0.672	&	0.896	&	0.843	&	0.714	\\
100	&	10	&	0.837	&	0.886	&	0.847	&	0.822	&	0.871	&	0.832	&	0.821	&	0.830	&	0.786	&	0.822	&	0.830	&	0.779	\\
200	&	5	&	0.833	&	0.881	&	0.858	&	0.823	&	0.860	&	0.844	&	0.875	&	0.850	&	0.641	&	0.884	&	0.842	&	0.584	\\
200	&	10	&	0.845	&	0.903	&	0.835	&	0.814	&	0.883	&	0.821	&	0.815	&	0.823	&	0.778	&	0.799	&	0.818	&	0.758	\\ \hline
25	&	50	&	0.803	&	0.858	&	0.858	&	0.815	&	0.890	&	0.886	&	0.804	&	0.851	&	0.850	&	0.806	&	0.859	&	0.857	\\
25	&	100	&	0.847	&	0.893	&	0.900	&	0.819	&	0.885	&	0.885	&	0.829	&	0.881	&	0.884	&	0.808	&	0.863	&	0.868	\\
50	&	50	&	0.819	&	0.874	&	0.872	&	0.836	&	0.907	&	0.907	&	0.808	&	0.859	&	0.861	&	0.826	&	0.878	&	0.876	\\
50	&	100	&	0.827	&	0.876	&	0.877	&	0.829	&	0.895	&	0.892	&	0.825	&	0.875	&	0.878	&	0.811	&	0.866	&	0.868	\\ \hline
\end{tabular}
\vspace{3mm}
\end{center}
\caption{\emph{Empirical coverage probabilities of the asymptotic Gaussian confidence interval for a nominal 90\% level. A location-scale shift model is considered. CS denotes cross-sectional pairs bootstrap, WB denotes wild bootstrap, PQR denotes the penalized estimator, and FE is the unpenalized fixed effects estimator.}}
\label{mc.table4}
\end{table}
\end{singlespace}

Tables \ref{mc.table1}, \ref{mc.table2}, \ref{mc.table3}, and \ref{mc.table4} present coverage probabilities for a nominal 90\% confidence interval for the slope parameter $\beta_{0}$. We present coverage probabilities using the empirical distribution of the bootstrap estimator (Tables \ref{mc.table1} and \ref{mc.table2}), as well as coverage probabilities of the asymptotic Gaussian confidence interval (Tables \ref{mc.table3} and \ref{mc.table4}). In the latter case, the coverage is constructed using the standard error of the corresponding bootstrap procedure. Tables \ref{mc.table1} and \ref{mc.table3} present results for the location shift model ($\zeta=0$), while Tables \ref{mc.table2} and \ref{mc.table4} present results for the location-scale shift model ($\zeta=0.5$). The tables present results for $\tau \in
\{0.50,0.75\}$, based on different combinations of $N\in \{25,50,100,200\}$ and $T\in \{5,10,50,100\}$. The number of
bootstrap repetitions is set to 400, and the results are obtained by using 1000 random samples.

The tables show results for two bootstrap methods. The cross-sectional pairs bootstrap (CS) samples over $i$ with replacement, keeping the entire block of
time series observations. The
wild bootstrap (WB) is implemented as discussed in Section \ref{subsec:wild}. We first obtain residuals $\hat{u}_{it}$ using the penalized quantile regression \eqref{penboot}, which is labeled `PQR' in the tables. The tuning parameter is obtained as $\hat{\lambda}_T = b_T \tilde{\lambda}$ where $\tilde{\lambda}$ is obtained by cross-validation and $b_T=0.5 T^{-\nu}$ controls the bias. The selection of $\nu = 1$ performed well in the simulations and it is consistent with Theorem \ref{thm:consistent}. As in the case of the wild bootstrap estimator proposed by Feng, He, and Hu (2011), 
a finite sample correction is recommended. We adopt an adjustment following closely the {\tt R} package {\tt quantreg} by Koenker
(2021)\nocite{rK21}. In our case, we adjust the residuals with the influence function and sign function following the Bahadur representation of the estimator derived in Theorem \ref{thm:AN}. Then, we generate $u^\ast_{it} = w_{it} | \hat{u}_{it}
|$, where $w_{it}$ is an i.i.d. random variable distributed as a two-point
distribution with probabilities $\tau$ and $1-\tau$ at $w_{it} = -2 \tau$ and
$w_{it} = 2 (1-\tau)$. Lastly, we generate the dependent variable as
$y_{it}^\ast = \hat{\alpha}_i + \hat{\beta} x_{it} + u^\ast_{it}$. The performance of the estimator \eqref{penbootthr} was similar and the results are not presented here to save space. Finally, we include the estimator \eqref{penboot} defined for $\lambda_T = 0$ and it is labeled `FE'. 

Following the result presented in Theorem \ref{thm:boot}, the coverage probabilities in Table \ref{mc.table1} are obtained considering the quantiles of the empirical distribution of $\sqrt{NT} (\beta^\ast  - \hat{\beta})$. As can be seen in the upper block of Table \ref{mc.table1}, the performance of the WB bootstrap estimators are excellent, and they are in general around the specified coverage probability. Furthermore, performance improves with $T$, and tends to be similar for both $0.5$ and $0.75$ quantiles. On the other hand, the performance of the CS estimator is poor, with estimates not approaching to specified nominal values. In the lower parts of the table, we present the performance of the estimators for different distributions $F_u$. The WB method continues to perform better than CS, and, as expected, the estimation of the higher quantile is more challenging in the $\chi_3^2$ case. In all the variations of the model considered in the table, the WB estimator performs much better than the CS estimator.

The results for the location-scale shift model presented in Table \ref{mc.table2} are similar. We continue to see that the WB bootstrap performs better than the CS method. This conclusion holds when we consider asymptotic Gaussian confidence intervals obtained using bootstrap standard errors $\mbox{se}(\beta^\ast)$ (see Tables \ref{mc.table3} and \ref{mc.table4}). Moreover, the tables confirm two results that were expected. First, as $T$ increases relative to $N$, the coverage of the WB improves. Second, the performance of WB in the case of $\lambda_T=0$ reveals that, in general, the procedure proposed in this paper is valid for approximating the distribution of the fixed effects estimator. 

\begin{figure}
\begin{center}
\centerline{\includegraphics[width=.75\textwidth]{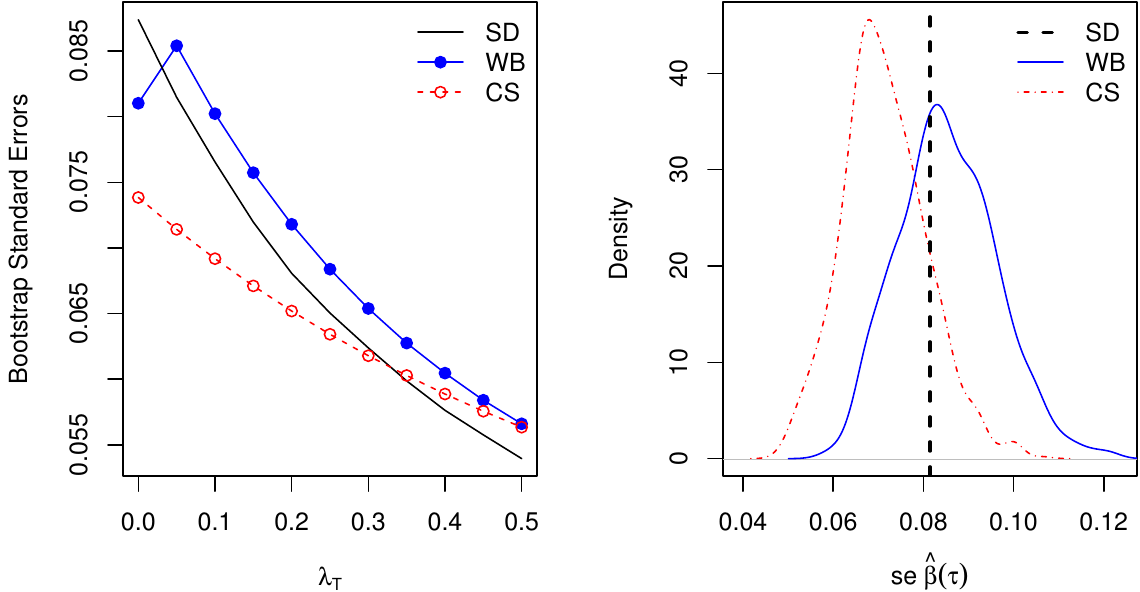}}
\caption{\emph{The performance of the bootstrap estimators as $\lambda_T$ increases. SD denotes standard deviation of the penalized estimator, CS denotes cross-sectional pair bootstrap, and WB denotes wild bootstrap estimator \eqref{penboot}.}}
\label{mc.figure0}
\end{center}
\end{figure}

We finish the section by briefly documenting the relative performance of the estimators of the standard errors. We generate data from a location-scale shift model ($\zeta=0.5$) when the error term $u_{it} \sim \mathcal{N}(0,1)$ and $\alpha_i \sim \mathcal{N}(0,1)$, by setting $N=100$, $T=10$, and $\tau = 0.5$. 
The left panel of Figure \ref{mc.figure0} shows CS and WB bootstrap estimates of the standard error, $\mbox{se}(\beta^\ast)$, and the standard deviation of the penalized estimator, $\mbox{sd}(\hat{\beta})$. The figure shows the advantage of the penalized estimator relative to the fixed effects estimator, as the standard deviation of the estimator is decreasing as $\lambda_T$ increases. We also see that the WB procedure performs better than CS when $\lambda_T$ is relatively small, and the performance of the WB estimator does not seem to change over the degree of shrinkage of the individual effects, as the bias appears to be roughly constant over $\lambda_T$. Using the right panel in Figure \ref{mc.figure0}, we explore further the difference in performance between approaches. The empirical distribution obtained by the CS procedure is not centered at the true value, and the distribution of the standard error of the WB is centered at $\mbox{se}(\hat{\beta}) = 0.081$ (with $\lambda_T = 0.05$). 

\section{Extensions}

In this section, we investigate the consistency of the wild bootstrap under different conditions. First, we extend the results of Theorems \ref{thm:consistent} and \ref{thm:AN} to allow for dependent data, and then we focus on the consistency of the wild bootstrap. In such case, we use the following assumptions:

\begin{assumptionC} \label{assume:stationary}
  The processes $\{(y_{it},\bm{x}_{it}), t \in 1, 2, \ldots\}$ are strictly stationary for each $i$ and $\beta$-mixing, and independent across $i$. Letting $\{\beta_i(j)\}_j$ denote the $\beta$-mixing coefficients, assume that there are constants $0 < a < 1$ and $B > 0$ such that $\sup_i \beta_i(j) \leq B a^j$ for all $j \geq 1$. 
\end{assumptionC}

\begin{assumptionC} \label{assume:density}
  The random vector $(u_{it}, u_{it+j})$ has a density conditional on $(\bm{x}_{it}, \bm{x}_{it+j})$ that is bounded uniformly over $i$ and $j \geq 1$.
\end{assumptionC}

\begin{assumptionC} \label{assume:Avar_dependent}
  Assume that the matrix $\bm{D}_N$ as defined in Assumption~\ref{assume:Avar} exists and is positive definite for all $N$ under Assumptions \ref{assume:stationary} and~\ref{assume:density} and that $\bm{D} = \lim_{N \rightarrow \infty} \bm{D}_N$ exists and is positive definite.  Also assume that
  \begin{equation*}
    \tilde{\bm{V}} = \lim_{N,T \to \infty} \frac{1}{NT} \sum_{i=1}^N \Var \left( \sum_{t=1}^T (\tau - I(y_{it} < \bm{x}_{it}'\bm{\beta}_0 + \alpha_{i0})) \left( \bm{x}_{it} - \varphi_i^{-1} \bm{E}_i \right) \right) 
  \end{equation*}
  is positive definite.
\end{assumptionC}

Theorem~\ref{thm:AN_dependent} presents both consistency and asymptotic normality results for the estimator with dependent error terms.
\begin{theorem} \label{thm:AN_dependent}
  Under Assumptions \ref{assume:stationary}-\ref{assume:Avar_dependent}, \ref{assume:xsupport} and \ref{assume:ID_weakconv}, if $\log(N)^2/T \to 0$ and $\lambda_T = o_p(T)$ as $N, T \to \infty$, the estimator $\hat{\bm{\beta}}$ is consistent. Moreover, if $N^2 (\log N)^3 / T \to 0$ and $\lambda_T = o_p( T^{1/2} (\log N)^{1/2} )$ as $N, T \rightarrow \infty$, then
  \begin{equation*}
    \sqrt{NT} (\hat{\bm{\beta}} - \bm{\beta}_0) \indist \mathcal{N}(\bm{0}, \tilde{\bm{\Omega}}),
  \end{equation*}
  where $\tilde{\bm{\Omega}} = \bm{D}^{-1} \tilde{\bm{V}} \bm{D}^{-1}$.
\end{theorem}

Theorem~\ref{thm:boot_dependent} shows consistency of the bootstrap distribution estimator in the case of dependent errors.  This more complex situation requires another assumption:

\begin{assumptionA} \label{assume:weight_dependent}
  Suppose that
  \begin{equation*} \label{joint_G_condition}
    \lim_{N, T \rightarrow \infty} \frac{1}{N} \sum_{i=1}^N \sum_{j=1}^{T-1} \left( 1 - \frac{j}{T} \right) \left( \text{P}^*\{w_{it} < 0, w_{it+j} < 0\} - \prob{u_{it} \leq 0, u_{it+j} \leq 0 | \bm{x}_{it}, \bm{x}_{it+j}} \right) = 0.
  \end{equation*}
\end{assumptionA}

Assumption~\ref{assume:weight_dependent} is a high-level assumption on the distribution of bootstrap weights.  The assumption guarantees that the variance of the bootstrap estimator is bounded and sufficiently close to the true variance, because the weights mimic the within-unit dependence structure of the errors. A feasible version could use a plug-in estimate of the average of the joint conditional CDFs of $(u_{it}, u_{it+j})$ to generate weights that satisfy the average probability.

\begin{theorem} \label{thm:boot_dependent}
  Suppose that the bootstrap weights satisfies assumptions~\ref{C1}-\ref{assume:weight_dependent} and the data satisfy assumptions \ref{assume:stationary}-\ref{assume:Avar_dependent}, \ref{assume:xsupport} and \ref{assume:ID_weakconv}. If $N^2 (\log N)^3 / T \to 0$ and $\lambda_T = o_p( T^{1/2} (\log N)^{1/2})$ as $N, T \rightarrow \infty$, then
  \begin{equation*}
    \sup_{b \in \RR^p} \left| \prob{\sqrt{NT}( \bm{\beta}^* - \hat{\bm{\beta}}) \leq b | \bm{S}} - \prob{\sqrt{NT}( \hat{\bm{\beta}} - \bm{\beta}_0) \leq b} \right| \inpr 0,
  \end{equation*}
  where $\bm{S}$ denotes the observed sample and $\bm{\beta}^\ast$ denotes the slope estimator defined by~\eqref{penboot}.
\end{theorem}

Finally, we investigate if the conditions on the size of $T$ relative to $N$ needed for the asymptotic normality in Theorem~\ref{thm:AN} can be improved, especially in the light of recent work by~\citet{AGalvao2020}. If instead of focusing on the stochastic order of the terms of the Bahadur representation of the penalized estimator, we focus on the expected values of the remainder terms, it is possible to show that the rates can be improved substantially. In order to show asymptotic normality, we employ the following assumption about the behavior of the penalty parameter.

\begin{assumptionB} \label{assume:lambda_tail}
  For some $\kappa \geq 2$, there exists a constant $K > 0$ such that $\prob{ \lambda_T > K T^{1/2} (\log T)^{1/2} } = O(T^{-\kappa})$.
\end{assumptionB}

Assumption~\ref{assume:lambda_tail} dictates the rate at which the probability of observing large a $\lambda_T$ becomes small asymptotically. As illustrated in remark \ref{lambdaT_bound}, it is needed to provide a tail bound for the distribution of individual effects, which figure in the remainder terms of the Bahadur representation used to find the asymptotic distribution of $\bm{\hat{\beta}}$ (such a bound holds naturally for terms related to minimizing the quantile regression objective function with bounded regressors, a fact used extensively in~\citet{AGalvao2020}).  In the theorem below, we require $\lambda_T = O_p(\log T) = o_p(T^{1/2} (\log T)^{1/2})$, so this assumption only mildly strengthens the other regularity conditions. 

\begin{theorem} \label{thm:rates}
  Under Assumptions \ref{assume:data} and \ref{assume:xsupport}-\ref{assume:lambda_tail}, if $N (\log T)^2 / T \to 0$ and $\lambda_T = O_p(\log T)$ as $N, T \rightarrow \infty$, then
  \begin{equation*}
    \sqrt{NT} (\hat{\bm{\beta}} - \bm{\beta}_0) \indist \mathcal{N}(\bm{0}, \bm{\Omega}),
  \end{equation*}
  where $\bm{\Omega} = \bm{D}^{-1} \bm{V} \bm{D}^{-1}$. 
\end{theorem}

The proof in Theorem \ref{thm:rates} uses an infeasible estimator $\tilde{\alpha}_i$ that is obtained considering $T$ observations $y_{it} - \bm{x}_{it}'\bm{\beta}_0$. The difference between $\tilde{\alpha}_i$ and $\hat{\alpha}_i$ converges to zero as the slope coefficient $\hat{\bm{\beta}}$ converges in probability towards $\bm{\beta}_0$, under the condition on $\lambda_T$. Therefore, the remainder terms of the corresponding Bahadur representations are sufficiently close, leading to the improvements in the rates first obtained in \citet{AGalvao2020} for the fixed effects estimator. We now show the consistency of the bootstrap distribution estimator under these relatively closer orders of $N$ and $T$.

\begin{theorem} \label{thm:boot_rates}
  Under Assumptions \ref{C1}-\ref{C3} and the conditions of Theorem \ref{thm:rates},
  \begin{equation*}
    \sup_{b \in \RR^p} \left| \prob{\sqrt{NT}( \bm{\beta}^* - \hat{\bm{\beta}}) \leq b | \bm{S}} - \prob{\sqrt{NT}( \hat{\bm{\beta}} - \bm{\beta}_0) \leq b} \right| \inpr 0.
  \end{equation*}
  where $\bm{S}$ denotes the observed sample and $\bm{\beta}^\ast$ denotes the slope estimator defined by~\eqref{penboot}.
\end{theorem}

\section{An Empirical Illustration}

In recent years, policy makers and the general public have been debating and re-evaluating several aspects of trade, including the benefits of trade agreements \citep[][among others]{mB2001, sH2016}. An important question is whether workers have been negatively affected by the North American Free Trade Agreement (NAFTA), which was signed by the governments of the United States of America, Canada, and Mexico in 1993. Hakobyan and McLaren (2016) find that the effect of NAFTA on \textit{average} wage growth in the period 1990-2000 was negative. In this section, we use similar data and apply our approach to study the distributional impact of NAFTA. Our findings suggest that the agreement increased wage inequality. Low-wage workers experienced significant negative wage growth, while high-wage workers experienced, in general, significant positive wage growth. Our results are similar to evidence on the effect of Chinese imports on low-wage American workers \citep{denisChet2016}.   

\subsection{Data}

Following Hakobyan and McLaren (2016), we use a 5\% sample from the U.S. Census. We employ two cross-sectional samples in the year 1990 and 2000, and therefore, workers in the sample are observed once. The longitudinal nature of the analysis comes from exploiting the fact that we observe multiple individuals in a given industry and location. The sample includes workers between 25 and 64 years of age who reported positive income. We have demographic information including age, gender, marital status, race, and educational attainment of the worker classified in four categories: high school dropout, high school graduate, some college, and college graduate. 

The data on U.S. tariffs and Mexico's revealed comparative advantage (RCA) are obtained from Hakobyan and McLaren (2016). Using their data, we have access to average U.S. tariffs by industry of employment of the worker and location (or Consistent Public-Use Microdata Area, abbreviated \emph{conspuma}) of residence of the worker. In 1990, the average tariff by industry in 1990 was 2.1\% percent (with a standard deviation of 3.9\%), while the average local tariff by conspuma level  was 1.03\% (with a standard deviation of 0.67\%). In the period 1990-2000, the tariffs decreased 1.7\% at the industry level and 0.9\% at the conspuma level. These descriptive statistics are used in the next section to estimate the percentage change in wages associated with the reduction in tariffs. We consider all industries with the exception of agriculture.

\subsection{Model}

To investigate the effect of NAFTA on the wages of American workers, we consider a specification that allows for the impact of the trade agreement to vary by industry, location, and educational attainment of the worker. To that end, we consider the following model as in Hakobyan and McLaren (2016):
\begin{equation}
y_{ijc} = \bm{\beta}_{1L}' \bm{L}_{ic} + \bm{\beta}_{2L}' \Delta \bm{L}_{ic} + \bm{\beta}_{1I}' \bm{I}_{ij} + \bm{\beta}_{2I}' \Delta \bm{I}_{ij} + \bm{X}_{ijc}' \bm{\Pi} + \alpha_{jc} + u_{ijc},  \label{main}
\end{equation}
where the response variable $y_{ijc}$ is the logarithm of wages for worker $i$, who is employed in industry $j$ and resides in conspuma $c$, $\bm{L}_{ic}$ and $\Delta \bm{L}_{ic}$ are location variables to be described below, $\bm{I}_{ij}$ and $\Delta \bm{I}_{ij}$ are industry variables, $\bm{X}_{ijc}$ is the vector of control variables considered in Hakobyan and McLaren (2016), and $\alpha_{jc}$ is a industry-conspuma effect. The error term is denoted by $u_{ijc}$. 

The location variables are defined as $\bm{L}_{ic} = (L_{ic,1},L_{ic,2},L_{ic,3},L_{ic,4})'$, where $L_{ic,k}$ is the product of an indicator for educational category $k$ of worker $i$, an indicator variable for whether $i$ is in the 2000 sample, and the average tariff in the conspuma of residence of worker $i$. Similarly, we can define $\Delta \bm{L}_{ic} = (\Delta L_{ic,1},\Delta L_{ic,2},\Delta L_{ic,3},\Delta L_{ic,4})'$, as the change in $\bm{L}_{ic}$ due to the change in tariffs between 1990 and 2000 in the conspuma of residence of worker $i$. In terms of the industry variables, $\bm{I}_{ij} = (I_{ij,1},I_{ij,2},I_{ij,3},I_{ij,4})'$, where $I_{ij,k}$ is the product of an indicator for educational category $k$, the RCA in industry $j$, an indicator variable for whether $i$ is in the 2000 sample, and the tariff of the industry that employs worker $i$. Similarly, we define $\Delta \bm{I}_{ij} = (\Delta I_{ij,1},\Delta I_{ij,2},\Delta I_{ij,3},\Delta I_{ij,4})'$, as the change in $\bm{I}_{ij,k}$ due to the tariff change between 1990 and 2000 in the industry that employs worker $i$.


Because industry latent factors and trends in some areas can affect wages and also the changes in tariffs, we employ the penalized estimator \eqref{pqr} to estimate a high-dimensional model with more than 84,000 parameters $\alpha_{jc}$. The parameters of interest in equation \eqref{main} are $\bm{\beta}_{1L}$, $\bm{\beta}_{2L}$, $\bm{\beta}_{1I}$, and $\bm{\beta}_{2I}$, which measure the initial effect of tariffs by location and industry ($\bm{\beta}_{1L}$ and $\bm{\beta}_{1I}$), and the impact effect of a reduction of tariffs by location and industry ($\bm{\beta}_{2L}$ and $\bm{\beta}_{2I}$). Using these parameters, it is possible to obtain the effect of the trade agreement on wages. For instance, for locations that lost all of their protection after the introduction of NAFTA, the effect of the local average tariff is measured by $\bm{\beta}_{1L} - \bm{\beta}_{2L}$. Similarly, for industries that lost all of their protection, the effect of the industry tariff is $\bm{\beta}_{1I} - \bm{\beta}_{2I}$. 

\begin{singlespace}
\begin{table}
\begin{center}\small
\begin{tabular}{l c c c c c c } \hline
 & 	Mean 	 &  \multicolumn{5}{c}{Quantiles} \\
 &  Effect &  0.1	&	0.25 &	0.5	& 0.75 &	0.9 \\  \hline
 &  \multicolumn{6}{c}{High school dropouts} \\  \hline
Initial tariff effect, $\beta_{1I,1}$	&	2.018	&	1.156	&	2.603	&	1.880	&	0.991	&	0.434	\\
	&(1.274)&(0.820)&(1.120)&(1.047)&(0.847)&(1.105)\\
Impact effect, $\beta_{2I,1}$ &3.569&3.082&4.625&3.245&1.666&0.600\\
&(1.544)&(0.945)&(1.314)&(1.191)&(1.024)&(1.290)\\
Industry effect: $\beta_{1I,1}-\beta_{2I,1}$&-1.551&-1.925&-2.022&-1.365&-0.675&-0.166\\
&[0.000]&[0.000]&[0.000]&[0.000]&[0.000]&[0.556]\\  \hline
 &  \multicolumn{6}{c}{High school graduates} \\  \hline
Initial tariff effect, $\beta_{1I,2}$ &1.081&5.015&2.224&0.426&-2.216&-2.933\\
&(0.870)&(0.523)&(0.626)&(0.747)&(0.515)&(0.436)\\
Impact effect,  $\beta_{2I,2}$ &2.315&9.259&4.318&1.337&-2.469&-3.855\\
&(1.086)&(0.595)&(0.736)&(0.873)&(0.618)&(0.543)\\
Industry effect: $\beta_{1I,2}-\beta_{2I,2}$&-1.234&-4.245&-2.094&-0.911&0.253&0.922\\
&[0.000]&[0.000]&[0.000]&[0.000]&[0.022]&[0.000]\\  \hline
 &  \multicolumn{6}{c}{Some college} \\  \hline
Initial tariff effect, $\beta_{1I,3}$ &-0.181&3.187&2.631&-0.921&-2.963&-3.765\\
&(1.146)&(0.820)&(1.172)&(1.151)&(0.779)&(0.879)\\
Impact effect, $\beta_{2I,3}$&1.070&7.360&4.889&-0.263&-3.452&-4.662\\
&(1.396)&(0.972)&(1.468)&(1.359)&(0.954)&(1.026)\\
Industry effect: $\beta_{1I,3}-\beta_{2I,3}$&-1.234&-4.245&-2.094&-0.911&0.253&0.922\\
&[0.000]&[0.000]&[0.000]&[0.000]&[0.022]&[0.000]\\  \hline
 &  \multicolumn{6}{c}{College graduate} \\  \hline
Initial tariff effect, $\beta_{1I,4}$&-2.438&7.623&-1.363&-6.538&-7.681&-8.688\\
&(1.839)&(1.826)&(1.362)&(1.856)&(1.041)&(1.181)\\
Impact effect, $\beta_{2I,4}$&-2.095&12.840&-0.024&-8.066&-9.828&-11.490\\
&(2.175)&(2.215)&(1.630)&(2.291)&(1.178)&(1.301)\\ 
Industry effect: $\beta_{1I,4}-\beta_{2I,4}$&-0.343&-5.217&-1.339&1.528&2.147&2.801\\
&[0.439]&[0.000]&[0.000]&[0.000]&[0.000]&[0.000]\\  \hline
Location variables       &	Yes & Yes	&	Yes	&	Yes	&	Yes	&	Yes	\\
Control variables       &	Yes & Yes	&	Yes	&	Yes	&	Yes	&	Yes	\\
Number of $\alpha_{jc}$ effects	&	84,266	&	84,266	&	84,266	&	84,266	&	84,266	&	84,266	\\
Observations	&	9,580,568	&	9,580,568	&	9,580,568	&	9,580,568	&	9,580,568	&	9,580,568	\\ \hline
\end{tabular}
\vspace{3mm}
\end{center}
\caption{\emph{Regression results for the industry effects by educational category of the worker. We present standard errors in parenthesis, and p-values of a test for the equality of initial and impact effects in brackets.}}
\label{table1.results}
\end{table}
\end{singlespace}

\subsection{Main empirical results}

Table \ref{table1.results} reports results for the coefficients $\bm{\beta}_{1I}$,
and $\bm{\beta}_{2I}$ for the four educational categories. The table also shows results for $\beta_{1I,k} - \beta_{2I,k}$ for each educational category $k$ and p-values (in brackets) of Wald-type tests for the null hypothesis $\mbox{H}_0: \beta_{1I,k} = \beta_{2I,k}$. The variance of the test is obtained using the proposed wild residual bootstrap procedure. The first column presents mean fixed effects regression results, that is, estimation of model \eqref{main} by least squares methods. The last five columns show penalized quantile regression (PQR) results with $\lambda_T$ selected by cross-validation. The standard errors are obtained by the proposed wild residual bootstrap procedure. To save space, we do not present results on the control variables included in the vector $\bm{L}_{ic}$, $\Delta \bm{L}_{ic}$, and $\bm{X}_{ijc}$, but the fixed effects results shown in the first column are similar to the results in Table 4 (column (2)) in Hakobyan and McLaren (2016).  

Looking at the first set of estimates in the first rows, we see that an initial tariff estimate equal to 2.02 and an impact effect of 3.57. Based on the standard deviation of tariffs at the industry level, a 1\% standard deviation increase in the initial industry tariff has an effect of reducing wages by $3.9\% \times -1.55$, or $-6.05\%$ in the period 1990-2000. This implies that, among industries with tariff declining after the introduction of NAFTA, average wage growth is negative for high school dropouts. The results, however, show that the average response does not summarize well the distributional impact of NAFTA. While the industry effect, which is measured as the difference between the initial effect and the impact effect, is negative ($-1.93$, or $-7.50\%$) and significant for high school dropouts at the 0.1 quantile, it is small ($-0.17$, or $-0.65\%$) and insignificant at the 0.9 quantile. Moreover, we find that the largest differences between the 0.1 and 0.9 effects are among college graduates in industries that lost all of their protection, suggesting that wage growth has been also unequal by educational attainment. 

\begin{singlespace}
\begin{figure}
\begin{center}
\centerline{\includegraphics[width=.6\textwidth]{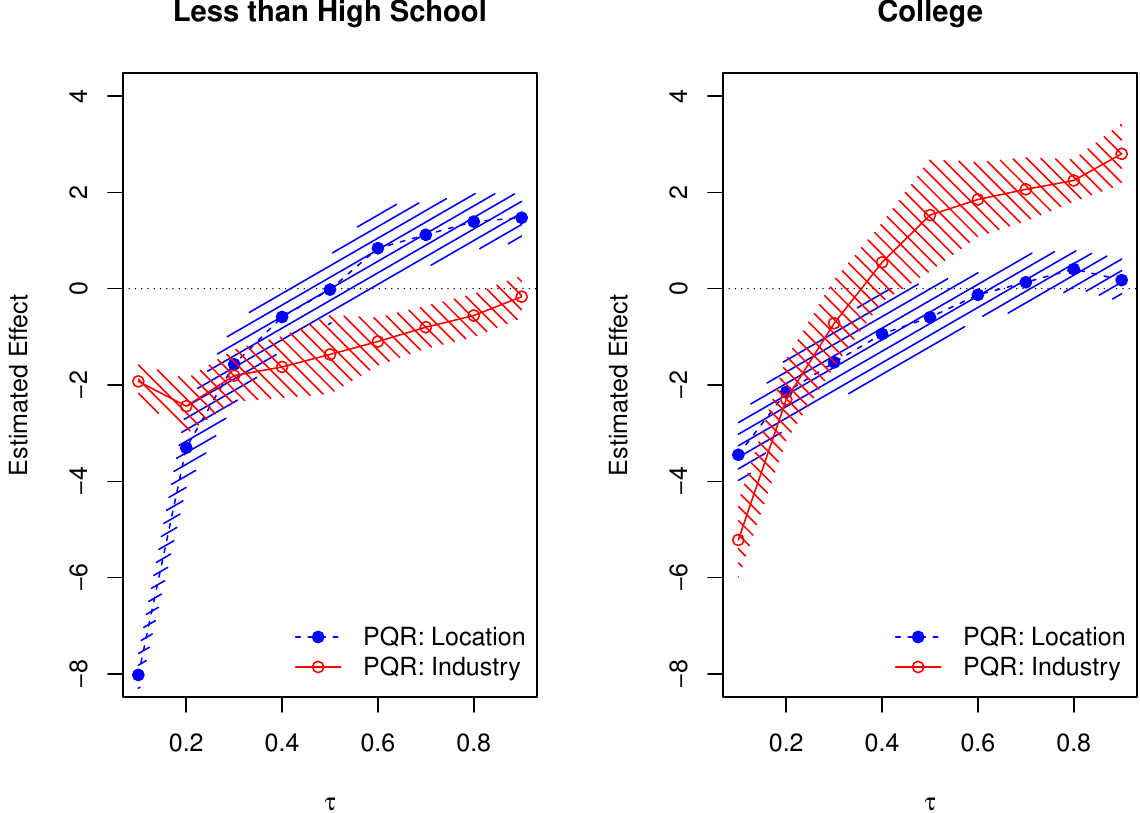}}
\caption{\emph{Conditional wage growth impacts. PQR denotes penalized quantile regression and the dashed areas are 95\% confidence intervals.}\label{nafta.figure}}
\end{center}
\end{figure}
\end{singlespace}

Lastly, using Figure \ref{nafta.figure}, we report point estimates and confidence intervals for the location and industry effects for high school dropouts and college graduates. 
The evidence reveals that inequality increased in the period after the implementation of the trade agreement. 

\section{Conclusion}

In this article, we address the problem of estimating the distribution of the penalized quantile regression estimator for longitudinal data using a wild residual bootstrap procedure. Originally introduced by Koenker (2004) as a convenient alternative to the quantile regression estimator with fixed effects, the practical use of the penalized estimator has been limited by challenges involving inference. We show that the wild bootstrap procedure is asymptotically valid for approximating the distribution of the penalized estimator. We derive a series of new asymptotic results and carry out a simulation study that indicates that the wild residual bootstrap performs better than an alternative bootstrap approach commonly used in practice for similar estimators that do not include a penalty term. 

Although the paper makes an important contribution by providing a valid method for statistical inference, there are several questions that remain to be answered. We believe that the procedure leads to valid inference in the case of $J$ quantiles estimated simultaneously, but we leave this to future research.  Moreover, under an assumption of sparsity as in other high-dimensional models, we expect changes in the consistency and asymptotic normality results. In terms of theoretical developments, we did not consider the case where $\alpha_i$ is a random effect. Lastly, the practical implementation of the wild bootstrap in the case of dependent data involves a few challenges. We hope to investigate these directions in future work.

\appendix

\section{Proof of main results}

{\bf Remarks on notation and definitions:} The estimators $\hat{\bm{\beta}}$ and $\bm{\beta}^\ast$ depend on $\tau$ and $\lambda_T$, but we suppress this dependency for notational simplicity. 
The proofs refer to Knight's (1998)\nocite{kK98} identity: $\rho_{\tau}(u-v) - \rho_{\tau}(u) = - v \psi_{\tau}(u) + \int_{0}^{v} (I(u \leq s) - I(u \leq 0)) \ud s$, where $\rho_{\tau} = u(\tau - I(u < 0))$ is the quantile regression check function and $\psi_\tau(u) = \tau - I(u < 0)$ is the associated score function.  Throughout the appendix, we define $\bm{\theta}_i = (\bm{\beta}', \alpha_i)'$ for each $i$, $\bm{\alpha} = (\alpha_1, \ldots, \alpha_N)$ and $\bm{\theta} = (\bm{\beta}', \bm{\alpha}')'$.

\begin{proof}[Proof of Theorem~\ref{thm:consistent}]
  Consistency follows from derivations analogous to those in \citet{kK12}, tailored to accommodate a penalty term.  Let $\hat{\bm{\theta}}$ be the minimizer of the normalized objective function
  \begin{equation*}
    \MM_{NT}(\bm{\theta}) = \frac{1}{NT} \sum_{i=1}^N \sum_{t=1}^T \rho_{\tau} \left( y_{it} - \bm{x}_{it}^\prime \bm{\beta} - \alpha_i \right) + \frac{\lambda_T}{NT} \sum_{i=1}^N |  \alpha_i  | .
  \end{equation*}
  Define the $i$-th contribution to the objective function
  \begin{equation*}
    \MM_{Ti}(\bm{\theta}_i) := \frac{1}{T} \sum_{t=1}^T \rho_{\tau} \left( y_{it} - \bm{x}_{it}^\prime \bm{\beta} - \alpha_i \right) + \frac{\lambda_T}{T} |  \alpha_i  |
  \end{equation*}
  and let $\Delta_{Ti}(\bm{\theta}_i) = \MM_{Ti}(\bm{\theta}_i)  - \MM_{Ti}(\bm{\theta}_{i0})$, that is,
  \begin{equation*}
    \Delta_{Ti}(\bm{\theta}_i) = \frac{1}{T} \sum_{t=1}^T \left\{ \rho_{\tau}\left( u_{it} - \bm{x}_{it}' (\bm{\beta} - \bm{\beta}_{0}) - (\alpha_i - \alpha_{i0}) \right) - \rho_{\tau}\left( u_{it} \right) \right\} + \frac{\lambda_T}{T} (|  \alpha_i  | - | \alpha_{i0}  |). 
  \end{equation*}

  By Knight's identity, $\Delta_{Ti}(\bm{\theta}_i) = \VV_{Ti}^{(1)}(\bm{\theta}_i) + \VV_{Ti}^{(2)}(\bm{\theta}_i)$, where
  \begin{align*}
    \VV_{Ti}^{(1)}(\bm{\theta}_i) &= -\frac{1}{T} \sum_{t=1}^T \left\{ \bm{x}_{it}' ( \bm{\beta} - \bm{\beta}_{0} ) + (\alpha_i - \alpha_{i0}) \right\} \psi_{\tau}( u_{it} ) + \frac{\lambda_T}{T} (|  \alpha_i  | - |  \alpha_{i0}  |), \\
    \VV_{Ti}^{(2)}(\bm{\theta}_i) &= \frac{1}{T} \sum_{t=1}^T  \int_{0}^{\bm{x}_{it}' (\bm{\beta} - \bm{\beta}_{0}) + (\alpha_i - \alpha_{i0})} \left( I( u_{it} \leq s ) - I( u_{it} \leq 0 ) \right) \ud s.
  \end{align*}

  We first show the consistency of $\hat{\bm{\beta}}$ for $\bm{\beta}_0$. For each $\phi > 0$, define the ball $\mathcal{B}_i(\phi) := \{ \bm{\theta}_i : \| \bm{\theta}_i - \bm{\theta}_{i0} \|_1 \leq \phi \}$ and the boundary $\partial \mathcal{B}_i(\phi) := \{ \bm{\theta}_i : \| \bm{\theta}_i - \bm{\theta}_{i0} \|_1  = \phi \}$. For each $\bm{\theta}_i \not \in \mathcal{B}_i(\phi)$, define $\bar{\bm{\theta}}_i  = r_i \bm{\theta}_i + (1 - r_i) \bm{\theta}_{i0}$ where $r_i = \phi / \| \bm{\theta}_i - \bm{\theta}_{i0} \|_1$. By construction, $r_i \in (0,1)$ and $\bar{\bm{\theta}}_i \in \partial \mathcal{B}_i(\phi)$. 

  Using the convexity of $\MM_{Ti}(\bm{\theta}_i)$,
  \begin{align}
    r_i \big( \MM_{Ti}(\bm{\theta}_i) & - \MM_{Ti}(\bm{\theta}_{i0}) \big) \geq \MM_{Ti}(\bar{\bm{\theta}}_i) - \MM_{Ti}(\bm{\theta}_{i0}) \notag \\
    {}  &=  \ex{\Delta_{Ti}(\bar{\bm{\theta}}_i)} + \left( \Delta_{Ti}(\bar{\bm{\theta}}_i) - \ex{\Delta_{Ti}(\bar{\bm{\theta}}_i)} \right).  \label{app-convex} 
  \end{align}

  Under Assumptions \ref{assume:data} and \ref{assume:ID}, we obtain, for $1 \leq i \leq N$, 
  \begin{align*}
    \ex{\Delta_{Ti}(\bm{\theta}_{i})} &= \frac{\lambda_T}{T} (|  \alpha_i  | - |  \alpha_{i0}  |) + \ex{\int_{0}^{\bm{x_{it}}' (\bm{\beta} - \bm{\beta}_{0}) + (\alpha_i - \alpha_{i0})}  (F_i ( s | \bm{x}_{i1} ) - \tau ) \ud s} \\
    {} &\geq \frac{\lambda_T}{T} (|  \alpha_i  | - |  \alpha_{i0}  |) + \epsilon_\phi 
  \end{align*}
  for some $\epsilon_\phi > 0$.  Using this in \eqref{app-convex} results in
  \begin{equation*}
    r_i \Delta_{Ti}(\bm{\theta}_i)  \geq \epsilon_\phi + \frac{\lambda_T}{T} ( | \bar{\alpha}_i | - |  \alpha_{i0}  |) + \left( \Delta_{Ti}(\bar{\bm{\theta}}_i) - \ex{\Delta_{Ti}(\bar{\bm{\theta}}_i)} \right).
  \end{equation*}

  By the definition of $\hat{\bm{\theta}}_i$ as the minimizer of $N^{-1} \sum_i \MM_{Ti}(\bm{\theta}_i)$, we have
  \begin{eqnarray*}
    \left\{ \| \hat{\bm{\theta}}_i - \bm{\theta}_{i0} \|_1  > \phi \right\} & \subseteq & \left\{ \exists i \in \{1, \ldots N\} : \hat{\bm{\theta}}_i \not \in  \mathcal{B}_i(\phi) \text{ and } \MM_{Ti}(\hat{\bm{\theta}}_i) \leq \MM_{Ti}(\bm{\theta}_{i0}) \right\} \\
    & \subseteq & \bigg\{ \max_{1 \leq i \leq N} \sup_{\bm{\theta}_i \in \mathcal{B}_i(\phi)}  \Big| (\lambda_T/T) (|  \alpha_i  | - |  \alpha_{i0}  |) + \Delta_{Ti}(\bm{\theta}_i) - \ex{\Delta_{Ti}(\bm{\theta}_i)} \Big| \geq \epsilon_\phi  \bigg\}.
  \end{eqnarray*}
  Therefore, it is sufficient to show that
  \begin{equation} \label{step1limit}
    \lim_{N \rightarrow \infty} \mathrm{P} \bigg\{ \max_{1 \leq i \leq N} \sup_{\bm{\theta}_i \in \mathcal{B}_i(\phi)}  \Big| (\lambda_T/T) (|  \alpha_i  | - |  \alpha_{i0}  |) + \Delta_{Ti}(\bm{\theta}_i) - \ex{\Delta_{Ti}(\bm{\theta}_i)} \Big| \geq \epsilon_\phi \bigg\} = 0,
  \end{equation}
  which is implied by
  \begin{equation} \label{step1bound}
    \max_{1 \leq i \leq N} \mathrm{P} \bigg\{ \sup_{\bm{\theta}_i \in \mathcal{B}_i(\phi)}  \Big| (\lambda_T/T) (|  \alpha_i  | - |  \alpha_{i0}  |) \Big| + \Big| \Delta_{Ti}(\bm{\theta}_i) - \ex{\Delta_{Ti}(\bm{\theta}_i)} \Big| \geq \epsilon_\phi \bigg\} = o(N^{-1}).
  \end{equation}

    Normalize $\bm{\theta}_{i0} = \bm{0}_{p+1}$ for $1 \leq i \leq N$, so that $\mathcal{B}_i(\phi) = \mathcal{B}(\phi)$ for all $1 \leq i \leq N$.  Let $h_{\bm{\theta}}(u,\bm{x}) := \rho_{\tau}(u - \bm{x}' \bm{\beta} - \alpha ) - \rho_{\tau}(u) + (\lambda_T/T) | \alpha |$.  By Assumption \ref{assume:xsupport} and the reverse triangle inequality, letting $\Lambda = \lambda_U / T$, for some constant $C$,
  \begin{equation*}
    |h_{\bm{\theta}} (u,\bm{x}) - h_{\bm{\theta}'}(u,\bm{x})| \leq 2 ( 1 + \| \bm{x} \| + \lambda_T / T ) \left( \| \bm{\beta} - \bm{\beta}' \|_1 + |\alpha - \alpha'| \right) \leq C(1 + M + \Lambda) \| \bm{\theta} - \bm{\theta}' \|_1.
  \end{equation*}
  
  For any $\phi > 0$, consider covering $\mathcal{B}(\phi)$, a compact set in $\RR^{p + 1}$, with $L_1$-balls of diameter $\epsilon$ over $B(\phi)$: generally $K = (\phi / \epsilon + 1)^{p + 1}$ such balls are required.  Cover $\mathcal{B}(\phi)$ with $K$ balls of diameter $\epsilon / 3\kappa$  where $\kappa = C(1 + M + \Lambda)$, and which have centers $\bm{\theta}^{(k)}$ for $k = 1, \ldots K$.  Then the number of balls required is $K \leq \left( \frac{3 \kappa \phi}{\epsilon} + 1 \right)^{p+1} = O(\epsilon^{-(p+1)})$.  Covering $B(\phi)$ with balls of this diameter implies that there is some $k \in \{1, \ldots K\}$ such that
  \begin{align*}
    \Big| \Delta_{Ti} (\bm{\theta}) &- \ex{\Delta_{Ti} (\bm{\theta})} - \Delta_{Ti} (\bm{\theta}^{(k)}) - \ex{\Delta_{Ti} (\bm{\theta}^{(k)})} \Big| \\
    & \leq \left| \Delta_{Ti} (\bm{\theta}) - \Delta_{Ti} (\bm{\theta}^{(k)}) \right| + \left| \ex{\Delta_{Ti} (\bm{\theta})} - \ex{\Delta_{Ti}(\bm{\theta}^{(k)})} \right| \leq  2 \kappa  \frac{\epsilon}{3 \kappa} = \frac{2}{3} \epsilon.
  \end{align*}
  Therefore for each $\bm{\theta} \in \mathcal{B}(\phi)$ there is a $k \in \{1,2,...,K\}$ such that
  \begin{equation*}
    \left| \Delta_{Ti} (\bm{\theta}) - \ex{\Delta_{Ti} (\bm{\theta})} \right| \leq \left| \Delta_{Ti} (\bm{\theta}^{(k)}) - \ex{\Delta_{Ti} (\bm{\theta}^{(k)})} \right| + \frac{2}{3} \epsilon,
  \end{equation*}
  and
  \begin{eqnarray*}
    \mathrm{P} \bigg\{ \sup_{\bm{\theta} \in \mathcal{B}(\phi)} \Big| \Delta_{Ti}(\bm{\theta}) - \ex{\Delta_{Ti}(\bm{\theta})} \Big| > \epsilon \bigg\} & \leq & \prob{ \max_{1 \leq k \leq K} \left| \Delta_{Ti}(\bm{\theta}^{(k)}) - \ex{\Delta_{Ti}(\bm{\theta}^{(k)})} \right| + \frac{2 \epsilon}{3} > \epsilon }  \\
    & \leq & \sum_{k=1}^K \prob{ \left| \Delta_{Ti}(\bm{\theta}^{(k)}) - \ex{\Delta_{Ti}(\bm{\theta}^{(k)})} \right| + \frac{2 \epsilon}{3} > \epsilon }  \\
    & = & \sum_{k=1}^K \prob{ \left| \Delta_{Ti}(\bm{\theta}^{(k)}) - \ex{\Delta_{Ti}(\bm{\theta}^{(k)})} \right| > \epsilon/3 }.
  \end{eqnarray*}
  For each term,
  \begin{multline*}
    \Delta_{Ti}(\bm{\theta}^{(k)}) - \ex{\Delta_{Ti}(\bm{\theta}^{(k)})} = \frac{1}{T} \sum_{t=1}^T \left( \rho_\tau(u_{it} - \bm{x}_{it}' \bm{\beta}^{(k)} - \alpha^{(k)}) - \rho_\tau(u_{it}) \right) \\
    - \ex{ \frac{1}{T} \sum_{t=1}^T \left( \rho_\tau(u_{it} - \bm{x}_{it}' \bm{\beta}^{(k)} - \alpha^{(k)}) - \rho_\tau(u_{it}) \right)},
  \end{multline*}
  because the terms involving the penalty depend on $\alpha^{(k)}$ and cancel.  Because each $\bm{\theta}^{(k)} \in \mathcal{B}(\phi)$, it can verified that $|\rho_\tau(u_{it} - \bm{x}_{it}' \bm{\beta}^{(k)} - \alpha^{(k)}) - \rho_\tau(u_{it})| \leq (1 + M) \phi$.  Hoeffding's inequality implies $\prob{ \left| \Delta_{Ti}(\bm{\theta}^{(k)}) - \ex{\Delta_{Ti}(\bm{\theta}^{(k)})} \right| > \epsilon/3 } \leq 2 \exp\left\{ - \frac{(\epsilon/3)^2 T}{2 (1 + M)^2 \phi^2} \right\}$.  Therefore for any $\epsilon > 0$,
  \begin{equation*}
    \prob{ \sup_{\bm{\theta} \in \mathcal{B}(\phi)} | \Delta_{Ti}(\bm{\theta}) - \ex{\Delta_{Ti}(\bm{\theta})} | > \epsilon / 2 } \leq 2 K \exp\{ - DT\}.
  \end{equation*}

  Considering the penalty term, $(\lambda_T/T) (|  \alpha_i  | - |  \alpha_{i0}  |) \leq (\lambda_T/T) |\alpha_i - \alpha_{i0}| = O_p(\lambda_T/T)$, assuming $|\alpha_i - \alpha_{i0}| = O_p(1)$.  Under the condition that $\lambda_T = o_p(T)$, $\lambda_T / T < \epsilon / 2$ with probability increasing to 1.  Therefore, consistency of $\hat{\bm{\beta}}$ is implied by the conditions $\log N = o(T)$ and $\lambda_T = o_p(T)$ as $N,T \to \infty$.

  The consistency of $\hat{\bm{\beta}}$ implies consistency of $\hat{\alpha}_i$.  Recall that $\hat{\alpha}_i = \arg \min \MM_{NT}(\hat{\bm{\beta}}, \alpha)$. Isolating the part that depends on $\alpha_i$, define the new ball $\mathcal{B}_i(\phi) := \{ \alpha \in \RR : | \alpha - \alpha_{i0} | \leq \phi \}$.  For any $\alpha_i$ is not in $\mathcal{B}_i(\phi)$ define $\bar{\alpha}_i = r \alpha_i + (1 - r_i) \alpha_{i0}$ where $r_i = \phi / (|\alpha_i - \alpha_{i0}|)$ for $\phi > 0$.  Because the objective function is convex
  \begin{eqnarray*}
    r_i \Big( \MM_{Ti}(\hat{\bm{\beta}}, \alpha_i) &-& \MM_{Ti}(\hat{\bm{\beta}}, \alpha_{i0}) \Big) \geq \MM_{Ti}(\hat{\bm{\beta}}, \bar{\alpha}_i) - \MM_{Ti}(\hat{\bm{\beta}}, \alpha_{i0}) \\
    & = &  \{ \MM_{Ti}(\hat{\bm{\beta}}, \bar{\alpha}_i) - \MM_{Ti}(\bm{\beta}_0, \alpha_{i0}) \} - \{ \MM_{Ti}(\hat{\bm{\beta}}, \alpha_{i0}) - \MM_{Ti}(\bm{\beta}_0, \alpha_{i0}) \}  \\
    & = &  \Delta_{Ti}(\hat{\bm{\beta}}, \bar{\alpha}_i) - \Delta_{Ti}(\hat{\bm{\beta}}, \alpha_{i0}) \\
    & = &  \{ \Delta_{Ti}(\hat{\bm{\beta}}, \bar{\alpha}_i) - \ex{ \Delta_{Ti}(\bm{\beta}, \bar{\alpha}_i) }\big|_{ \bm{\beta} = \hat{\bm{\beta}}} \} + \ex{ \Delta_{Ti}(\bm{\beta}, \bar{\alpha}_i) } \big|_{ \bm{\beta} = \hat{\bm{\beta}}} \\ 
    &  & - \{ \Delta_{Ti}(\hat{\bm{\beta}}, \alpha_{i0}) - \ex{ \Delta_{Ti}(\bm{\beta}, \alpha_{i0}) }\big|_{ \bm{\beta} = \hat{\bm{\beta}}} \} - \ex{ \Delta_{Ti}(\bm{\beta}, \alpha_{i0}) } \big|_{ \bm{\beta} = \hat{\bm{\beta}}} \\
    & = &  \{ \Delta_{Ti}(\hat{\bm{\beta}}, \bar{\alpha}_i) - \ex{ \Delta_{Ti}(\bm{\beta}, \bar{\alpha}_i) }\big|_{ \bm{\beta} = \hat{\bm{\beta}}} \} - \{ \Delta_{Ti}(\hat{\bm{\beta}}, \alpha_{i0}) \\
    &  & - \ex{ \Delta_{Ti}(\bm{\beta}, \alpha_{i0}) }\big|_{ \bm{\beta} = \hat{\bm{\beta}}} \} + \{ \ex{ \Delta_{Ti}(\bm{\beta}, \bar{\alpha}_i) } \big|_{ \bm{\beta} = \hat{\bm{\beta}}} - \ex{ \Delta_{Ti}(\bm{\beta}_0, \bar{\alpha}_i) }  \} \\ 
    &  & - \{ \ex{ \Delta_{Ti}(\bm{\beta}, \alpha_{i0}) } \big|_{ \bm{\beta} = \hat{\bm{\beta}}} - \ex{ \Delta_{Ti}(\bm{\beta}_0, \alpha_{i0}) } \} + \ex{ \Delta_{Ti}(\bm{\beta}_0, \bar{\alpha}_{i}) } 
  \end{eqnarray*}

  Note that the last term $\ex{ \Delta_{Ti}(\bm{\beta}_0, \bar{\alpha}_{i}) } \geq (\lambda_T/T) \ex{ | \bar{\alpha}_i | - |  \alpha_{i0}  |} + \epsilon_\phi $ for some $\epsilon_\phi > 0$ by Assumption \ref{assume:ID}.  Thus, using similar calculations as before, we have
  \begin{eqnarray*}
    \big\{ \exists i &\in& \{1, \ldots, N\} : | \hat{\alpha}_i - \alpha_{i0} |  > \phi \big\} \\ & \subseteq & \left\{ \max_{1 \leq i \leq N} \sup_{\alpha_i \in \mathcal{B}_i(\phi)}  \left( (\lambda_T/T) |  |  \alpha_i  | - |  \alpha_{i0} | | + \left| \Delta_{Ti}(\hat{\bm{\beta}}, \alpha) - \ex{\Delta_{Ti}(\bm{\beta}, \alpha)}\big|_{ \bm{\beta} = \hat{\bm{\beta}}}  \right| \right) \geq \frac{\epsilon_\phi}{4}  \right\} \\
    & \cup  & \left\{ \max_{1 \leq i \leq N} \sup_{\alpha_i \in \mathcal{B}_i(\phi)}  \left| \ex{ \Delta_{Ti}(\hat{\bm{\beta}}, \alpha) }\big|_{ \bm{\beta} = \hat{\bm{\beta}}} - \ex{\Delta_{Ti}(\bm{\beta}_0, \alpha)} \right| \geq \frac{\epsilon_\phi}{4}  \right\} =: \mathcal{A}_{1N} \cup \mathcal{A}_{2N}.
  \end{eqnarray*}

  Because of the convexity of the objective function, the term involving $\bar{\alpha}_i$ is finite, and the entire bias term is $O_p(\lambda_T/T) = o_p(1)$ under the assumption on $\lambda_T$.  By the consistency of $\hat{\bm{\beta}}$ and equation~\eqref{step1bound}, $\prob{\mathcal{A}_{1N}} \to 0$. Moreover, by Assumption \ref{assume:xsupport} and the reverse triangle inequality, $|\ex{\Delta_{Ti}(\bm{\beta}, \alpha)} - \ex{\Delta_{Ti}(\bm{\beta}_0, \alpha)}| \leq C M \|\bm{\beta} - \bm{\beta}_0\|_1$ (due to cancellation of the penalty terms), $\hat{\bm{\beta}} - \bm{\beta}_0 \to 0$ implies $\prob{\mathcal{A}_{2N}} \to 0$.
  \end{proof}

  \begin{proof}[Proof of Theorem~\ref{thm:AN}]
  Define the scores with respect to $\bm{\beta}$ and $\alpha_i$ for the $i$-th contribution to the objective function by
  \begin{eqnarray*}
    \mathbb{H}_{Ti}^{(\beta)}(\bm{\theta}_i) & := & \frac{1}{T} \sum_{t=1}^T \bm{x}_{it} \psi_{\tau}(y_{it} - \bm{x}_{it}' \bm{\beta} - \alpha_i) \\
    \mathbb{H}_{Ti}^{(\alpha)}(\bm{\theta}_i) & := & \frac{1}{T} \sum_{t=1}^T \psi_{\tau}(y_{it} - \bm{x}_{it}' \bm{\beta} - \alpha_i) + \frac{\lambda_T}{T} \sign( \alpha_{i} )
  \end{eqnarray*}
  and define $H_{N}^{(\beta)}(\bm{\theta}_i) := \ex{\mathbb{H}_{N}^{(\beta)}(\bm{\theta}_i)}$ and $H_{Ti}^{(\alpha)}(\bm{\theta}_i) := \ex{\mathbb{H}_{Ti}^{(\alpha)}(\bm{\theta}_i)}$, that is,
  \begin{eqnarray*}
    H_{Ti}^{(\beta)}(\bm{\theta}_i)  & = & \ex{ \bm{x}_{i1} \left(\tau - F_i( \bm{x}_{i1}' (\bm{\beta} - \bm{\beta}_0) + (\alpha_i - \alpha_{i0}) | \bm{x}_{i1})} \right)  \\
    H_{Ti}^{(\alpha)}(\bm{\theta}_i) & = & \ex{ \tau - F_i( \bm{x}_{i1}' (\bm{\beta} - \bm{\beta}_0) + (\alpha_i - \alpha_{i0}) | \bm{x}_{i1})}  + (\lambda_T/T) \sign( \alpha_i ).
  \end{eqnarray*}

  First, derive a Bahadur representation for $(\hat{\bm{\beta}} - \bm{\beta}_0)$. For each $i$ write
  \begin{equation} \label{eq:Halphalong}
    \mathbb{H}_{Ti}^{(\alpha)}(\hat{\bm{\theta}}_i) =  \mathbb{H}_{Ti}^{(\alpha)}(\bm{\theta}_{i0}) + \left( \mathbb{H}_{Ti}^{(\alpha)}(\hat{\bm{\theta}}_{i}) - H_{Ti}^{(\alpha)}(\hat{\bm{\theta}}_i) - \mathbb{H}_{Ti}^{(\alpha)}(\bm{\theta}_{i0}) + H_{Ti}^{(\alpha)}(\bm{\theta}_{i0}) \right) + H_{Ti}^{(\alpha)}(\hat{\bm{\theta}}_{i}) - H_{Ti}^{(\alpha)}(\bm{\theta}_{i0}).
  \end{equation}
  Recalling the definitions made in Assumption~\ref{assume:Avar} and the bounds in Assumption~\ref{assume:ID_weakconv}, expand the differentiable part of $H_{Ti}^{(\alpha)}$ around $\bm{\theta}_{i0}$ to find
  \begin{multline*}
    H_{Ti}^{(\alpha)}(\hat{\bm{\theta}}_i) - H_{Ti}^{(\alpha)}(\bm{\theta}_{i0}) = - \bm{E}_i' (\hat{\bm{\beta}} - \bm{\beta}_0) - \varphi_i (\hat{\alpha}_i - \alpha_{i0}) \\
    + O_p \left( \| \hat{\bm{\beta}} - \bm{\beta}_0 \|^2 \right) + O_p \left( (\hat{\alpha}_i - \alpha_{i0})^2 \right) + (\lambda_T/T) \left( \sign( \hat{\alpha}_i ) - \sign( \alpha_{i0} ) \right).
  \end{multline*}
    Using the last expression and equation \eqref{eq:Halphalong}, solve for $\hat{\alpha}_i - \alpha_{i0}$ to find
  \begin{multline} \label{eq:bahaduralpha}
    \hat{\alpha}_i - \alpha_{i0} = - \varphi_i^{-1} \bm{E}_i' (\hat{\bm{\beta}} - \bm{\beta}_0) + \varphi_i^{-1} \left( \mathbb{H}_{Ti}^{(\alpha)}(\bm{\theta}_{i0}) - \frac{\lambda_T}{T} \sign(\alpha_{i0}) \right) \\
    + \varphi_i^{-1} \left( \mathbb{H}_{Ti}^{(\alpha)}(\hat{\bm{\theta}}_{i}) - H_{Ti}^{(\alpha)}(\hat{\bm{\theta}}_i) - \mathbb{H}_{Ti}^{(\alpha)}(\bm{\theta}_{i0}) + H_{Ti}^{(\alpha)}(\bm{\theta}_{i0}) \right) \\
    - \varphi_i^{-1} \left( \mathbb{H}_{Ti}^{(\alpha)}(\hat{\bm{\theta}}_i) - \frac{\lambda_T}{T} \sign(\hat{\alpha}_i) \right) + O_p \left( \| \hat{\bm{\beta}} - \bm{\beta}_0 \|^2 \right) + O_p \left( (\hat{\alpha}_i - \alpha_{i0}) ^2 \right).
  \end{multline}

  Similarly, expand $H_{Ti}^{(\beta)}$ around $\bm{\theta}_{i0}$, noting $H_{Ti}^{(\beta)}(\bm{\theta}_{i0}) = \zero_p$, to find
  \begin{equation}
    H_{Ti}^{(\beta)}(\hat{\bm{\theta}}_i) = - \bm{J}_i (\hat{\bm{\beta}} - \bm{\beta}_0) - \bm{E}_i (\hat{\alpha}_i - \alpha_{i0}) + o_p (\| \hat{\bm{\beta}} - \bm{\beta}_0 \|) + O_p \left( (\hat{\alpha}_i - \alpha_{i0})^2 \right). \label{eq:H2}
  \end{equation}
  Substituting \eqref{eq:bahaduralpha} in equation \eqref{eq:H2}, after simplification, we obtain
  \begin{multline}
    H_{Ti}^{(\beta)}(\hat{\bm{\theta}}_i) = - (\bm{J}_i - \varphi_i^{-1} \bm{E}_i \bm{E}_i') (\hat{\bm{\beta}} - \bm{\beta}_0) - \varphi_i^{-1} \bm{E}_i \left( \mathbb{H}_{Ti}^{(\alpha)}(\bm{\theta}_{i0}) - \frac{\lambda_T}{T} \sign(\alpha_{i0}) \right) \\
    - \varphi_i^{-1} \bm{E}_i \left( \mathbb{H}_{Ti}^{(\alpha)}(\hat{\bm{\theta}}_{i}) - H_{Ti}^{(\alpha)}(\hat{\bm{\theta}}_i) - \mathbb{H}_{Ti}^{(\alpha)}(\bm{\theta}_{i0}) + H_{Ti}^{(\alpha)}(\bm{\theta}_{i0}) \right) \\
    + \varphi_i^{-1} \bm{E}_i \left( \mathbb{H}_{Ti}^{(\alpha)}(\hat{\bm{\theta}}_i) - \frac{\lambda_T}{T} \sign(\hat{\alpha}_i) \right) + o_p(\| \hat{\bm{\beta}} - \bm{\beta} \| ) + O_p \left( (\hat{\alpha}_i - \alpha_{i0}) ^2 \right) \label{Hn_intermediate}
  \end{multline}
  Once again, for each $i$ we have
  \begin{equation} \label{comp_beta}
    \mathbb{H}_{Ti}^{(\beta)}(\hat{\bm{\theta}}_i) =  \mathbb{H}_{Ti}^{(\beta)}(\bm{\theta}_{i0}) + \left( \mathbb{H}_{Ti}^{(\beta)}(\hat{\bm{\theta}}_i) - H_{Ti}^{(\beta)}(\hat{\bm{\theta}}_i) - \mathbb{H}_{Ti}^{(\beta)}(\bm{\theta}_{i0}) \right) + H_{Ti}^{(\beta)}(\hat{\bm{\theta}}_i).
  \end{equation}
    Substitute \eqref{comp_beta} into the left-hand side of \eqref{Hn_intermediate} and solve for $\hat{\bm{\beta}} - \bm{\beta}_0$.  Rearrange to find
    \begin{multline} \label{big_mess}
      (\bm{J}_i - \varphi_i^{-1} \bm{E}_i \bm{E}_i') (\hat{\bm{\beta}} - \bm{\beta}_0) + o_p(\| \hat{\bm{\beta}} - \bm{\beta} \| ) = - \varphi_i^{-1} \bm{E}_i \left( \mathbb{H}_{Ti}^{(\alpha)}(\bm{\theta}_{i0}) - \frac{\lambda_T}{T} \sign(\alpha_{i0}) \right) + \mathbb{H}_{Ti}^{(\beta)}(\bm{\theta}_{i0}) \\
    - \varphi_i^{-1} \bm{E}_i \left( \mathbb{H}_{Ti}^{(\alpha)}(\hat{\bm{\theta}}_{i}) - H_{Ti}^{(\alpha)}(\hat{\bm{\theta}}_i) - \mathbb{H}_{Ti}^{(\alpha)}(\bm{\theta}_{i0}) + H_{Ti}^{(\alpha)}(\bm{\theta}_{i0}) \right) \\
    + \left( \mathbb{H}_{Ti}^{(\beta)}(\hat{\bm{\theta}}_{i}) - H_{Ti}^{(\beta)}(\hat{\bm{\theta}}_i) - \mathbb{H}_{Ti}^{(\beta)}(\bm{\theta}_{i0}) + H_{Ti}^{(\beta)}(\bm{\theta}_{i0}) \right) \\
    + \varphi_i^{-1} \bm{E}_i \left( \mathbb{H}_{Ti}^{(\alpha)}(\hat{\bm{\theta}}_i) - \frac{\lambda_T}{T} \sign(\hat{\alpha}_i) \right) - \mathbb{H}_{Ti}^{(\beta)}(\hat{\bm{\theta}}_i) + O_p \left( (\hat{\alpha}_i - \alpha_{i0})^2 \right).
  \end{multline}

    It can be verified that for each $i$, $\left| \mathbb{H}_{Ti}^{(\alpha)}(\hat{\bm{\theta}}_i) \right| \leq 1/T$ (for $\lambda_T < \min\{\tau, 1-\tau\} T$).  That implies the $i$-th individual effect estimate $\hat{\alpha}_i$ is between the $(\tau - (\lambda_T + 1) / T)$-th and $(\tau + (\lambda_T + 1) / T)$-th sample quantiles of the unit-$i$ observations $\{y_{it} - \bm{x}_{it}'\hat{\bm{\beta}}\}_{t=1}^T$. Therefore
  \begin{equation} \label{opt_error}
    \mathbb{H}_{Ti}^{(\alpha)}(\hat{\bm{\theta}}_i) - \frac{\lambda_T}{T} \sign(\hat{\alpha}_i) = \frac{1}{T} \sum_{t=1}^T \left( \tau - I(y_{it} - \bm{x}_{it}' \hat{\bm{\beta}} \leq \hat{\alpha}_i) \right) = O_p(\lambda_T / T).
  \end{equation}
  Similarly, $\mathbb{H}_{Ti}^{(\beta)}(\hat{\bm{\theta}}_i) = O_p(\lambda_T / T)$.  Now define $\mathbb{K}_{Ti}^{(\theta)}(\bm{\theta}_i) = \mathbb{H}_{Ti}^{(\beta)}(\bm{\theta}_i) - \varphi_i^{-1} \bm{E}_i (\mathbb{H}_{Ti}^{(\alpha)}(\bm{\theta}_i) - (\lambda_T/T) \sign(\alpha_i))$, $K_{Ti}^{(\theta)}(\bm{\theta}_i) = \ex{ \mathbb{K}_{Ti}^{(\theta)}(\bm{\theta}_i) }$ and $\bm{D}_N = \frac{1}{N} \sum_{i=1}^N (\bm{J}_i - \varphi_i^{-1} \bm{E}_i \bm{E}_i')$.  Averaging equation~\eqref{big_mess} over $i$ and using the above definitions and~\eqref{opt_error} we have
  \begin{multline} \label{eq:big_eqn}
    \hat{\bm{\beta}} - \bm{\beta}_0 + o_p \left( \|\hat{\bm{\beta}} - \bm{\beta}_0\| \right) = \bm{D}_N^{-1} \frac{1}{N} \sum_{i=1}^N \mathbb{K}_{Ti}^{(\theta)}(\bm{\theta}_{i0}) \\
    + \bm{D}_N^{-1} \frac{1}{N} \sum_{i=1}^N \left( \mathbb{K}_{Ti}^{(\theta)}(\hat{\bm{\theta}}_{i}) - K_{Ti}^{(\theta)}(\hat{\bm{\theta}}_i) - \mathbb{K}_{Ti}^{(\theta)}(\bm{\theta}_{i0}) + K_{Ti}^{(\theta)}(\bm{\theta}_{i0}) \right)  \\
    + O_p\left( \lambda_T / T \right) + O_p \left( \sup_i (\hat{\alpha}_i - \alpha_{i0}) ^2 \right).
  \end{multline}

  Next, we establish the rates of convergence for the estimators.  Step 2 of the proof of Theorem 3.2 of \citet{kK12} shows that if $\sup_i |\hat{\alpha}_i - \alpha_{i0}| \vee \| \hat{\bm{\beta}} - \bm{\beta}_0 \| = O_p(\delta_N)$, then for $d_{NT} = (|\log \delta_N| / T) \vee (\delta_N |\log \delta_N| / T)^{1/2}$,
  \begin{equation} \label{eq:bound_procdiff}
    \left\| \frac{1}{N} \sum_{i=1}^N \mathbb{K}_{Ti}^{(\theta)}(\hat{\bm{\theta}}_{i}) - K_{Ti}^{(\theta)}(\hat{\bm{\theta}}_i) - \mathbb{K}_{Ti}^{(\theta)}(\bm{\theta}_{i0}) + K_{Ti}^{(\theta)}(\bm{\theta}_{i0}) \right\| = O_p(d_{NT}) = o_p(T^{-1/2}),
  \end{equation}
  where the second equality follows from the consistency of $\hat{\bm{\theta}}$.  The first term on the right-hand side of~\eqref{eq:big_eqn} is $O_p((NT)^{-1/2}) = o_p(T^{-1/2})$.  Then, we have
  \begin{equation} \label{eq:beta_order1}
    \| \hat{\bm{\beta}} - \bm{\beta}_0 \| = o_p(T^{-1/2}) + O_p(T^{-1} \lambda_T) + O_p\left( \sup_i ( \hat{\alpha}_i - \alpha_{i0} )^2 \right).
  \end{equation}

  Using~\eqref{eq:beta_order1}, we find that with probability approaching 1, there is some $K$ such that
  \begin{multline*}
    \sup_i |\hat{\alpha}_i - \alpha_{i0}| \leq K \sup_i \left| \mathbb{H}_{Ti}^{(\alpha)}(\bm{\theta}_{i0}) - \frac{\lambda_T}{T} \sign(\alpha_{i0}) \right| \\
    + K \sup_i \left\| \mathbb{H}_{Ti}^{(\alpha)}(\hat{\bm{\alpha}}_{i}) - H_{Ti}^{(\alpha)}(\hat{\bm{\alpha}}_i) - \mathbb{H}_{Ti}^{(\alpha)}(\bm{\alpha}_{i0}) + H_{Ti}^{(\alpha)}(\bm{\alpha}_{i0}) \right\| + O_p(T^{-1} \lambda_T) + o_p(T^{-1/2}).
  \end{multline*}
  The first term in the above sum is mean zero and bounded.  Hoeffding's inequality implies that for some $K$,
  \begin{align*}
    \mathrm{P} \Big\{ \sup_i \Big| \mathbb{H}_{Ti}^{(\alpha)}(\bm{\theta}_{i0}) - \frac{\lambda_T}{T} \sign(\alpha_{i0}) \Big| &> T^{-1/2} (\log N)^{1/2} K \Big\} \\
    {} &\leq \sum_{i=1}^N \prob{ \left| \mathbb{H}_{Ti}^{(\alpha)}(\bm{\theta}_{i0}) - \frac{\lambda_T}{T} \sign(\alpha_{i0}) \right| > T^{-1/2} (\log N)^{1/2} K } \leq 2 N^{1 - K^2 / 2},
  \end{align*}
  so that $\sup_i |\mathbb{H}_{Ti}^{(\alpha)}(\bm{\theta}_{i0}) - (\lambda_T / T) \sign(\alpha_{i0}) | = O_p(T^{-1/2}(\log N)^{1/2})$.  Step 3 of the proof of Theorem~3.2 of \citet{kK12} implies that
  \begin{equation*}
    \sup_i \left\| \mathbb{H}_{Ti}^{(\alpha)}(\hat{\bm{\alpha}}_{i}) - H_{Ti}^{(\alpha)}(\hat{\bm{\alpha}}_i) - \mathbb{H}_{Ti}^{(\alpha)}(\bm{\alpha}_{i0}) + H_{Ti}^{(\alpha)}(\bm{\alpha}_{i0}) \right\| = o_p(T^{-1/2} (\log N)^{1/2}).
  \end{equation*}
  Together, these estimates imply that if $\lambda_T = o_p(T^{1/2} (\log N)^{1/2})$, then
  \begin{equation} \label{alpha_rate}
    \sup_i | \hat{\alpha}_i - \alpha_{i0} | = O_p \left( T^{-1/2} (\log N)^{1/2} \right),
  \end{equation}
  and via~\eqref{eq:beta_order1} that
  \begin{equation}
    \| \hat{\bm{\beta}} - \bm{\beta}_0 \| = o_p \left( T^{-1/2} (\log N)^{1/2} \right).
  \end{equation}

  The condition on $\lambda_T$ and the argument of \citet{kK12} implies that if $T^{-1} N^2(\log N)^3 \rightarrow 0$, we may rewrite equation~(\ref{eq:big_eqn}) as
  \begin{equation*}
    \sqrt{NT} (\hat{\bm{\beta}} - \bm{\beta}_0) = \bm{D}_N^{-1} \frac{1}{\sqrt{NT}} \sum_{i=1}^N \sum_{t=1}^T ( \bm{x}_{it} - \varphi_i^{-1} \bm{E}_i ) \psi_{\tau}(y_{it} - \bm{x}_{it}' \bm{\beta}_0 - \alpha_{i0}) + o_p(1), \label{eq:bahadurpen}
  \end{equation*}
  and the Lyapunov Central Limit Theorem implies that $\sqrt{NT} (\hat{\bm{\beta}} - \bm{\beta}_0) \indist \mathcal{N}(\bm{0}, \bm{\Omega})$.
\end{proof}

\begin{proof}[Proof of Theorem~\ref{thm:boot}]
  In this proof, the notation $\bm{X}^* \stackrel{p^*}{\rightarrow} \bm{X}$ denotes convergence in probability of $\bm{X}^*$ to $\bm{X}$ under the resampling distribution, conditional on the observed sample $\bm{S}$.  Similarly, let $\exs{\cdot} = \ex{ \cdot | \bm{S}}$ and $\mathrm{P}^*\{\cdot\} = P\{\cdot | \bm{S}\}$ denote the expected value operator and probability calculated conditional on the data, and stochastic order symbols $O_{p^*}(\cdot)$ and $o_{p^*}(\cdot)$ are interpreted conditional on the observed sample.  The proof is divided in two parts. The first part of the proof shows consistency by demonstrating that feasible and infeasible versions of the wild residual bootstrap estimator are equivalent as $N$ and $T \to \infty$. The second part of the proof establishes asymptotic normality of $\bm{\theta}^\ast$.

  For all $i$ and $t$ let $y_{it}^* = \bm{x}_{it}'\hat{\bm{\beta}} + \hat{\alpha}_i + w_{it} | \hat{u}_{it} |$, and let $\bm{\theta}^* = (\bm{\beta}^{*'},\bm{\alpha}^{*'})'$ be the solution of $\min_{\bm{\theta}} \MM_{NT}^*(\bm{\theta})$ where
  \begin{equation*}
    \MM_{NT}^*(\bm{\theta}) = \frac{1}{NT} \sum_{i=1}^N \sum_{t=1}^T \rho_{\tau} \left( y_{it}^* - \bm{x}_{it}^\prime \bm{\beta} - \alpha_i \right) + \frac{\lambda_T}{NT} \sum_{i=1}^N |  \alpha_i  |.
  \end{equation*}
  Also define the $i$-th contribution to the objective function by $\MM_{Ti}^*$,
  \begin{equation*}
    \MM_{Ti}^*(\bm{\theta}_i) = \frac{1}{T} \sum_{t=1}^T \rho_{\tau} \left( y_{it}^* - \bm{x}_{it}^\prime \bm{\beta} - \alpha_i \right) + \frac{\lambda_T}{T} |  \alpha_i  |,
  \end{equation*}
  so that $\MM_{NT}^* = \frac{1}{N} \sum_i \MM_{Ti}^*$. Before examining $\MM_{NT}^*$ and $\bm{\theta}^*$, consider an infeasible resampled objective function using the true error terms instead of the estimated residuals: let $y_{it}^\circ = \bm{x}_{it}'\bm{\beta}_0 + \alpha_{i0} + w_{it} | u_{it} |$ and define
  \begin{equation*}
    \MM_{Ti}^\circ(\bm{\theta}_i) = \frac{1}{T} \sum_{t=1}^T \rho_{\tau} \left( y_{it}^\circ - \bm{x}_{it}^\prime \bm{\beta} - \alpha_i \right) + \frac{\lambda_T}{T} |  \alpha_i  |.
  \end{equation*}
  Let $\bm{\theta}^\circ$ be the minimizer of $\frac{1}{N} \sum_i \MM_{Ti}^\circ(\bm{\theta}_i)$.  As in the proof of Theorem \ref{thm:consistent}, we define $\Delta_{Ti}^\circ(\bm{\theta}_i) = \MM_{Ti}^\circ(\bm{\theta}_i) - \MM_{Ti}^\circ(\bm{\theta}_{i0})$. Note that $\exs{\Delta_{Ti}^\circ(\bm{\theta}_i)}$ is minimized at $\bm{\theta}_{i0}$.

  Define the $\phi$ ball $\mathcal{B}_i(\phi) := \{ \bm{\theta}_i : \| \bm{\theta}_i - \bm{\theta}_{i0} \|_1 \leq \phi \}$ around $\bm{\theta}_{i0}$ and for $\bm{\theta}_i$ outside of the ball, define the weight $r_i = \phi / \| \bm{\theta}_i - \bm{\theta}_{i0} \|_1$ and midpoint $\bar{\bm{\theta}}_i  = r_i \bm{\theta}_i + (1 - r_i) \bm{\theta}_{i0}$. Then
  \begin{equation*}
    r_i(\MM_{Ti}^\circ(\bm{\theta}_i) - \MM_{Ti}^\circ(\bm{\theta}_{i0}) \geq \exs{\Delta_{Ti}^\circ(\bar{\bm{\theta}}_i)} + \left( \Delta_{Ti}^\circ(\bar{\bm{\theta}}_i) - \exs{\Delta_{Ti}^\circ(\bar{\bm{\theta}}_i)} \right),
  \end{equation*}
  Similarly to the consistency proof, we have
  \begin{align} \label{exp_delcirc}
    \exs{ \Delta^\circ_{Ti}(\bm{\theta}_i) }  = & \frac{\lambda_T}{T} \left\{ \left| \alpha_i \right| - \left| \alpha_{i0} \right| \right\} \notag \\ 
    & + \exs{\frac{1}{T} \sum_{t=1}^T \int_0^{l_{it}(\bm{\theta}_i)} ( I(w_{it}|u_{it}| \leq s) - I(w_{it}|u_{it}| \leq 0) ) \ud s}
  \end{align} 
  where $l_{it}(\bm{\theta}_i) = \bm{x}_{it}' (\bm{\beta} - \bm{\beta}_{0}) + (\alpha_i - \alpha_{i0})$. By Lemma~\ref{L3}, equation \eqref{exp_delcirc} can be rewritten
  \begin{equation*}
    \exs{ \Delta^\circ_{Ti}(\bm{\theta}_i) } = \frac{\lambda_T}{T} \left\{ \left| \alpha_i \right| - \left| \alpha_{i0} \right| \right\} + \frac{1}{T} \sum_{t=1}^T f_i(0 | \bm{x}_{it}) l_{it}(\bm{\theta}_i)' l_{it}(\bm{\theta}_i) + o_p(\sup_t \|l_{it}(\bm{\theta}_i)\|^2).
  \end{equation*}
  For $\bar{\bm{\theta}}_i$ on the $\phi$ ball around $\theta_{i0}$, there is some $\epsilon_\phi > 0$ such that
  \begin{equation*}
    \exs{ \Delta^\circ_{Ti}(\bar{\bm{\theta}}_i) } \geq \frac{\lambda_T}{T} \left\{ \left| \bar{\alpha}_i \right| - \left| \alpha_{i0} \right| \right\} + \epsilon_\phi
  \end{equation*}
  where $\epsilon_\phi > 0$ and
  \begin{equation} \label{delcirc_lower}
  r_i \Delta_{Ti}^\circ(\bm{\theta}_i) \geq \epsilon_\phi + \frac{\lambda_T}{T} ( | \bar{\alpha}_i | - |  \alpha_{i0}  |) + \left( \Delta_{Ti}^\circ(\bar{\bm{\theta}}_i) - \exs{\Delta_{Ti}^\circ(\bar{\bm{\theta}}_i)} \right).
  \end{equation}

  Then, similarly to the proof of consistency of $\hat{\bm{\theta}}$, the minimizer $\bm{\theta}^\circ$ is consistent if the following probability is satisfied:
  \begin{multline} \label{delcirc_consistent}
  \sup_{1 \leq i \leq N} \mathrm{P}^* \bigg\{ \sup_{\bm{\theta}_i, \in \mathcal{B}_i(\phi)}  \Big| (\lambda/T) (|  \alpha_i  | - |  \alpha_{i0}  |) \Big| + \Big| \Delta_{Ti}^\circ(\bm{\theta}_i) - \exs{\Delta_{Ti}^\circ(\bm{\theta}_i)} \Big| \geq \epsilon_\phi \bigg\} = o_p(N^{-1}).
\end{multline}
  The steps to show that $\bm{\theta}^\circ \stackrel{p^*}{\rightarrow} \bm{\theta}_0$ from this point on are identical to those in Theorem \ref{thm:consistent}.

  Having established the consistency of the infeasible estimator $\bm{\theta}^\circ$, denoting
  \begin{equation*}
    \Delta_{Ti}^*(\bm{\theta}_i) = \MM_{Ti}^*(\bm{\theta}_i) - \MM_{Ti}^*(\hat{\bm{\theta}}_{i}) = \frac{1}{T} \sum_{t=1}^T \left( \rho_{\tau}(y_{it}^* - \bm{x}_{it}' \bm{\beta} - \alpha_i) - \rho_{\tau}(y_{it}^* - \bm{x}_{it}' \hat{\bm{\beta}} - \hat{\alpha}_i) \right) + \frac{\lambda_T}{T} ( | \alpha_i | - | \hat{\alpha}_i | ),
  \end{equation*}
  we consider $\sup_{\bm{\theta}_i \in \Theta} |\Delta_{Ti}^*(\bm{\theta}_i) - \Delta_{Ti}^\circ(\bm{\theta}_i)|$.  Notice that for each $i$,
  \begin{multline}
    \sup_{\bm{\theta}_i \in \RR^{p+1}} |\Delta_{Ti}^*(\bm{\theta}_i) - \Delta_{Ti}^\circ(\bm{\theta}_i)| = \Bigg| \frac{1}{T} \sum_{t=1}^T \Big( \rho_\tau \left( w_{it} |\hat{u}_{it}| - \bm{x}_{it}'(\bm{\beta} - \hat{\bm{\beta}}) - (\alpha_i - \hat{\alpha}_i) \right) \\
    - \rho_\tau \left( w_{it} |u_{it}| - \bm{x}_{it}'(\bm{\beta} - \bm{\beta}_0) - (\alpha_i - \alpha_{i0}) \right) \\
    - \left( \rho_\tau(w_{it} |\hat{u}_{it}|) - \rho_\tau(w_{it} |u_{it}|) \right) \Big) + \frac{\lambda_T}{T} \left( |  \hat{\alpha}_i  | - |  \alpha_{i0}  | \right) \Bigg| \\
    \leq M \left( 1 + \frac{2}{T} \sum_{t=1}^T |w_{it}| \right) \left\| \hat{\bm{\beta}} - \bm{\beta}_0 \right\| + \left( 1 + \frac{\lambda_T}{T} + \frac{2}{T} \sum_{t=1}^T |w_{it}| \right) \left| \hat{\alpha}_i - \alpha_{i0} \right|.
  \end{multline}
  Using the consistency of $\hat{\bm{\theta}}_i$ for all $i$ and as long as $\lambda_T / T = O_p(1)$, the average of these differences over $i$ is $o_{p^*}(1)$ as $N,T \rightarrow \infty$ and also
  \begin{equation*}
    \sup_{\bm{\theta}_i \in \RR^{p+1}} \left| \Delta_{Ti}^*(\bm{\theta}_i) - \exs{\Delta_{Ti}^*(\bm{\theta}_i)} - \left\{ \Delta_{Ti}^\circ(\bm{\theta}_i) - \exs{\Delta_{Ti}^\circ(\bm{\theta}_i)} \right\} \right| = o_{p^*}(1)
  \end{equation*}
  as $N, T \rightarrow \infty$.  Finally, replacing the $\Delta^\circ$ terms with $\Delta^*$ terms in~\eqref{delcirc_lower} and~\eqref{delcirc_consistent} and approximating the inequalities with $\Delta^\circ$ terms implies that $\bm{\theta}^* \stackrel{p^*}{\rightarrow} \bm{\theta}^\circ$. Therefore, the wild residual bootstrap estimator $\bm{\theta}^*$ is consistent because, as demonstrated above, $\bm{\theta}^\circ \stackrel{p^*}{\rightarrow} \bm{\theta}_0$.

  Next consider the weak convergence of the estimator. Define the $i$-th contribution to the scores for $\MM_{NT}^*$ with respect to $\bm{\beta}$ and $\alpha_i$,
  \begin{equation*}
    \mathbb{H}_{Ti}^{(\beta)*}(\bm{\theta}_i) = \frac{1}{T} \sum_{t=1}^T \bm{x}_{it} \psi_{\tau}(u^\ast_{it} - \bm{x}_{it}' ( \bm{\beta} - \hat{\bm{\beta}} ) - ( \alpha_i - \hat{\alpha}_i ))
  \end{equation*}
  and
  \begin{equation*}
    \mathbb{H}_{Ti}^{(\alpha)*}(\bm{\theta}_i) = \frac{1}{T} \sum_{t=1}^T \psi_{\tau}(u^\ast_{it} - \bm{x}_{it}' ( \bm{\beta} - \hat{\bm{\beta}} ) - ( \alpha_i - \hat{\alpha}_i ) ) + \frac{\lambda_T}{T} \sign( \alpha_i ),
  \end{equation*}
  where $u_{it}^\ast = w_{it} | \hat{u}_{it} |$.

  Write
  \begin{multline} \label{alstar_exp}
    \mathbb{H}_{Ti}^{(\alpha)*}(\bm{\theta}^*_i) = \mathbb{H}_{Ti}^{(\alpha)*}(\hat{\bm{\theta}}_i) + \left( \mathbb{H}_{Ti}^{(\alpha)*}(\bm{\theta}^*_i) - \mathbb{H}_{Ti}^{(\alpha)*}(\hat{\bm{\theta}}_i) - \exs{\mathbb{H}_{Ti}^{(\alpha)*}(\bm{\theta}^*_i) - \mathbb{H}_{Ti}^{(\alpha)*}(\hat{\bm{\theta}}_i)} \right) \\
    + \exs{\mathbb{H}_{Ti}^{(\alpha)*}(\bm{\theta}^*_i) - \mathbb{H}_{Ti}^{(\alpha)*}(\hat{\bm{\theta}}_i)} 
  \end{multline}

  For the next part, make the following definitions, which are sample analogs to quantities defined in Assumption~\ref{assume:Avar}.  Let $\bar{\varphi}_i = \frac{1}{T} \sum_t f_i(0 | \bm{x}_{it})$, $\bar{\bm{E}}_i = \frac{1}{T} \sum_t f_i(0 | \bm{x}_{it}) \bm{x}_{it}$, $\bar{\bm{J}}_i = \frac{1}{T} \sum_t f_i(0 | \bm{x}_{it}) \bm{x}_{it} \bm{x}_{it}'$ and $\bar{\bm{D}}_N = \frac{1}{N} \sum_i (\bar{\bm{J}}_i - \bar{\varphi}_i^{-1} \bar{\bm{E}}_i \bar{\bm{E}}_i')$.

  Part 2 of Lemma \ref{L3} and $\bm{\theta}^*_i \stackrel{p^*}{\rightarrow} \hat{\bm{\theta}}_i$ imply that for all $1 \leq i \leq N$,
  \begin{multline} \label{Hstar_alpha_exp}
    \exs{\mathbb{H}_{Ti}^{(\alpha)*}(\bm{\theta}^*_i) - \mathbb{H}_{Ti}^{(\alpha)*}(\hat{\bm{\theta}}_i)} = -\frac{1}{T} \sum_{t=1}^T f_i(0 | \bm{x}_{it}) \left( \bm{x}_{it}' (\bm{\beta}^* - \hat{\bm{\beta}}) + (\alpha_i^* - \hat{\alpha}_i) \right) + \frac{\lambda_T}{T} ( \sign(\alpha_i^*) - \sign(\hat{\alpha}_i) ) \\
    + O_{p^*} \left( (\alpha_i^* - \hat{\alpha}_i)^2 \vee \|\bm{\beta}^* - \hat{\bm{\beta}}\|^2 \right) + O_p \left( (\hat{\alpha}_i - \alpha_{i0})^2 \vee \|\hat{\bm{\beta}} - \bm{\beta}_0\|^2 \right).
  \end{multline}
  Rewrite~\eqref{alstar_exp} using the above equation as
  \begin{multline} \label{bahadur_alstar}
    \alpha_i^* - \hat{\alpha}_i = -\bar{\varphi}_i^{-1} \bar{\bm{E}}_i' (\bm{\beta}^* - \hat{\bm{\beta}}) + \bar{\varphi}_i^{-1} \left( \mathbb{H}_{Ti}^{(\alpha)*}(\hat{\bm{\theta}}_i) - \frac{\lambda_T}{T} \sign(\hat{\alpha}_i) \right) \\
    + \bar{\varphi}_i^{-1} \left( \mathbb{H}_{Ti}^{(\alpha)*}(\bm{\theta}^*_i) - \mathbb{H}_{Ti}^{(\alpha)*}(\hat{\bm{\theta}}_i) - \exs{\mathbb{H}_{Ti}^{(\alpha)*}(\bm{\theta}^*_i) - \mathbb{H}_{Ti}^{(\alpha)*}(\hat{\bm{\theta}}_i)} \right) \\
    - \bar{\varphi}_i^{-1} \left( \mathbb{H}_{Ti}^{(\alpha)*}(\bm{\theta}^*_i) - \frac{\lambda_T}{T} \sign(\alpha_i^*) \right) + O_{p^*} \left( (\alpha_i^* - \hat{\alpha}_i)^2 \vee \| \bm{\beta}^* - \hat{\bm{\beta}} \|^2 \right) + O_p \left( (\hat{\alpha}_i - \alpha_{i0}) ^2 \vee \| \hat{\bm{\beta}} - \bm{\beta}_0 \|^2 \right).
  \end{multline}

  Similarly,
  \begin{multline} \label{bestar_exp}
    \mathbb{H}_{Ti}^{(\beta)*}(\bm{\theta}^*_i) = \mathbb{H}_{Ti}^{(\beta)*}(\hat{\bm{\theta}}_i) + \left( \mathbb{H}_{Ti}^{(\beta)*}(\bm{\theta}^*_i) - \mathbb{H}_{Ti}^{(\beta)*}(\hat{\bm{\theta}}_i) - \exs{\mathbb{H}_{Ti}^{(\beta)*}(\bm{\theta}^*_i) - \mathbb{H}_{Ti}^{(\beta)*}(\hat{\bm{\theta}}_i)} \right) \\
    + \exs{\mathbb{H}_{Ti}^{(\beta)*}(\bm{\theta}^*_i) - \mathbb{H}_{Ti}^{(\beta)*}(\hat{\bm{\theta}}_i)}.
  \end{multline}
  Lemma~\ref{L3} can be used again to calculate the estimate
  \begin{multline} \label{Hstar_beta_exp}
    \exs{\mathbb{H}_{Ti}^{(\beta)*}(\bm{\theta}^*_i) - \mathbb{H}_{Ti}^{(\beta)*}(\hat{\bm{\theta}}_i)} = -\bar{\bm{J}}_i (\bm{\beta}^* - \hat{\bm{\beta}}) - \bar{\bm{E}}_i (\alpha_i^* - \hat{\alpha}_i) \\
    + o_{p^*} \left( \|\bm{\beta}^* - \hat{\bm{\beta}}\| \right) + o_p \left( \|\hat{\bm{\beta}} - \bm{\beta}_0\| \right) + O_{p^*} \left( \sup_i (\alpha_i^* - \hat{\alpha}_i)^2 \right) + O_p \left( \sup_i (\hat{\alpha}_i - \alpha_{i0})^2 \right).
  \end{multline}
  Now analogous to the proof Theorem~\ref{thm:AN}, define
  \begin{equation} \label{Kstar_def}
    \mathbb{K}_{Ti}^{(\theta)*}(\bm{\theta}_i) = \mathbb{H}_{Ti}^{(\beta)*}(\bm{\theta}_i) - \bar{\varphi}_i^{-1} \bar{\bm{E}}_i \left( \mathbb{H}_{Ti}^{(\alpha)*}(\bm{\theta}_i) - \frac{\lambda_T}{T} \sign(\alpha_i) \right)
  \end{equation}
  and note that $\mathbb{K}_{Ti}^{(\theta)*}(\bm{\theta}_i^*) = O_{p^*}(\lambda_T / T)$.  Then equation \eqref{bestar_exp} can be rewritten as
  \begin{multline}
    \left( \bar{\bm{J}}_i - \bar{\varphi}_i^{-1} \bar{\bm{E}}_i \bar{\bm{E}}_i' \right) ( \bm{\beta}^* - \hat{\bm{\beta}} ) + o_{p^*} \left( \|\bm{\beta}^* - \hat{\bm{\beta}}\| \right) + o_p \left( \|\hat{\bm{\beta}} - \bm{\beta}_0\| \right) = \mathbb{K}_{Ti}^{(\theta)*}(\hat{\bm{\theta}}_i) \\
    + \left( \mathbb{K}_{Ti}^{(\theta)*}(\bm{\theta}^*_i) - \mathbb{K}_{Ti}^{(\theta)*}(\hat{\bm{\theta}}_i) - \exs{\mathbb{K}_{Ti}^{(\theta)*}(\bm{\theta}^*_i) - \mathbb{K}_{Ti}^{(\theta)*}(\hat{\bm{\theta}}_i)} \right) \\
    + O_{p^*}(T^{-1} \lambda_T) + O_{p^*} \left( \sup_i (\alpha_i^* - \hat{\alpha}_i)^2 \right) + O_p \left( \sup_i (\hat{\alpha}_i - \alpha_{i0})^2 \right).
  \end{multline}

  Rearrange and average over $i$ to find
  \begin{multline} \label{betastar_expansion}
    \bm{\beta}^* - \hat{\bm{\beta}} + o_{p^*} \left( \|\bm{\beta}^* - \hat{\bm{\beta}}\| \right) + o_p \left( \|\hat{\bm{\beta}} - \bm{\beta}_0\| \right) = \bar{\bm{D}}_N^{-1} \frac{1}{N} \sum_{i=1}^N \mathbb{K}_{Ti}^{(\theta)*}(\hat{\bm{\theta}}_i) \\
    + \bar{\bm{D}}_N^{-1} \frac{1}{N} \sum_{i=1}^N \left( \mathbb{K}_{Ti}^{(\theta)*}(\bm{\theta}_{i}^*) - \mathbb{K}_{Ti}^{(\theta)*}(\hat{\bm{\theta}}_i) - \exs{\mathbb{K}_{Ti}^{(\theta)*}(\bm{\theta}^*_i) + \mathbb{K}_{Ti}^{(\theta)*}(\hat{\bm{\theta}}_i)} \right)  \\
    + O_{p^*}(T^{-1} \lambda_T) + O_{p^*} \left( \sup_i (\alpha_i^* - \hat{\alpha}_i)^2 \right) + O_p \left( \sup_i (\hat{\alpha}_i - \alpha_{i0})^2 \right).
  \end{multline}

  Next we find the stochastic order of the second term on the right-hand side of~\eqref{betastar_expansion}.  With $\bm{X}_{it} = (\bm{x}_{it}', 1)'$, let $\bm{X}_{it}' \bm{\Delta} = \bm{x}_{it}' (\hat{\bm{\beta}} - \bm{\beta}_0) + (\hat{\alpha}_i - \alpha_{i0})$ and $\bm{X}_{it}' \bm{\delta} = \bm{x}_{it}' (\bm{\beta}^* - \hat{\bm{\beta}}) + (\alpha_i^* - \hat{\alpha}_{i})$ and write $\hat{u}_{it} = u_{it} + \bm{X}_{it}' \bm{\Delta}$.  Define the functions $g_{\bm{\delta}}(w, u, \bm{X}, \bm{\Delta}) = I(w|u + \bm{X}'\bm{\Delta}| - \bm{X}'\bm{\delta} < 0) - I(w|u + \bm{X}'\bm{\Delta}| < 0)$.  The class of functions $g_{\bm{\delta}} - \exs{g_{\bm{\delta}}}$ is a bounded, mean-zero VC-subgraph class of functions.  Finally, letting $\underline{c} = \min\{c_1, c_2\}$, where $c_1, c_2$ were used in \ref{C2}, the unconditional second moment of $g_\delta$ satisfies
  \begin{align}
    \ex{(g_{\bm{\delta}}(\bm{Z}_{it}))^2} &= \ex{ I( |w_{it}| |u_{it} + \bm{X}_{it}'\bm{\Delta}| < |\bm{X}_{it}'\bm{\delta}| ) } \notag \\
    {} &\leq \ex{ I( |u_{it} + \bm{X}_{it}'\bm{\Delta}| < |\bm{X}_{it}'\bm{\delta}| / \underline{c}) } \notag \\
    {} &= \ex{ F_i(-\bm{X}_{it}'\bm{\Delta} + |\bm{X}_{it}'\bm{\delta}| / \underline{c} | \bm{X}_{it}) - F_i(-\bm{X}_{it}'\bm{\Delta} - |\bm{X}_{it}'\bm{\delta}| / \underline{c} | \bm{X}_{it}) }  \notag \\
    {} &\leq K (M+1) \|\bm{\delta}\|, \label{gstar_var}
  \end{align}
  the last inequality holding due to Assumption~\ref{assume:ID_weakconv}.  This implies that $\exs{(g_{\bm{\delta}}(\bm{Z}_{it}) - \exs{g_{\bm{\delta}}(\bm{Z}_{it})})^2} \leq K(M+1)\|\bm{\delta}\|$ with probability approaching 1.  Then Proposition B.1 of \citet{kK12} implies that with $\delta^*_N = \sup_i |\alpha_i^* - \hat{\alpha}_i| + \|\bm{\beta}^* - \hat{\bm{\beta}}\|$ and $d_{NT}^* = |\log \delta^*_N| / T \vee \sqrt{\delta_N^* |\log \delta_N^*| / T}$,
  \begin{equation} \label{bootstrap_steq}
    \bar{\bm{D}}_N^{-1} \frac{1}{N} \sum_{i=1}^N \left( \mathbb{K}_{Ti}^{(\theta)*}(\bm{\theta}_{i}^*) - \mathbb{K}_{Ti}^{(\theta)*}(\hat{\bm{\theta}}_i) - \exs{\mathbb{K}_{Ti}^{(\theta)*}(\bm{\theta}^*_i) + \mathbb{K}_{Ti}^{(\theta)*}(\hat{\bm{\theta}}_i)} \right) = O_{p^*}(d_{NT}^*) = o_{p^*}(T^{-1/2}),
  \end{equation}
  where the last equality comes from the consistency of $\bm{\theta}^*$.

  Combine~\eqref{betastar_expansion}, \eqref{bootstrap_steq}, the fact that the first term on the right-hand side of~\eqref{betastar_expansion} is $O_{p^*}((NT)^{-1/2}) = o_{p^*}(T^{-1/2})$ and $\sup_i | \hat{\alpha}_i - \alpha_{i0} | = O_p \left( T^{-1/2} (\log N)^{1/2} \right) = o_p(T^{-1/2})$ to write
  \begin{equation} \label{bstar_approx}
    \|\bm{\beta}^* - \hat{\bm{\beta}}\| = O_{p^*} \left( \sup_i (\alpha_i^* - \hat{\alpha}_i)^2 \right) + O_{p^*}(T^{-1} \lambda_T) + o_{p^*}\left( T^{-1/2} \right) + o_p \left( T^{-1/2} \right).
  \end{equation}
  Then the preliminary rates of convergence of the coordinates of $\bm{\theta}^*_i$ can be established similarly to the proof of asymptotic normality of $\hat{\bm{\theta}}_i$.  For example, using~\eqref{bahadur_alstar},
  \begin{multline} \label{bootstrap_supalpha_order}
    \sup_i |\alpha_i^* - \hat{\alpha}_i| \leq K \Bigg\{ \sup_i \left| \mathbb{H}_{Ti}^{(\alpha)*}(\hat{\bm{\theta}}_i) - \frac{\lambda_T}{T} \sign(\hat{\alpha}_i) \right| \\
    + \sup_i \left| \mathbb{H}_{Ti}^{(\alpha)*}(\bm{\theta}_{i}) - \mathbb{H}_{Ti}^{(\alpha)*}(\hat{\bm{\theta}}_i) - \exs{\mathbb{H}_{Ti}^{(\alpha)*}(\bm{\theta}^*_i) + \mathbb{H}_{Ti}^{(\alpha)*}(\hat{\bm{\theta}}_i)} \right| \Bigg\} \\
    + O_{p^*}(T^{-1} \lambda_T) + o_{p^*}(T^{-1/2}) + o_p(T^{-1/2})
  \end{multline}
  with probability approaching 1.  These terms can be bounded by following the calculations similar to the asymptotic normality proof, conditional on the data, using the functions $g_{\bm{\delta}}(\bm{Z})$ defined earlier, resulting in $\sup_i |\alpha_i^* - \hat{\alpha}_i| = O_{p^*}(T^{-1/2} (\log N)^{1/2})$.  Using~\eqref{bstar_approx}, this implies $\|\bm{\beta}^* - \hat{\bm{\beta}}\| = o_{p^*}(T^{-1/2} (\log N)^{1/2})$.  The rest of the proof proceeds as in the proof of asymptotic normality of $\hat{\bm{\theta}}$, with the addition of the moment conditions on the $w_{it}$ and the convergence of $\bar{\varphi}_i$, $\bar{\bm{E}}_i$ and $\bar{\bm{J}}_i$ to their population counterparts for all $i$ using the law of large numbers as $N, T \rightarrow \infty$.
\end{proof}

\begin{lemma} \label{lem:qmoment}
  Suppose that Assumptions~\ref{C1}-\ref{C3} and~\ref{assume:data}-\ref{assume:xsupport} hold.  If $\bm{\theta}_i$ for $1 \leq i \leq N$ lie in a compact set and $\sup_{N,T} \ex{| \sqrt{N} \lambda_T / \sqrt{T} |^q} < \infty$ for $q > 2$, then $\sup_{N,T} \exs{\| \sqrt{NT} (\bm{\beta}^* - \hat{\bm{\beta}}) \|^q} < \infty$.
\end{lemma}

\begin{proof}[Proof of Lemma~\ref{lem:qmoment}]
  Follow the steps in the expansions used in Theorem~\ref{thm:boot} but write out the remainder terms explicitly.  Specifically, rewrite~\eqref{Hstar_alpha_exp} as
  \begin{equation*}
    \exs{\mathbb{H}_{Ti}^{(\alpha)*}(\bm{\theta}^*_i) - \mathbb{H}_{Ti}^{(\alpha)*}(\hat{\bm{\theta}}_i)} = -\bar{\bm{E}}_i' (\bm{\beta}^* - \hat{\bm{\beta}}) - \bar{\varphi}_i (\alpha_i^* - \hat{\alpha}_i) + (\lambda_T/T) \left( \sign( \alpha_i^* ) - \sign( \hat{\alpha}_i ) \right) + R_i^{(\alpha)*}
  \end{equation*}
  where
  \begin{multline*}
    R_i^{(\alpha)*} := \exs{ \frac{1}{T} \sum_{t=1}^T \psi_{\tau}(w_{it} |\hat{u}_{it}| - \bm{x}_{it}' ( \bm{\beta}^* - \hat{\bm{\beta}} ) - ( \alpha_i^* - \hat{\alpha}_i )) - \frac{1}{T} \sum_{t=1}^T \psi_\tau(w_{it} |\hat{u}_{it}|) } \\
    + \bar{\bm{E}}_i' (\bm{\beta}^* - \hat{\bm{\beta}}) + \bar{\varphi}_i (\alpha_i^* - \hat{\alpha}_i).
  \end{multline*}
  Similarly define
  \begin{equation*}
    R_i^{(\beta)*} := \exs{\mathbb{H}_{Ti}^{(\beta)*}(\bm{\theta}^*_i) - \mathbb{H}_{Ti}^{(\beta)*}(\hat{\bm{\theta}}_i)} + \bar{\bm{J}}_i (\bm{\beta}^* - \hat{\bm{\beta}}) + \bar{\bm{E}}_i (\alpha_i^* - \hat{\alpha}_i)
  \end{equation*}
  which was represented by error terms in equation~\eqref{Hstar_beta_exp} in the proof of Theorem~\ref{thm:boot}.  Then~\eqref{betastar_expansion} can be equivalently written
  \begin{multline} \label{big_eqn}
    \bar{\bm{D}}_N ( \bm{\beta}^* - \hat{\bm{\beta}} ) - \frac{1}{N} \sum_{i=1}^N \left( R_i^{(\beta)*} - \bar{\varphi}_i^{-1} \bar{\bm{E}}_i R_i^{(\alpha)*} \right) + O_{p^*}(\lambda_T / T) = \frac{1}{N} \sum_{i=1}^N \mathbb{K}_{Ti}^{(\theta)*}(\hat{\bm{\theta}}_i) \\
    + \frac{1}{N} \sum_{i=1}^N \left( \mathbb{K}_{Ti}^{(\theta)*}(\bm{\theta}_i^*) - \mathbb{K}_{Ti}^{(\theta)*}(\hat{\bm{\theta}}_i) - \exs{ \mathbb{K}_{Ti}^{(\theta)*}(\bm{\theta}_i^*) - \mathbb{K}_{Ti}^{(\theta)*}(\hat{\bm{\theta}}_i) } \right).
  \end{multline}
  The left-hand side includes the remainder terms, which are functions of the difference between bootstrap parameter estimate and original-sample parameter estimate.  Assuming the parameters lie in a compact set implies that the remainder terms are uniformly bounded and have $q$-th moment.  The $q$-th moment of the other remainder is finite by assumption.  The rest of the proof shows that the right hand side is uniformly $q$-integrable.

  Consider the first term on the right-hand side of~\eqref{big_eqn}, scaled by $\sqrt{NT}$:
  \begin{equation} \label{part1}
    \sqrt{NT} \mathbb{K}_{Ti}^{(\theta)*}(\hat{\bm{\theta}}_i) = \frac{1}{\sqrt{NT}} \sum_{i=1}^N \sum_{t=1}^T (\bm{x}_{it} - \bar{\varphi}_i^{-1} \bar{\bm{E}}_i) (\tau - I(w_{it} | \hat{u}_{it} | \leq 0)).
  \end{equation}
  The bounds on $\bm{x}_{it}$ and the density of the errors imply that $\bar{\varphi}_i^{-1} \bar{\bm{E}}_i$ are bounded.  Given the conditions on $G_W$, the expected value of each summand is zero conditional on the data.

  Let $\bm{X}_{it} = (\bm{x}_{it}', 1)'$, $\bm{Z}_{it} = (w_{it}, u_{it}, \bm{X}_{it})$ and $\bm{\Delta} \in \RR^{p+1}$.  Define the class of functions $\mathcal{H} = \{ h_{\bm{\Delta}}(\bm{Z}) := I(w|u + \bm{X}'\bm{\Delta}| < 0) : \bm{\Delta} \in \RR^{p+1} \}$.  This class of indicators is a VC subgraph class.  To see this, first rewrite
  \begin{equation*}
    \{ w|u + \bm{X}'\bm{\Delta}| < 0 \} = \\
    \{ w|u + \bm{X}'\bm{\Delta}| < 0 \} \cap \{w < 0\} \cup \{ w|u + \bm{X}'\bm{\Delta}| < 0 \} \cap \{w > 0\}.
  \end{equation*}
  For $w$ positive (the opposite case is analogous), the class of sets $\{w|u + \bm{X}'\bm{\Delta}| < 0\} \cap \{ w > 0 \}$ is equivalent to the class $\{ |u + \bm{X}'\bm{\Delta}| < 0 \} = \{ u + \bm{X}'\bm{\Delta} < 0 \} \cap \{ -u - \bm{X}'\bm{\Delta} < 0 \}$.  Each of these sets forms a VC class \citep[Problem 2.6.14]{vanderVaartWellner96} and the class of their intersections is also a VC class \citep[Lemma 2.6.17]{vanderVaartWellner96}.  Then the class of unions of sets formed in this way is also a VC class, and $\mathcal{H}$ is a VC subgraph class.

  Because of the fact that the indicators in equation~\eqref{part1} are a VC subgraph class bounded by 1, their uniform covering number satisfies $\sup_Q N(\epsilon, \mathcal{H}, L_2(Q)) \leq A \left( \frac{1}{\epsilon} \right)^v$ for some $A, v$ and $0 < \epsilon < 1$ and $Q$ a probability measure.  This implies that
  \begin{equation} \label{Jint}
    J(1, \mathcal{H}) := \sup_Q \int_0^1 \sqrt{ 1 + \log N(\epsilon \| F \|_{Q,2}, \mathcal{H}, L_2(Q)) } \ud \epsilon < \infty
  \end{equation}
  where the supremum is taken over all discrete probability measures $Q$ \citep[p. 239]{vanderVaartWellner96}.  Then Theorem 2.14.1 of \citet{vanderVaartWellner96} implies that there exists a constant $C$ such that
  \begin{equation*}
    \max_{1 \leq i \leq N} \sup_{T \geq 1} \ex{ \left\| \frac{1}{\sqrt{T}} \sum_{t=1}^T (\bm{x}_{it} - \bar{\varphi}_i^{-1} \bar{\bm{E}}_i) (\tau - I(w_{it} | \hat{u}_{it} | \leq 0)) \right\|^q } \leq C J(1, \mathcal{H}).
  \end{equation*}
  Then \citet[Problem 2.3.4]{vanderVaartWellner96} implies that
  \begin{equation*}
    \sup_{N,T} \ex{ \left\| \frac{1}{\sqrt{NT}} \sum_{i=1}^N \sum_{t=1}^T (\bm{x}_{it} - \bar{\varphi}_i^{-1} \bar{\bm{E}}_i) (\tau - I(w_{it} | \hat{u}_{it} | \leq 0)) \right\|^q } < \infty.
  \end{equation*}

  For the second term, it is sufficient to consider, for any $i$,
  \begin{multline*}
    \sup_{T \geq 1} \mathrm{E} \bigg[ \bigg| \frac{1}{\sqrt{T}} \sum_{t=1}^T \psi_\tau \left( w_{it} |\hat{u}_{it}| - \bm{x}_{it}'(\bm{\beta}^* - \hat{\bm{\beta}}) - (\alpha_i^* - \hat{\alpha}_i) \right) - \psi_\tau \left( w_{it} |\hat{u}_{it}| \right) \\
    - \exs{\psi_\tau \left( w_{it} |\hat{u}_{it}| - \bm{x}_{it}'(\bm{\beta}^* - \hat{\bm{\beta}}) - (\alpha_i^* - \hat{\alpha}_i) \right) - \psi_\tau \left( w_{it} |\hat{u}_{it}| \right)} \bigg|^q \bigg] .
  \end{multline*}

  Let $\bm{X}_{it} = (\bm{x}_{it}', 1)'$, $\bm{Z}_{it} = (w_{it}, u_{it}, \bm{X}_{it})$ and $\bm{\xi} = (\bm{\Delta}', \bm{\delta}')' \in \RR^{2(p+1)}$.  Define the functions $g_{\bm{\xi}}(\bm{Z}) = I(w|u + \bm{X}'\bm{\Delta}| - \bm{X}'\bm{\delta} < 0) - I(w|u + \bm{X}'\bm{\Delta}| < 0)$ and the class of functions $\mathcal{G} = \{g_{\bm{\xi}} - \ex{g_{\bm{\xi}}} : \bm{\xi} \in \RR^{2(p+1)} \}$.  Then the above display is finite if
  \begin{equation*}
    \ex{ \left| \sup_{\bm{\xi}} \frac{1}{\sqrt{T}} \sum_{t=1}^T \left( g_{\bm{\xi}}(\bm{Z}_{it}) - \exs{g_{\bm{\xi}}(\bm{Z}_{it})} \right) \right|^q } < \infty.
  \end{equation*}
  However, manipulations similar to the previous step show that $\mathcal{G}$ is also a VC-subgraph class, and therefore, using~\eqref{Jint} for the class $\mathcal{G}$, we have for another constant $C$ that
  \begin{equation*}
    \max_{1 \leq i \leq N} \sup_{T \geq 1} \ex{ \left\| \sqrt{T} \left( \mathbb{H}_{Ti}^{(\alpha)*}(\bm{\theta}_{i}) - \mathbb{H}_{Ti}^{(\alpha)*}(\hat{\bm{\theta}}_i) - \exs{\mathbb{H}_{Ti}^{(\alpha)*}(\bm{\theta}^*_i) + \mathbb{H}_{Ti}^{(\alpha)*}(\hat{\bm{\theta}}_i)} \right) \right\|^q } \leq C J(1, \mathcal{G}) < \infty.
  \end{equation*}
  This implies
  \begin{equation*}
    \sup_{N,T} \ex{ \left\| \sqrt{NT} \frac{1}{N} \sum_{i=1}^N \left( \mathbb{K}_{Ti}^{(\theta)*}(\bm{\theta}_{i}) - \mathbb{K}_{Ti}^{(\theta)*}(\hat{\bm{\theta}}_i) - \exs{\mathbb{K}_{Ti}^{(\theta)*}(\bm{\theta}^*_i) + \mathbb{K}_{Ti}^{(\theta)*}(\hat{\bm{\theta}}_i)} \right) \right\|^q } < \infty.
  \end{equation*}

  The $c_r$ inequality implies that the right-hand side of~\eqref{big_eqn} is uniformly $q$-integrable.  Under the assumption that $\bar{\bm{D}}_N$ is invertible, $\hat{\bm{\beta}} - \bm{\beta}_0$ must be as well.
\end{proof}

\begin{proof}[Proof of Theorem~\ref{thm:var}]
  This proof is similar to Theorem 3.2, part (i) of \citet{Andreas2017}.  Let $\bm{Z}_{NT}^* = \sqrt{NT}(\bm{\beta}^* - \hat{\bm{\beta}})$.  Theorem~\ref{thm:boot} shows that $\bm{Z}_{NT}^* \indist \bm{Z}$ in probability, where $\bm{Z}$ is defined by the condition $\sqrt{NT}(\hat{\bm{\beta}} - \bm{\beta}_0) \indist \bm{Z}$.  $\exs{\bm{Z}_{NT}^* \bm{Z}_{NT}^{*'}} \inpr \ex{\bm{Z} \bm{Z}'}$ if and only if each coordinate converges in probability, so assume that $p = 1$ and we may deal with the 1-dimensional random variables $Z_{NT}^*$ and $Z$.  For any $K > 0$, write
  \begin{multline*}
    |\exs{Z_{NT}^{*2}} - \ex{Z^2}| \leq \exs{Z_{NT}^{*2}} - \exs{\min\{ Z_{NT}^{*2}, K \}} \\
    {} + \left| \exs{\min\{ Z_{NT}^{*2}, K \}} - \ex{\min\{ Z^2, K \}} \right| + \left| \ex{\min\{ Z^2, K \}} - \ex{Z^2} \right|.
  \end{multline*}
  The portmanteau lemma \citep{vanderVaart98} implies that $\exs{g(Z_{NT}^*)} \inpr \ex{g(Z)}$ for all continuous and bounded functions $g$, and that the second term on the right-hand side converges in probability to zero.  The first and third terms on the right-hand side are similar; consider just the first term.  Note that $\exs{Z_{NT}^{*2}} - \exs{\min\{ Z_{NT}^{*2}, K \}} \leq \exs{ Z_{NT}^{*2} I(Z_{NT}^{*2} > K) }$.  For any $\epsilon > 0$,
  \begin{equation*}
    \ex{\exs{ Z_{NT}^{*2} I(Z_{NT}^{*2} > K)}} \leq \sup_{N,T} \ex{Z_{NT}^{*2(1+\epsilon)}} K^{-\epsilon}
  \end{equation*}
  where the expectation on the right-hand side is taken with respect to all the random variables.  Lemma~\ref{lem:qmoment} (letting $q = 2(1 + \epsilon)$ there) implies that the expectation on the right-hand side is finite, so the right-hand side converges to zero as $K \rightarrow \infty$.  The Markov inequality implies the result.
\end{proof}

\begin{proof}[Proof of Theorem~\ref{thm:AN_dependent}]
  The proof of this theorem requires minor modifications to that of Theorems~\ref{thm:consistent} and \ref{thm:AN}.  Therefore we only specify the differences here.

  To show consistency, first note that Assumption~\ref{assume:ID_weakconv} implies Assumption~\ref{assume:ID}, used towards the beginning of the consistency proof.  Next, the bound using Hoeffding's inequality must be replaced.  Imposing the condition on $\lambda_T$ and choosing $q = [\sqrt{T}]$ and $s = 2 \log N$, Corollary C.1 of~\citet{kK12} implies that
  \begin{equation*}
    \max_{1 \leq i \leq N} \mathrm{P} \bigg\{ \sup_{\bm{\theta}_i \in \mathcal{B}_i(\phi)}  \Big| (\lambda_T/T) (|  \alpha_i  | - |  \alpha_{i0}  |) \Big| + \Big| \Delta_{Ti}(\bm{\theta}_i) - \ex{\Delta_{Ti}(\bm{\theta}_i)} \Big| \geq \epsilon_\phi \bigg\} = o(N^{-1}),
  \end{equation*}
  which implies (along with the rest of the argument in Theorem~\ref{thm:consistent}) consistency of the estimator.

  To show asymptotic normality, there are several terms that should be bounded under the dependent error condition.  The proof follows that of Theorem~\ref{thm:AN} until equation~\eqref{eq:big_eqn}.  The arguments leading to an analog of equation~\eqref{alpha_rate} are as in the proof of Theorem~5.1 of \citet{kK12}~--- specifically, use Corollary C.1 and Lemma C.1 with $q = [T^c]$ for some sufficiently small $0 < c < 1$ and $s = 2 \log N$ to show that
  \begin{align*}
    \left\| \frac{1}{N} \sum_{i=1}^N \mathbb{K}_{Ti}^{(\theta)}(\hat{\bm{\theta}}_{i}) - K_{Ti}^{(\theta)}(\hat{\bm{\theta}}_i) - \mathbb{K}_{Ti}^{(\theta)}(\bm{\theta}_{i0}) + K_{Ti}^{(\theta)}(\bm{\theta}_{i0}) \right\| &= O_p(T^{-1/2} \delta_{N}^{1/4} (\log N)^{1/2} \vee T^{c-1} \log N) \\
    {} &= o_p(T^{-1/2} (\log N)^{1/2}) \\
    \intertext{and similarly, using the same $q = [T^c]$ and $s = 2\log N$,}
    \sup_i \left| \mathbb{H}_{Ti}^{(\alpha)}(\bm{\theta}_{i0}) - \frac{\lambda_T}{T} \sign(\alpha_{i0}) \right| &= O_p(T^{-1/2} (\log N)^{1/2}) \\
    \sup_i \left\| \mathbb{H}_{Ti}^{(\alpha)}(\hat{\bm{\alpha}}_{i}) - H_{Ti}^{(\alpha)}(\hat{\bm{\alpha}}_i) - \mathbb{H}_{Ti}^{(\alpha)}(\bm{\alpha}_{i0}) + H_{Ti}^{(\alpha)}(\bm{\alpha}_{i0}) \right\| &= o_p(T^{-1/2} (\log N)^{1/2}).
  \end{align*}

  To show the asymptotic normality of the term analogous to the final sum in the proof of Theorem~\ref{thm:AN}, note that all the $\mathbb{K}_{Ti}^{(\theta)}(\bm{\theta}_{i0})$ which are defined below equation \eqref{opt_error} are independent by Assumption \ref{assume:stationary}.  For a given $i$, $(\tau - I(y_{it} \leq \bm{x}_{it}' \bm{\beta}_0 + \alpha_{i0})) (\bm{x}_{it} - \varphi_i^{-1} \bm{E}_i)$ are uniformly bounded so $\sup_i \ex{ | \mathbb{K}_{Ti}^{(\theta)}(\bm{\theta}_{i0}) |^3 } < \infty$, while $\lim_{N \rightarrow \infty} \frac{1}{N} \sum_{i=1}^N \Var ( \mathbb{K}_{Ti}^{(\theta)}(\bm{\theta}_{i0}) ) = \tilde{\bm{V}}$ which is positive definite by assumption.
\end{proof}

\begin{proof}[Proof of Theorem~\ref{thm:boot_dependent}]
The proof is a modification of the proof of Theorem \ref{thm:boot}, following the developments in Theorem \ref{thm:AN_dependent}. To save space and avoid repetition, we concentrate our attention on the modifications of the proof.

For consistency we need a bootstrap equivalent of \eqref{delcirc_consistent}.  Apply the Bernstein inequality for $\beta$-mixing sequences in Corollary C.1 of~\citet{kK12}, choosing $q = [\sqrt{T}]$ and $s = 2\log N$.  Because of the condition on $\lambda_T$ we concentrate on the second term of \eqref{delcirc_consistent}. Under Assumption \ref{assume:stationary}, we have that
\begin{equation*}
  \max_{1 \leq i \leq N} \mathrm{P}^* \bigg\{ \sup_{\bm{\theta}_i, \in \mathcal{B}_i(\phi)} \Big| \Delta_{Ti}^\circ(\bm{\theta}_i) - \exs{\Delta_{Ti}^\circ(\bm{\theta}_i)} \Big| \geq \epsilon_\phi \bigg\} = o_p\left(N^{-1}\right).
\end{equation*}

  The proof of asymptotic normality is analogous to that of Theorem~\ref{thm:boot} through expansion \eqref{betastar_expansion}. The condition on $\lambda_T$ and Theorem~\ref{thm:AN} imply that several of the remainder terms are small, and we need only make one order estimate in~\eqref{bootstrap_steq} and two estimates in~\eqref{bootstrap_supalpha_order} under the $\beta$-mixing assumption.

  First, find an expression similar to~\eqref{bootstrap_steq} under Assumption~\ref{assume:stationary}.  The calculations in Theorem~\ref{thm:boot} leading up to~\eqref{gstar_var} imply that $\exs{(g_{\bm{\delta}}(\bm{Z}_{it}))^2} \leq C\|\bm{\delta}\|$ with probability approaching 1, and the Cauchy-Schwarz inequality implies similarly that for any $\bm{\delta}_1, \bm{\delta}_2$, $\exs{ |g_{\bm{\delta}_1}(\bm{Z}_{it}) \cdot g_{\bm{\delta}_2}(\bm{Z}_{it})|^2} \leq C\|\bm{\delta}_1\|\|\bm{\delta}_2\|$ with probability approaching 1.  Therefore Lemma~C.1 of \citet{kK12} implies that for any positive integer $q$, with $\delta^*_N = \sup_i |\alpha_i^* - \hat{\alpha}_i| + \|\bm{\beta}^* - \hat{\bm{\beta}}\|$,
  \begin{equation*}
    \text{Var}^* \left( \frac{1}{\sqrt{q}} \sum_{t=1}^q g_{\bm{\delta}}(\bm{Z}_{it}) \right) \leq (\delta_N^*)^{1/2}
  \end{equation*}
  with probability approaching 1.  For some $c \in (0, 1)$ let $q = [T^c]$ and $s = 2 \log N$, and apply Corollary C.1 of \citet{kK12} to find 
  \begin{multline} \label{bootstrap_steq_stationary}
    \bar{\bm{D}}_N^{-1} \frac{1}{N} \sum_{i=1}^N \left( \mathbb{K}_{Ti}^{(\theta)*}(\bm{\theta}_{i}^*) - \mathbb{K}_{Ti}^{(\theta)*}(\hat{\bm{\theta}}_i) - \exs{\mathbb{K}_{Ti}^{(\theta)*}(\bm{\theta}^*_i) + \mathbb{K}_{Ti}^{(\theta)*}(\hat{\bm{\theta}}_i)} \right) \\
    = O_{p^*} \left( T^{-1/2} (\delta_N^*)^{1/4} (\log N)^{1/2} \vee T^{c-1} \log N \right).
  \end{multline}

  Second, consider the expansion~\eqref{bootstrap_supalpha_order} under Assumption~\ref{assume:stationary}.  The second term is $o_{p^*}(T^{-1/2})$ using the result from the previous paragraph.  The first term is an average of (under the bootstrap measure) mean-zero terms.  It can be verified directly that $\exs{\psi_\tau^2(u_{it}^*)} = \tau (1 - \tau)$ and $\exs{|\psi_\tau(u_{it}^*) \psi_\tau(u_{is}^*)|^2} \leq \tau^2(1 - \tau)^2$.  Then Corollary~C.1 of \citet{kK12} implies that (using $s = 2 \log N$ and $q = [T^c]$)
  \begin{equation} \label{bootstrap_supalpha_stationary_order}
    \sup_i \left| \mathbb{H}_{Ti}^{(\alpha)*}(\hat{\bm{\theta}}_i) - \frac{\lambda_T}{T} \sign(\hat{\alpha}_i) \right| = O_{p^*} \left( T^{-1/2} (\log N)^{1/2} \vee T^{c-1} \log N \right).
  \end{equation}

  Now using~\eqref{bootstrap_steq_stationary} and~\eqref{bootstrap_supalpha_order} along with the rate condition on $N$ and $T$ and the condition on $\lambda_T$, we have (recalling definition~\eqref{Kstar_def})
  \begin{align}
    \sqrt{NT} (\bm{\beta}^* - \hat{\bm{\beta}}) &= \bar{\bm{D}}_N^{-1} \sqrt{NT} \frac{1}{N} \sum_{i=1}^N \mathbb{K}_{Ti}^{(\theta)*}(\hat{\bm{\theta}}_i) + o_{p^*}(1) \label{baha_dep} 
  \end{align}
  Under conditions \ref{C1}-\ref{C3}, it is clear that $\exs{\mathbb{K}_{Ti}^{(\theta)*}(\hat{\bm{\theta}}_i)} = 0$ for all $i$, and
  \begin{multline*}
    \mbox{Var}^\ast \left( \mathbb{K}_{Ti}^{(\theta)*}(\hat{\bm{\theta}}_i) \right) = \frac{1}{T} \sum_{t=1}^T (\bm{x}_{it} - \bar{\varphi}_i^{-1} \bar{\bm{E}}_i) (\bm{x}_{it} - \bar{\varphi}_i^{-1} \bar{\bm{E}}_i)' \exs{ \psi_\tau^2(u_{it}^*) } \\
    + 2 \sum_{j=1}^{T-1} (1 - j/T) (\bm{x}_{it} - \bar{\varphi}_i^{-1} \bar{\bm{E}}_i) (\bm{x}_{it+j} - \bar{\varphi}_i^{-1} \bar{\bm{E}}_i)' \exs{ \psi_\tau(u_{it}^*) \psi_\tau(u_{it+j}^*) }.
  \end{multline*}
  It can be calculated directly that the expected values in the first sum on the right-hand side are all $\tau(1- \tau)$.  Therefore (given the convergence in probability of $\bar{\varphi}_i^{-1} \bar{\bm{E}}_i$ to $\varphi_i^{-1} \bm{E}_i$) for consistent variance estimation it is sufficient to show that,
  \begin{equation} \label{covariances_converge}
    \plim_{N, T \rightarrow \infty} \frac{1}{N} \sum_{i=1}^N \sum_{j=1}^{T-1} (1 - j/T) \left( \exs{ \psi_\tau(u_{it}^*) \psi_\tau(u_{it+j}^*) } - \ex{ \psi_\tau(u_{it}) \psi_\tau(u_{it+j}) | \bm{x}_{it}, \bm{x}_{it+j} } \right) = 0.
  \end{equation}

  For any $(i, t)$ we have
  \begin{align*}
    \exs{ \psi_\tau(u_{it}^*) \psi_\tau(u_{it+j}^*) } &= \exs{ (\tau - I(u_{it}^* < 0)) (\tau - I(u_{it+j}^* < 0)) } \\
    {} &= \exs{ (\tau - I(w_{it} < 0)) (\tau - I(w_{it+j} < 0)) } = \tau - 2\tau^2 + \text{P}^* \{I(w_{it} < 0, w_{it+j} < 0)\}.
  \end{align*}
  Similarly,
  \begin{align*}
    \ex{ \psi_\tau(u_{it}) \psi_\tau(u_{it+j}) | \bm{x}_{it}, \bm{x}_{it+j} } &= \ex{ (\tau - I(u_{it} < 0)) (\tau - I(u_{it+j} < 0)) | \bm{x}_{it}, \bm{x}_{it+j} } \\
    {} &= \tau - 2\tau^2 + \prob{ u_{it} < 0, u_{it+j} < 0 | \bm{x}_{it} \bm{x}_{it+j} }.
  \end{align*}
  Inserting these expressions in~\eqref{covariances_converge}, it can be seen that \ref{assume:weight_dependent} implies the variance is correctly estimated.

  Finally, we apply a CLT for dependent sequences to~\eqref{baha_dep}. As in Theorem \ref{thm:AN_dependent}, we check a Lyapunov condition on the sum of the $ \mathbb{K}_{Ti}^{(\theta)*}(\hat{\bm{\theta}}_i)$ terms by \ref{assume:stationary}. By Assumptions \ref{assume:xsupport}, \ref{assume:ID_weakconv}, and \ref{assume:Avar}, $\psi_{\tau}( u_{it}^\ast ) (\bm{x}_{it} - \bar{\varphi}_i^{-1} \bar{\bm{E}}_i)$ is uniformly bounded. Moreover, under conditions \ref{assume:stationary} and \ref{assume:density} and the conditions on $w_{it}$, $\sup_i \exs{ | \mathbb{K}_{Ti}^{(\theta)*}(\hat{\bm{\theta}}_i) |^3 } = O_p(1)$ and $\sum_{i=1}^{N} \exs{ | \mathbb{K}_{Ti}^{(\theta)*}(\hat{\bm{\theta}}_i) |^3 } = o_p(N^{3/2})$.  This implies the result.
\end{proof}

\begin{proof}[Proof of Theorem~\ref{thm:rates}]
  The proof of this result is identical to the proof of Theorem~\ref{thm:AN} through equation~\eqref{eq:big_eqn}.  Lemma~\ref{lem:alpha_rate} in the supplementary appendix shows that
  $$\sup_i |\hat{\alpha}_i - \alpha_{i0}| = O_p\left( \| \hat{\bm{\beta}} - \bm{\beta}_0 \| + T^{-1/2} (\log T)^{1/2} + T^{-1} \lambda_T \right).$$
  Rewriting~\eqref{eq:big_eqn} using this result (and given that $T^{-2} \lambda_T = o(T^{-1} \lambda_T)$), 
  \begin{multline} \label{eq:big_eqn_versionA}
    \hat{\bm{\beta}} - \bm{\beta}_0 = \bm{D}_N^{-1} \frac{1}{N} \sum_{i=1}^N \mathbb{K}_{Ti}^{(\theta)}(\bm{\theta}_{i0}) + \bm{D}_N^{-1} \frac{1}{N} \sum_{i=1}^N \left( \mathbb{K}_{Ti}^{(\theta)}(\hat{\bm{\theta}}_{i}) - K_{Ti}^{(\theta)}(\hat{\bm{\theta}}_i) - \mathbb{K}_{Ti}^{(\theta)}(\bm{\theta}_{i0}) + K_{Ti}^{(\theta)}(\bm{\theta}_{i0}) \right)  \\
    + O_p(T^{-1} \lambda_T) + O_p\left( \| \hat{\bm{\beta}} - \bm{\beta}_0 \|^2 + T^{-1} \log T \right).
  \end{multline}

  To show the asymptotic normality of the first term, note that all the $\mathbb{K}_{Ti}^{(\theta)}(\bm{\theta}_{i0})$ are independent across $i$. For a given $i$, $(\tau - I(y_{it} \leq \bm{x}_{it}' \bm{\beta}_0 + \alpha_{i0})) (\bm{x}_{it} - \varphi_i^{-1} \bm{E}_i)$ are uniformly bounded so $\sup_i \ex{ | \mathbb{K}_{Ti}^{(\theta)}(\bm{\theta}_{i0}) |^3 } < \infty$, while $\lim_{N \rightarrow \infty} \frac{1}{N} \sum_{i=1}^N \Var ( \mathbb{K}_{Ti}^{(\theta)}(\bm{\theta}_{i0}) ) = \bm{V}$ which is positive definite by assumption. These conditions are sufficient to imply that a central limit theorem can be applied to the first term of~\eqref{eq:big_eqn_versionA}. Therefore this term is $O_p((NT)^{-1/2})$. 

  Lemma~\ref{lem:4k} shows that
  \begin{multline}
    \bm{D}_N^{-1} \frac{1}{N} \sum_{i=1}^N \left( \mathbb{K}_{Ti}^{(\theta)}(\hat{\bm{\theta}}_{i}) - K_{Ti}^{(\theta)}(\hat{\bm{\theta}}_i) - \mathbb{K}_{Ti}^{(\theta)}(\bm{\theta}_{i0}) + K_{Ti}^{(\theta)}(\bm{\theta}_{i0}) \right) \\
    = O_p \left( \|\hat{\bm{\beta}} - \bm{\beta}_0\|^{1/2} T^{-1/2} (\log T)^{1/2} + T^{-1} \log T + T^{-2/3} N^{-1/2} + T^{-1} (\log T)^{1/2} \lambda_T^{1/2} \right).
  \end{multline}
  Then
  \begin{multline*}
    \| \hat{\bm{\beta}} - \bm{\beta}_0 \| = O_p \left( (NT)^{-1/2} \right) + O_p \left( \| \hat{\bm{\beta}} - \bm{\beta}_0 \|^{1/2} T^{-1/2} (\log T)^{1/2} \right) + O_p \left( T^{-1} \log T \right) \\
    + O_p(T^{-1} \lambda_T) + O_p \left( T^{-1} (\log T)^{1/2} \lambda_T^{1/2} \right) .
  \end{multline*}
  Using the fact that $0 \leq \delta \leq a + b \delta^{1/2} \Rightarrow 0 \leq \delta \leq 4\max\{a, b^2\}$ (``fact 1'' from \citet{AGalvao2020}), we may shorten this to
  \begin{equation*}
    \| \hat{\bm{\beta}} - \bm{\beta}_0 \| = O_p \left( (NT)^{-1/2} \right) + O_p \left( T^{-1} \log T \right) + O_p(T^{-1} \lambda_T) + O_p \left( T^{-1} (\log T)^{1/2} \lambda_T^{1/2} \right) .
  \end{equation*}
  If $\lambda_T = O_p(\log T)$, the final three remainder terms have the same order. If $N T^{-1} (\log T)^2 \rightarrow 0$ then the asymptotically normal term dominates, implying the result.
\end{proof}

\begin{proof}[Proof of Theorem~\ref{thm:boot_rates}]
  The proof of this theorem is identical to that of Theorem~\ref{thm:boot} up to~\eqref{betastar_expansion}, reprinted here for convenience with some remainder terms changed using the assumption that $\lambda_T = O_p(\log T)$ and what is known of $\hat{\bm{\theta}}$ from previous theorems:
  \begin{multline} \label{betastar_expansion_reprint}
    \bm{\beta}^* - \hat{\bm{\beta}} + o_{p^*} \left( \|\bm{\beta}^* - \hat{\bm{\beta}}\| \right) + o_p \left( (NT)^{-1/2} \right) = \bar{\bm{D}}_N^{-1} \frac{1}{N} \sum_{i=1}^N \mathbb{K}_{Ti}^{(\theta)*}(\hat{\bm{\theta}}_i) \\
    + \bar{\bm{D}}_N^{-1} \frac{1}{N} \sum_{i=1}^N \left( \mathbb{K}_{Ti}^{(\theta)*}(\bm{\theta}_{i}^*) - \mathbb{K}_{Ti}^{(\theta)*}(\hat{\bm{\theta}}_i) - \exs{\mathbb{K}_{Ti}^{(\theta)*}(\bm{\theta}^*_i) + \mathbb{K}_{Ti}^{(\theta)*}(\hat{\bm{\theta}}_i)} \right)  \\
      + O_{p^*}(T^{-1} \lambda_T) + O_{p^*} \left( \sup_i (\alpha_i^* - \hat{\alpha}_i)^2 \right) + O_{p^*} \left( T^{-1} \log T \right).
  \end{multline}

  The inequalities of Lemma S.1.3 of \citet{ChaoVolgushevCheng17} do not apply to the functions in this expression because of the bootstrap weights in the functions.  However, the results of their subsection S.2.1 (which draw on \citet{Koltchinskii6} and \citet{Massart0}) may be used to tailor appropriate concentration inequalities.

  Rewriting \eqref{bahadur_alstar} with what is known thus far,
  \begin{multline} \label{alstar_exp_other}
    \alpha_i^* - \hat{\alpha}_i = O_{p^*} \left( \|\bm{\beta}^* - \hat{\bm{\beta}}\| \right) + \bar{\varphi}_i^{-1} \left( \mathbb{H}_{Ti}^{(\alpha)*}(\hat{\bm{\theta}}_i) - \frac{\lambda_T}{T} \sign(\hat{\alpha}_i) \right) \\
    + \bar{\varphi}_i^{-1} \left( \mathbb{H}_{Ti}^{(\alpha)*}(\bm{\theta}^*_i) - \mathbb{H}_{Ti}^{(\alpha)*}(\hat{\bm{\theta}}_i) - \exs{\mathbb{H}_{Ti}^{(\alpha)*}(\bm{\theta}^*_i) - \mathbb{H}_{Ti}^{(\alpha)*}(\hat{\bm{\theta}}_i)} \right) \\
    + O_{p^*}(\lambda_T / T) + O_p \left( T^{-1} \log T \right).
  \end{multline}
  Noting that
  \begin{equation*}
    \mathbb{H}_{Ti}^{(\alpha)*}(\hat{\bm{\theta}}_i) - \frac{\lambda_T}{T} \sign(\hat{\alpha}_i) = \frac{1}{T} \sum_{t=1}^T \psi_\tau( w_{it} | \hat{u}_{it} |),
  \end{equation*}
  this is a sum of 
	mean-zero functions with variance bounded by $\tau(1 - \tau)$ and that are members of a VC-subgraph class as described in Lemma~\ref{lem:qmoment}.  Therefore equations S.2.2 and S.2.3 of \citet{ChaoVolgushevCheng17} may be combined with the union bound to find that
  \begin{equation*}
    \sup_i \left| \mathbb{H}_{Ti}^{(\alpha)*}(\hat{\bm{\theta}}_i) - \frac{\lambda_T}{T} \sign(\hat{\alpha}_i) \right|  = O_{p^*} \left( T^{-1/2} (\log T)^{1/2} \right).
  \end{equation*}
  Similarly, the terms in the second line of~\eqref{alstar_exp_other} were described as $g_{\bm{\delta}}$ in the proof of Theorem~\ref{thm:boot}.  When $T^{-1}$ is smaller than the maximal variance of the $g_{\bm{\delta}}$ in this class, that is, when $\|C \bm{\delta}\| > T^{-1}$, S.2.2 and S.2.3 of \citet{ChaoVolgushevCheng17} may be used again with the union bound to find that, using the notation in the proof of Theorem~\ref{thm:boot},
  \begin{equation*}
    \sup_i \frac{1}{T} \sum_{t=1}^T \left( g_{\bm{\delta}}(\bm{Z}_{it}) - \exs{ g_{\bm{\delta}}(\bm{Z}_{it}) } \right) = O_{p^*} \left( \|\bm{\delta}\|^{1/2} T^{-1/2} (\log T)^{1/2} + T^{-1} \log T \right),
  \end{equation*}
  which in turn imply that
  \begin{multline*}
    \sup_i \left| \mathbb{H}_{Ti}^{(\alpha)*}(\bm{\theta}^*_i) - \mathbb{H}_{Ti}^{(\alpha)*}(\hat{\bm{\theta}}_i) - \exs{\mathbb{H}_{Ti}^{(\alpha)*}(\bm{\theta}^*_i) - \mathbb{H}_{Ti}^{(\alpha)*}(\hat{\bm{\theta}}_i)} \right| \\
    = O_{p^*} \left( \left( \|\bm{\beta}^* - \hat{\bm{\beta}}\| + \sup_i |\alpha_i^* - \hat{\alpha}_i| \right)^{1/2} T^{-1/2} (\log T)^{1/2} + T^{-1} \log T \right) \\
    = o_{p^*}(T^{-1/2} (\log T)^{1/2}).
  \end{multline*}
  These stochastic orders imply that
  \begin{equation*}
    \sup_i |\alpha_i^* - \hat{\alpha}_i| = O_{p^*} \left( \|\bm{\beta}^* - \hat{\bm{\beta}}\| + T^{-1/2}(\log T)^{1/2} + T^{-1} \lambda_T \right).
  \end{equation*}
  More lengthy calculations that are analogs to Lemmas~\ref{lem:4k} and~\ref{lem:atilde} in the supplemental appendix (conditional on the observations) imply that
  \begin{multline*}
    \frac{1}{N} \sum_{i=1}^N \left( \mathbb{K}_{Ti}^{(\theta)*}(\bm{\theta}_{i}^*) - \mathbb{K}_{Ti}^{(\theta)*}(\hat{\bm{\theta}}_i) - \exs{\mathbb{K}_{Ti}^{(\theta)*}(\bm{\theta}^*_i) + \mathbb{K}_{Ti}^{(\theta)*}(\hat{\bm{\theta}}_i)} \right) \\
    = O_{p^*} \left( \|\bm{\beta}^* - \hat{\bm{\beta}}\| T^{-1/2} (\log T)^{1/2} + T^{-1} \log T + T^{-2/3} N^{-1/2} + T^{-1} (\log T)^{1/2} \lambda_T^{1/2} \right).
  \end{multline*}
  Then the rest of the proof goes as in Theorem~\ref{thm:rates}, implying the result.
\end{proof}

\newpage

\vspace{10mm}

\begin{center}

\long\def\symbolfootnote[1]{\begingroup
  \def\thefootnote{\fnsymbol{footnote}}\footnote[1]{This draft: \today.}
\endgroup}

{\bf SUPPLEMENTARY APPENDIX TO \\ ``WILD BOOTSTRAP INFERENCE FOR PENALIZED QUANTILE REGRESSION FOR LONGITUDINAL DATA''}\symbolfootnote[1] 

\long\def\symbolfootnote[2]{\begingroup 
\def\thefootnote{\fnsymbol{footnote}}\footnote[2]{Carlos Lamarche: Department of Economics, University of Kentucky, 223G Gatton College of Business \& Economics, Lexington, KY 40506. Email: clamarche@uky.edu. Thomas Parker: Department of Economics, University of Waterloo, 200 University Ave. West, Waterloo, ON, Canada N2L 3G1. Email: tmparker@uwaterloo.ca}
\endgroup}

\vspace{2.5mm}

{\small CARLOS LAMARCHE AND THOMAS PARKER}\symbolfootnote[2]

\end{center}

\vspace{2.5mm}

\onehalfspacing

\section{Additional Theoretical Results} 

\setcounter{lemma}{0}
\renewcommand{\thelemma}{S.\arabic{lemma}}
\renewcommand{\theequation}{S.\arabic{equation}}

Lemma~\ref{lem:upperbnd} below implies a natural upper bound for $\lambda_T$.  If we consider the $\alpha_i$ as parameters associated with indicator functions for individual $i$ in the design matrix, then the column associated with each $i$ has $L_1$ norm equal to $T$.  In the text we set $\lambda_U = \max\{\tau, 1-\tau\}T$, because otherwise all the individual effects would be set to zero.

\begin{lemma} \label{lem:upperbnd}
  Subdivide the covariates for the $i$-th observation as $(\bm{X}_i', x_{pi})' \in \RR^p$.  Suppose that the conformable vector of estimates $(\hat{\bm{a}}, \hat{b})$ is defined by
  \begin{equation}
    (\hat{\bm{a}}, \hat{b}) = \argmin_{\bm{a}, b \in \RR^p} \sum_{i=1}^N \rho_\tau(y_i - \bm{X}_i' \bm{a} - bx_{pi}) + \lambda (\|\bm{a}\|_1 + |b|).
  \end{equation}
  Then letting $\bm{x}_p$ denote the $p$-th column of the design matrix,
  \begin{equation*}
    \max\{\tau, 1-\tau\} \|\bm{x}_p\|_1 < \lambda \quad \Rightarrow \quad \hat{b} = 0.
  \end{equation*}
\end{lemma}

\begin{proof}[Proof of Lemma~\ref{lem:upperbnd}]
  Note that if 
  \begin{equation*}
    \min_{\bm{a},b} \left( \sum_{i=1}^N \rho_\tau(y_i - \bm{X}_i' \bm{a} - bx_{pi}) + \lambda(\|\bm{a}\|_1 + |b|) \right) - \min_{\bm{a}} \left( \sum_{i=1}^N \rho_\tau(y_i - \bm{X}_i' \bm{a}) + \lambda \|\bm{a}\|_1 \right) > 0,
  \end{equation*}
  then it is optimal to set $\hat{b} = 0$.  Note that (using the definition of the full solution $(\hat{\bm{a}}, \hat{b})$)
  \begin{multline*}
    \min_{\bm{a},b} \left( \sum_{i=1}^N \rho_\tau(y_i - \bm{X}_i' \bm{a} - bx_{pi}) + \lambda(\|\bm{a}\|_1 + |b|) \right) - \min_{\bm{a}} \left( \sum_{i=1}^N \rho_\tau(y_i - \bm{X}_i' \bm{a}) + \lambda \|\bm{a}\|_1 \right) \geq \\
    \sum_{i=1}^N \rho_\tau(y_i - \bm{X}_i' \hat{\bm{a}} - \hat{b}x_{pi}) + \lambda(\|\hat{\bm{a}}\|_1 + |\hat{b}|) - \sum_{i=1}^N \rho_\tau(y_i - \bm{X}_i' \hat{\bm{a}}) - \lambda \|\hat{\bm{a}}\|_1 \\
    = \sum_{i=1}^N \left( \rho_\tau(y_i - \bm{X}_i' \hat{\bm{a}} - \hat{b}x_{pi}) - \rho_\tau(y_i - \bm{X}_i' \hat{\bm{a}}) \right) + \lambda |\hat{b}|.
  \end{multline*}
  Therefore if
  \begin{equation*}
    \sum_{i=1}^N \left( \rho_\tau(y_i - \bm{X}_i' \hat{\bm{a}}) - \rho_\tau(y_i - \bm{X}_i' \hat{\bm{a}} - \hat{b}x_{pi}) \right) < \lambda |\hat{b}|,
  \end{equation*}
  then $\hat{b} \neq 0$ is not optimal.  Applying Lemma~\ref{triangle} to the left-hand side of the above expression, we have
  \begin{align*}
    \sum_{i=1}^N \left( \rho_\tau(y_i - \bm{X}_i' \hat{\bm{a}}) - \rho_\tau(y_i - \bm{X}_i' \hat{\bm{a}} - \hat{b}x_{pi}) \right) &\leq \max\{\tau, 1-\tau\} \sum_{i=1}^N |\hat{b}x_{pi}| \\
    {} &\leq \max\{\tau, 1-\tau\} |\hat{b}| \|\bm{x}_p\|_1.
  \end{align*}
  Therefore for any $b \neq 0$, the condition
  \begin{equation*}
    \max\{\tau, 1-\tau\} |b| \|\bm{x}_p\|_1 < \lambda |b| \quad \Leftrightarrow \quad \max\{\tau, 1-\tau\} \|\bm{x}_p\|_1 < \lambda
  \end{equation*}
  implies that that $b$ is not an optimizer of the objective function.
\end{proof}

The following lemma collects together two results on expansions that are related to the wild bootstrap method described in the main text.
\begin{lemma}\label{L3}
  Let $u_{it}$ have conditional distribution $F_i$ and density functions $f_i$ as described in Assumptions~\ref{assume:ID} and~\ref{assume:ID_weakconv}, and suppose that Assumption~\ref{assume:xsupport} is satisfied.  Let $w_{it} \sim G_W$ be independent of $(u_{it}, \bm{x}_{it})$ and suppose its distribution satisfies Assumptions~\ref{C1}-\ref{C3}.  Then letting $\bm{X}_{it} = (\bm{x}_{it}', 1)'$, under either Assumption~\ref{assume:data} or~\ref{assume:stationary}:
  \begin{enumerate}
    \item For each $i$,
      \begin{equation*}
        \frac{1}{T} \sum_{t=1}^T \exs{ \int_0^{\bm{X}_{it}'\bm{\Delta}} (\psi_\tau(w_{it}|u_{it}| - s) - \psi_\tau(w_{it}|u_{it}|)) \ud s } = -f_i(0) \frac{1}{T} \sum_{t=1}^T \bm{\Delta}'\bm{X}_{it} \bm{X}_{it}'\bm{\Delta} + o_p(\|\bm{\Delta}\|^2).
      \end{equation*}
    \item For each $i$,
    \begin{multline*}
      \frac{1}{T} \sum_{t=1}^T \exs{ \psi_\tau( w_{it} |u_{it} + \bm{X}_{it}'\bm{\Delta}| - \bm{X}_{it}'\bm{\delta}) - \psi_\tau( w_{it} |u_{it} + \bm{X}_{it}'\bm{\Delta}|) } \\
      = -f_i(0) \frac{1}{T} \sum_{t=1}^T \bm{X}_{it}'\bm{\delta} + O_p((\|\bm{\Delta}\| + \|\bm{\delta}\|)^2).
    \end{multline*}
  \end{enumerate}
\end{lemma}

\begin{proof}
  Both parts of this proof use the identity
  \begin{equation} \label{wild_equiv}
    \psi_\tau(u - s) - \psi_\tau(u) = I(s < u < 0)I(s < 0) - I(0 < u < s)I(s \geq 0).
  \end{equation}
  First we show part 1.  Use~\eqref{wild_equiv} to write
  \begin{equation*}
    \psi_\tau( w_{it} |u_{it}| - s) - \psi_\tau( w_{it} |u_{it}| ) = I(s < w_{it}|u_{it}| < 0) I(s < 0) - I(0 < w_{it}|u_{it}| < s) I(s \geq 0).
  \end{equation*}
  Then rewrite
  \begin{multline}
    \exs{ \int_0^{\bm{X}_{it}'\bm{\Delta}} \psi_\tau(w_{it} |u_{it}| - s) - \psi_\tau(w_{it} |u_{it}|) \ud s } = \\
    \exs{ \int_{\bm{X}_{it}'\bm{\Delta}}^0 I(s < w_{it} |u_{it}| < 0) \ud s } I(\bm{X}_{it}'\bm{\Delta} < 0) - \exs{ \int_0^{\bm{X}_{it}'\bm{\Delta}} I(0 < w_{it} |u_{it}| < s) \ud s } I(\bm{X}_{it}'\bm{\Delta} > 0).
  \end{multline}
  Now focusing on just the first expectation,
  \begin{align*}
    \ex{ \exs{ \int_{\bm{X}_{it}'\bm{\Delta}}^0 I(s < w_{it} |u_{it}| < 0) \ud s } \bigg\rvert \bm{X}_{it} } &= \int_{-\infty}^0 \int_{\bm{X}_{it}'\bm{\Delta}}^0 (F_i(-s / w) - F_i(s / w)) \ud s \ud G_W(w) \\
    {} &= \int_{-\infty}^0 \int_{\bm{X}_{it}'\bm{\Delta}}^0 (f_i(\bar{u}) + f_i(\tilde{u})) (s / w) \ud s \ud G_W(w)
  \end{align*}
  where $\bar{u}$ is between $\bm{X}_{it}'\bm{\Delta}$ and 0 and $\tilde{u}$ is between $-\bm{X}_{it}'\bm{\Delta}$ and 0.  Using Fubini's theorem and the properties of the distribution of $w_{it}$,
  \begin{align*}
    \ex{ \exs{ \int_{\bm{X}_{it}'\bm{\Delta}}^0 I(s < w_{it} |u_{it}| < 0) \ud s } \bigg\rvert \bm{X}_{it} } I(\bm{X}_{it}'\bm{\Delta} < 0) &= - \int_{\bm{X}_{it}'\bm{\Delta}}^0 \frac{f(\bar{u}) + f_i(\tilde{u})}{2} s \ud s I(\bm{X}_{it}'\bm{\Delta} < 0) \\
    {} &= -\left( f_i(0) + O(|\bm{X}_{it}'\bm{\Delta}|) \right) \bm{\Delta}'\bm{X}_{it} \bm{X}_{it}'\bm{\Delta} I(\bm{X}_{it}'\bm{\Delta} < 0).
  \end{align*}
  An analogous result holds for the other integral, with $I(\bm{X}_{it}'\bm{\Delta} > 0)$.  Combining the two results and averaging over $t$ for a given $i$ (under Assumption~\ref{assume:xsupport} and either Assumption~\ref{assume:data} or~\ref{assume:stationary}) implies the first assertion.

  To show the next part, again use~\eqref{wild_equiv} to write
  \begin{align*}
    {} &\phantom{=} \psi_\tau( w_{it} |u_{it} + \bm{X}_{it}'\bm{\Delta}| - \bm{X}_{it}'\bm{\delta}) - \psi_\tau( w_{it} |u_{it} + \bm{X}_{it}'\bm{\Delta}| ) \\
    {} &= I(\bm{X}_{it}'\bm{\delta} < w_{it}|u_{it} + \bm{X}_{it}'\bm{\Delta}| < 0) I(\bm{X}_{it}'\bm{\delta} < 0) - I(0 < w_{it}|u_{it} + \bm{X}_{it}'\bm{\Delta}| < \bm{X}_{it}'\bm{\delta}) I(\bm{X}_{it}'\bm{\delta} \geq 0).
  \end{align*}
  We have, using Assumption~\ref{C2},
  \begin{multline*}
    \ex{ \exs{ I(- \bm{X}_{it}'\bm{\Delta} - \bm{X}_{it}'\bm{\delta}/w_{it} < u_{it} < - \bm{X}_{it}'\bm{\Delta} + \bm{X}_{it}'\bm{\delta}/w_{it}) I(w_{it} < 0) } | \bm{X}_{it} } I(\bm{X}_{it}'\bm{\delta} < 0) \\
    = \int_{-\infty}^0 \left( F_i(-\bm{X}_{it}'\bm{\Delta} + \bm{X}_{it}'\bm{\delta}/w) - F_i(-\bm{X}_{it}'\bm{\Delta} - \bm{X}_{it}'\bm{\delta}/w) \right) \ud G_W(w) I(\bm{X}_{it}'\bm{\delta} < 0).
  \end{multline*}
  Expand the terms inside this integral around $(\bm{\Delta}, \bm{\delta}) = \zero$:
  \begin{align*}
    F_i(-\bm{X}_{it}'\bm{\Delta} + \bm{X}_{it}'\bm{\delta}/w) &= F_i(0) + f_i(\bar{u}) (-\bm{X}_{it}'\bm{\Delta} + \bm{X}_{it}'\bm{\delta}/w) \\
    F_i(-\bm{X}_{it}'\bm{\Delta} - \bm{X}_{it}'\bm{\delta}/w) &= F_i(0) + f_i(\tilde{u}) (-\bm{X}_{it}'\bm{\Delta} - \bm{X}_{it}'\bm{\delta}/w),
  \end{align*}
  where $\bar{u}$ is between $-\bm{X}_{it}'\bm{\Delta} + \bm{X}_{it}'\bm{\delta}/w$ and $0$ and $\tilde{u}$ is between $-\bm{X}_{it}'\bm{\Delta} - \bm{X}_{it}'\bm{\delta}/w$ and $0$.  Using Assumptions~\ref{assume:ID} and~\ref{C1}-\ref{C3},
  \begin{multline} \label{lower_approx}
    \int_{-\infty}^0 \left( f_i(\bar{u}) (-\bm{X}_{it}'\bm{\Delta} + \bm{X}_{it}'\bm{\delta}/w) - f_i(\tilde{u}) (-\bm{X}_{it}'\bm{\Delta} - \bm{X}_{it}'\bm{\delta}/w) \right) \ud G_W(w) I(\bm{X}_{it}'\bm{\delta} < 0) \\
    {} = \int_{-\infty}^0 \left( (f_i(\bar{u}) - f_i(\tilde{u})) (\bm{X}_{it}'\bm{\Delta}) - w^{-1} \left( f_i(\bar{u}) + f_i(\tilde{u}) \right) (\bm{X}_{it}'\bm{\delta}) \right) \ud G_W(w) I(\bm{X}_{it}'\bm{\delta} < 0) \\
    {} = \left( -f_i(0) (\bm{X}_{it}'\bm{\delta}) + O((|\bm{X}_{it}'\bm{\Delta}| + |\bm{X}_{it}'\bm{\delta}|)^2) \right) I(\bm{X}_{it}'\bm{\delta} < 0).
  \end{multline}
  Analogous computations imply
  \begin{multline} \label{upper_approx}
    \ex{ -\exs{ I(0 < w_{it}|u_{it} + \bm{X}_{it}'\bm{\Delta}| < \bm{X}_{it}'\bm{\delta}) } | \bm{X}_{it} } I(\bm{X}_{it}'\bm{\delta} \geq 0) \\
    {} = \left( -f_i(0) (\bm{X}_{it}'\bm{\delta}) + O((|\bm{X}_{it}'\bm{\Delta}| + |\bm{X}_{it}'\bm{\delta}|)^2) \right) I(\bm{X}_{it}'\bm{\delta} \geq 0).
  \end{multline}
  Combine equations~\eqref{lower_approx} and~\eqref{upper_approx}, average over $t$ for a given and use Assumption~\ref{assume:xsupport} and either of Assumptions~\ref{assume:data} or~\ref{assume:stationary} to find the second result.
\end{proof}

For the next lemmas let
\begin{equation*}
  \| \mathbb{P}_{Ti} - P_i \|_{\mathcal{G}} = \sup_{g \in \mathcal{G}} \left| \frac{1}{T} \sum_{t=1}^T \left( g(y_{it}, \bm{X}_{it}) - \E[g(y_{it}, \bm{X}_{it})] \right) \right|
\end{equation*}
and as in \citet{AGalvao2020}, define
\begin{equation} \label{def:g1}
  \mathcal{G}_1 = \big\{ (y, \bm{X}) \mapsto \bm{a}' \bm{X} (I(y \leq \bm{b}'\bm{X}) - \tau) I(\|\bm{X}\| \leq M) : \bm{b} \in \RR^{p+1}, \bm{a} \in \mathcal{S}^{p+1} \big\},
\end{equation}
where $\bm{X} = (\bm{x}', 1)'$, and
\begin{multline} \label{def:g2}
  \mathcal{G}_2(\delta) = \big\{ (y, \bm{X}) \mapsto \bm{a}' \bm{X} (I(y \leq \bm{X}' \bm{b}_1) - I(y \leq \bm{X}' \bm{b}_2)) I(\|\bm{X}\| \leq M) : \\
  \bm{b}_1, \bm{b}_2 \in \RR^{p+1}, \|\bm{b}_1 - \bm{b}_2\| \leq \delta, \bm{a} \in \mathcal{S}^{p+1} \big\}.
\end{multline}
Some lemmas below rely on an infeasible estimate of $\alpha_{i0}$. For each $i$, let
\begin{equation} \label{atil_def}
  \tilde{\alpha}_i = \argmin_a \sum_{t=1}^T \rho_\tau (y_{it} - \bm{x}_{it}' \bm{\beta}_0 - a) + \lambda_T |a|.
\end{equation}
The $\{\tilde{\alpha}_i\}$ differ from $\{\hat{\alpha}_i\}$ because the latter are all solutions to optimization problems like~\eqref{atil_def} but with $\hat{\bm{\beta}}$ in the place of $\bm{\beta}_0$.

\begin{lemma} \label{lem:alpha_rate}
  Under Assumptions~\ref{assume:data} and \ref{assume:xsupport}-\ref{assume:Avar},
  \begin{equation} \label{eq:alpha_iid}
    \sup_i |\hat{\alpha}_i - \alpha_{i0}| = O_p\left( \| \hat{\bm{\beta}} - \bm{\beta}_0 \| + T^{-1/2} (\log T)^{1/2} + T^{-1} \lambda_T  \right).
  \end{equation}
\end{lemma}

\begin{proof}[Proof of Lemma~\ref{lem:alpha_rate}]
  Equation~\eqref{eq:bahaduralpha} from the proof of Theorem~\ref{thm:AN} implies (under Assumption~\ref{assume:ID_weakconv} and using~\eqref{opt_error})
  \begin{multline} \label{eq:bahadur_alpha_order}
    \sup_i |\hat{\alpha}_i - \alpha_{i0}| = O_p\left( \| \hat{\bm{\beta}} - \bm{\beta}_0 \| \right) + O_p \left( \sup_i \left( \mathbb{H}_{Ti}^{(\alpha)}(\bm{\theta}_{i0}) - \frac{\lambda_T}{T} \sign(\alpha_{i0}) \right) \right) \\
    + O_p \left(  \sup_i \left( \mathbb{H}_{Ti}^{(\alpha)}(\hat{\bm{\theta}}_{i}) - H_{Ti}^{(\alpha)}(\hat{\bm{\theta}}_i) - \mathbb{H}_{Ti}^{(\alpha)}(\bm{\theta}_{i0}) + H_{Ti}^{(\alpha)}(\bm{\theta}_{i0}) \right) \right) + O_p \left( T^{-1} \lambda_T \right).
  \end{multline}

  Note that the expected value of $\mathbb{H}_{Ti}^{(\alpha)}(\bm{\theta}_{i0}) - (\lambda_T / T) \sign(\alpha_{i0}) = \frac{1}{T} \sum_{t} \psi_\tau(y_{it} - \bm{x}_{it}'\bm{\beta}_0 - \alpha_{i0})$ is zero for all $i$.  Setting (their notation first, ours second) $m = p + 1$, $n = T$ and $\xi_m = M + 1$, and using $\kappa_n = C \log T$ with $C > 1$, Lemma S.1.3 of \citet{ChaoVolgushevCheng17} and the union bound imply that the right-hand side of~\eqref{eq:bahadur_alpha_order} satisfies
  \begin{equation} \label{eq:g1_bound_lem}
    \sup_i \left| \mathbb{H}_{Ti}^{(\alpha)}(\bm{\theta}_{i0}) - (\lambda_T / T) \sign(\alpha_{i0}) \right| = O_p \left( \sup_i \| \mathbb{P}_{Ti} - P_i \|_{\mathcal{G}_1} \right) = O_p \left( T^{-1/2} (\log T)^{1/2} \right).
  \end{equation}
  Next, Lemma S.1.3 from \citet{ChaoVolgushevCheng17} may be used again (with the same constants) to find
  \begin{align}
    \sup_i \big| \mathbb{H}_{Ti}^{(\alpha)}(\hat{\bm{\theta}}_{i}) &- H_{Ti}^{(\alpha)}(\hat{\bm{\theta}}_i) - \mathbb{H}_{Ti}^{(\alpha)}(\bm{\theta}_{i0}) + H_{Ti}^{(\alpha)}(\bm{\theta}_{i0}) \big| \notag \\
    {} &= O_p \left( \sup_i \| \mathbb{P}_{Ti} - P_i \|_{\mathcal{G}_2(\|\hat{\bm{\beta}} - \bm{\beta}_0\| + \sup_i |\hat{\alpha}_i - \alpha_{i0}|)} \right) \notag \\
    {} &= O_p \left( (\|\hat{\bm{\beta}} - \bm{\beta}_0\| + \sup_i |\hat{\alpha}_i - \alpha_{i0}|)^{1/2} T^{-1/2} (\log T)^{1/2} + T^{-1} \log T \right) \notag \\
    {} &= o_p(T^{-1/2} (\log T)^{1/2}) \label{eq:g2_bound_lem}
  \end{align}
  by the consistency of $\hat{\bm{\theta}}_i$.  Using~\eqref{eq:g1_bound_lem} and~\eqref{eq:g2_bound_lem} in~\eqref{eq:bahadur_alpha_order} implies the result.
\end{proof}


\begin{lemma} \label{lem:4k}
  Under Assumptions~\ref{assume:data} and \ref{assume:xsupport}-\ref{assume:lambda_tail},
  \begin{multline}
    \bm{D}_N^{-1} \frac{1}{N} \sum_{i=1}^N \left( \mathbb{K}_{Ti}^{(\theta)}(\hat{\bm{\theta}}_{i}) - K_{Ti}^{(\theta)}(\hat{\bm{\theta}}_i) - \mathbb{K}_{Ti}^{(\theta)}(\bm{\theta}_{i0}) + K_{Ti}^{(\theta)}(\bm{\theta}_{i0}) \right) \\
    = O_p \left( \|\hat{\bm{\beta}} - \bm{\beta}_0\|^{1/2} T^{-1/2} (\log T)^{1/2} + T^{-1} \log T + T^{-2/3} N^{-1/2} + T^{-1} (\log T)^{1/2} \lambda_T^{1/2} \right).
  \end{multline}
\end{lemma}

\begin{proof}[Proof of Lemma~\ref{lem:4k}]
  First, for ease of notation define
  \begin{equation*}
    \frac{1}{N} \sum_{i=1}^N \left( \mathbb{K}_{Ti}^{(\theta)}(\bm{\theta}_{i}) - K_{Ti}^{(\theta)}(\bm{\theta}_i) - \mathbb{K}_{Ti}^{(\theta)}(\bm{\theta}_i') + K_{Ti}^{(\theta)}(\bm{\theta}_i') \right) := \frac{1}{N} \sum_{i=1}^N \mathcal{K}_i(\bm{\theta}_i, \bm{\theta}_i').
  \end{equation*}
  Given the assumed positive definiteness of $\bm{D}_N$, we may focus on the stochastic order of this average.  Recalling that $\tilde{\alpha}_i$ was defined in~\eqref{atil_def}, write
  \begin{equation} \label{4k_twopart}
    \frac{1}{N} \sum_{i=1}^N \mathcal{K}_i(\hat{\bm{\theta}}_i, \bm{\theta}_{i0}) = \frac{1}{N} \sum_{i=1}^N \mathcal{K}_i(\hat{\bm{\theta}}_i, (\bm{\beta}_0, \tilde{\alpha}_i)) + \frac{1}{N} \sum_{i=1}^N \mathcal{K}_i((\bm{\beta}_0, \tilde{\alpha}_i), \bm{\theta}_{i0}).
  \end{equation}

  Suppose that the assumptions of Theorem~\ref{thm:AN} are satisfied.  Recalling the definition of $\mathcal{G}_2(\delta)$ in~\eqref{def:g2},
  \begin{align*}
    \sup_i \mathcal{K}_i(\hat{\bm{\theta}}_i, (\bm{\beta}_0, \tilde{\alpha}_i)) &= O_p \left( \sup_i \| \mathbb{P}_{Ti} - P_i \|_{\mathcal{G}_2(\|\hat{\bm{\beta}} - \bm{\beta}_0\| + \sup_i |\hat{\alpha}_i - \tilde{\alpha}_i|)} \right) \\
    {} &= O_p \left( (\|\hat{\bm{\beta}} - \bm{\beta}_0\| + \sup_i |\hat{\alpha}_i - \tilde{\alpha}_i|)^{1/2} T^{-1/2} (\log T)^{1/2} + T^{-1} \log T \right),
  \end{align*}
  where the second estimate is a result of Lemma S.1.3 of \citet{ChaoVolgushevCheng17} with $m = p+1$, $\xi_m = M$ and $\kappa_n = C \log T$, using the union bound for the supremum.  Therefore Lemma~\ref{lem:atilde} implies that 
  \begin{equation} \label{K_part1}
    \sup_i \mathcal{K}_i(\hat{\bm{\theta}}_i, (\bm{\beta}_0, \tilde{\alpha}_i)) = O_p \left( \|\hat{\bm{\beta}} - \bm{\beta}_0\|^{1/2} T^{-1/2} (\log T)^{1/2} + T^{-1} \log T + T^{-1} (\log T)^{1/2} \lambda_T^{1/2} \right).
  \end{equation}

  Next we require the stochastic order of $\sup_i \mathcal{K}_i((\bm{\beta}_0, \tilde{\alpha}_i), \bm{\theta}_{i0})$.  Note that the $\{\mathcal{K}_i((\bm{\beta}_0, \tilde{\alpha}_i), \bm{\theta}_{i0})\}_i$ are independent and that
  \begin{equation*}
    \mathcal{K}_i ((\bm{\beta}_0, \tilde{\alpha}_i), \bm{\theta}_{i0}) = O_p \left( \| \mathbb{P}_{Ti} - P_i \|_{\mathcal{G}_2(|\tilde{\alpha}_i - \alpha_{i0}|)} \right).
  \end{equation*}
  Lemma~3 of \citet{AGalvao2020} shows that $\ex{\frac{1}{N} \sum_i \mathcal{K}_i((\bm{\beta}_0, \tilde{\alpha}_i), \bm{\theta}_{i0})} = O_p(T^{-1} \log T)$.  Consider bounding the order of the variance of this average.  By Assumption~\ref{assume:xsupport}, $\sup_i \| \mathcal{K}_i((\bm{\beta}_0, \tilde{\alpha}_i), \bm{\theta}_{i0}) \| \leq 4M$.  In addition, we have
  \begin{multline*}
    \prob{ \sup_i \| \mathcal{K}_i((\bm{\beta}_0, \tilde{\alpha}_i), \bm{\theta}_{i0}) \| > T^{-2/3} } \\
    \leq \prob{ \sup_i |\tilde{\alpha}_i - \alpha_{i0}| > cT^{-1/2} (\log T)^{1/2} } + \prob{ C \| \mathbb{P}_{Ti} - P_i \|_{\mathcal{G}_2(c T^{-1/2} (\log T)^{1/2})} > T^{-2/3} } \\
    = O(T^{-2}).
  \end{multline*}
  The above order estimate uses Lemma~\ref{lem:altil_exp} with $\kappa = 2$ for the first term.  It uses Lemma S.1.3 of \citet{ChaoVolgushevCheng17} for the second, setting $\xi_n = M + 1$, $m = p+1$, $\kappa_n = 2 \log T$ and $\delta_n = cT^{-1/2} (\log T)^{1/2}$ (their notation first, ours second), noting that $T^{-3/4} (\log T)^{3/4} = o(T^{-2/3})$.  Then the variance of one term in the average, writing $\mathcal{K}_i = \mathcal{K}_i((\bm{\beta}_0, \tilde{\alpha}_i), \bm{\theta}_{i0})$, is bounded by
  \begin{align*}
    \sup_i \Var( \mathcal{K}_i ) &\leq \sup_i \ex{ \mathcal{K}_i^2 I( |\mathcal{K}_i| > T^{-2/3} ) + \mathcal{K}_i^2 I( |\mathcal{K}_i| \leq T^{-2/3} ) } \\
    {} &\leq 16 M^2 \sup_i \prob{ |\mathcal{K}_i| > T^{-2/3} } + T^{-4/3} = O(T^{-4/3}).
  \end{align*}

  Then using independence over $i$ and $\ex{|X|} \leq |\ex{X}| + \sqrt{\Var(X)}$,
  \begin{equation} \label{K_part2}
    \frac{1}{N} \sum_{i=1}^N \mathcal{K}_i((\bm{\beta}_0, \tilde{\alpha}_i), \bm{\theta}_{i0}) = O_p \left( T^{-1} \log T + T^{-2/3} N^{-1/2} \right).
  \end{equation}
  Use~\eqref{K_part2} and~\eqref{K_part1} in~\eqref{4k_twopart} to find the result.
\end{proof}

\begin{lemma} \label{lem:atilde}
  Recall the definition of $\tilde{\alpha}_i$ from~\eqref{atil_def}. Under Assumptions~\ref{assume:data} and \ref{assume:xsupport}-\ref{assume:Avar},
  \begin{equation}
    \sup_i |\hat{\alpha}_i - \tilde{\alpha}_i| = O_p \left( \|\hat{\bm{\beta}} - \bm{\beta}_0\| + T^{-1} \log T + T^{-1} \lambda_T \right).
  \end{equation}
\end{lemma}

\begin{proof}[Proof of Lemma~\ref{lem:atilde}]
  For any value of $\bm{\beta}$ define the empirical CDF of $\{ y_{it} - \bm{x}_{it}' \bm{\beta} \}_t$ for unit $i$ by
  \begin{equation*}
    \hat{\mathbb{F}}_{iT}(y, \bm{\beta}) = \frac{1}{T} \sum_{t=1}^T I(y_{it} - \bm{x}_{it}' \bm{\beta} \leq y).
  \end{equation*}

  Given any value of $\bm{\beta}$, the solution to $\min_a \sum_{t=1}^T \rho_\tau(y_{it} - \bm{x}_{it}'\bm{\beta} - a) + \lambda_T |a|$ is a penalized sample quantile from $\{y_{it} - \bm{x}_{it}'\bm{\beta}\}_{t=1}^T$: the solution $a_i^*$ satisfies
  \begin{equation}
    \left| \hat{\mathbb{F}}_{iT}(a_i^*, \bm{\beta}) - \tau + (\lambda_T / T) \sign(a_i^*) \right| \leq 1/T \; a.s.
  \end{equation}
  That is, $a_i^*$ lies between the $(\tau - (\lambda_T + 1)/T)$-th and $(\tau + (\lambda_T + 1)/T)$-th sample quantiles of $\{y_{it} - \bm{x}_{it}'\bm{\beta}\}_t$.  Therefore
  \begin{equation*}
    \left| \hat{\mathbb{F}}_{iT}(\hat{\alpha}_i, \hat{\bm{\beta}}) - \hat{\mathbb{F}}_{iT}(\tilde{\alpha}_i, \bm{\beta}_0) \right| = O_p(T^{-1} \lambda_T).
  \end{equation*}
  Given this, the rest of the proof follows the same steps as the proof of Lemma 7 in \citet{AGalvao2020}, leading to
  \begin{equation*}
    \sup_i | \hat{\alpha}_i - \tilde{\alpha}_i | = O_p\left( \| \hat{\bm{\beta}} - \bm{\beta}_0 \| + T^{-1} \log T \right) + O_p(T^{-1} \lambda_T).
  \end{equation*}
\end{proof}

The following lemma about penalized sample quantile estimates is analogous to classical results about sample quantiles as in~\citet{Serfling80}.
\begin{lemma} \label{lem:altil_exp}
  Suppose that Assumptions~\ref{assume:data} and \ref{assume:xsupport}-\ref{assume:lambda_tail} hold.  Then there is a constant $c>0$ not depending on $i$, $N$ or $T$ such that
  \begin{equation*}
    \prob{ | \tilde{\alpha}_i - \alpha_{i0} | > c \kappa^{1/2} T^{-1/2} (\log T)^{1/2} } = O(T^{-\kappa}).
  \end{equation*}
\end{lemma}

\begin{proof}[Proof of Lemma~\ref{lem:altil_exp}]
  As in the proof of Lemma~\ref{lem:atilde}, let $\hat{\mathbb{F}}_{iT}(y, \bm{\beta}) = \frac{1}{T} \sum_{t=1}^T I(y_{it} - \bm{x}_{it}' \bm{\beta} \leq y)$.  Furthermore let $F_{iT}(y, \bm{\beta}) = \ex{\hat{\mathbb{F}}_{iT}(y, \bm{\beta})}$.  Given $\bm{\beta}_0$, the solution $\tilde{\alpha}_i$ for sufficiently large $T$ (assuming $\lambda_T = o_p(T)$) satisfies
  \begin{equation} \label{atil_foc}
    \left| \hat{\mathbb{F}}_{iT}(\tilde{\alpha}_i, \bm{\beta}_0) - \tau + (\lambda_T / T) \sign(\tilde{\alpha}_i) \right| \leq 1/T \; a.s.
  \end{equation}
  Fix $\epsilon > 0$ and note that $\prob{ |\tilde{\alpha}_i - \alpha_{i0}| > \epsilon } = \prob{ \tilde{\alpha}_i > \alpha_{i0} + \epsilon } + \prob{ \tilde{\alpha}_i < \alpha_{i0} - \epsilon }$.  Since~\eqref{atil_foc} implies that $\hat{\mathbb{F}}_{iT}(\tilde{\alpha}_i, \bm{\beta}_0) \leq \tau + (\lambda_T + 1) / T$, we may write
  \begin{align}
    \prob{ \tilde{\alpha}_i > \alpha_{i0} + \epsilon } &= \prob{ \hat{\mathbb{F}}_{iT}(\tilde{\alpha}_i, \bm{\beta}_0) > \hat{\mathbb{F}}_{iT}(\alpha_{i0} + \epsilon, \bm{\beta}_0) } \\
    {} &\leq \prob{ T \tau + (\lambda_T + 1) > \sum_{t=1}^T I(y_{it} - \bm{x}_{it}'\bm{\beta}_0 \leq \alpha_{i0} + \epsilon) } \notag \\
    {} &= \prob{ \sum_{t=1}^T I(y_{it} - \bm{x}_{it}'\bm{\beta}_0 > \alpha_{i0} + \epsilon) + (\lambda_T + 1) > T(1 - \tau) }. \notag
    \intertext{Letting $v_{it} = I(y_{it} - \bm{x}_{it}' \bm{\beta}_0 > \alpha_{i0} + \epsilon)$, rewrite this as}
    {} &= \prob{ \sum_{t=1}^T (v_{it} - \ex{v_{it}}) + (\lambda_T + 1) > TF_{iT}(\alpha_{i0} + \epsilon, \bm{\beta}_0) - T\tau } \notag \\ 
    {} &\leq \prob{ \sum_{t=1}^T (v_{it} - \ex{v_{it}}) > TF_{iT}(\alpha_{i0} + \epsilon, \bm{\beta}_0) - T\tau } \notag \\
    {} &\phantom{=} \qquad \qquad + \prob{ \lambda_T + 1 > TF_{iT}(\alpha_{i0} + \epsilon, \bm{\beta}_0) - T\tau }. \label{hoef1}
  \end{align}
  An analogous argument with $\tilde{v}_{it} = I(y_{it} - \bm{x}_{it}'\bm{\beta}_0 \leq \alpha_{i0} - \epsilon)$ implies that
  \begin{align}
    \prob{ \tilde{\alpha}_i < \alpha_{i0} - \epsilon } &\leq \prob{ \sum_{t=1}^T (\tilde{v}_{it} - \ex{\tilde{v}_{it}}) > T \tau - T F_{iT}(\alpha_{i0} - \epsilon, \bm{\beta}_0) } \notag \\
     {} &\phantom{=} \qquad \qquad + \prob{ \lambda_T + 1 > T \tau - T F_{iT}(\alpha_{i0} - \epsilon, \bm{\beta}_0) }. \label{hoef2}
  \end{align}
  Define $\delta_{iT} = \delta_{iT}(\epsilon)$ by
  \begin{equation*}
    \delta_{iT} = \min\left\{ F_{iT}(\alpha_{i0} + \epsilon, \bm{\beta}_0) - \tau, \tau - F_{iT}(\alpha_{i0} - \epsilon, \bm{\beta}_0) \right\}.
  \end{equation*}
  Applying Hoeffding's inequality to both~\eqref{hoef1} and~\eqref{hoef2} implies
  \begin{equation} \label{hoef}
    \prob{ |\tilde{\alpha}_i - \alpha_{i0}| > \epsilon } \leq 2 e^{-2T\delta_{iT}^2} + 2\prob{ \lambda_T + 1 > T \delta_{iT} }.
  \end{equation}

  Next, given $\kappa$ in \ref{assume:lambda_tail}, define $\epsilon_T = \underline{f}^{-1} \kappa^{1/2} T^{-1/2} (\log T)^{1/2}$ and consider bounding $\prob{ |\tilde{\alpha}_i - \alpha_{i0}| > \epsilon_T }$.  Note that $F_{iT}(\alpha_{i0} + u, \bm{\beta}_0) = \ex{F_{u_{it} | \bm{x}_{it}}(u | \bm{x}_{it})}$, and Assumption~\ref{assume:ID_weakconv} implies that $\underline{f} > 0$ exists.  As $T$ grows large, again under Assumption~\ref{assume:ID_weakconv}, $F_{iT}(\alpha_{i0} + \epsilon_T, \bm{\beta}_0) - \tau = \ex{ f_i(0 | \bm{x}_{it}) } \epsilon_T + o(\epsilon_T)$, implying that for given constant $c$, for all $T$ large enough,
  \begin{equation*}
    F_{iT}(\alpha_{i0} + \epsilon_T, \bm{\beta}_0) - \tau \geq c \kappa^{1/2} T^{-1/2} (\log T)^{1/2}
  \end{equation*}
  and similarly
  \begin{equation*}
    \tau - F_{iT}(\alpha_{i0} - \epsilon_T, \bm{\beta}_0) \geq c \kappa^{1/2} T^{-1/2} (\log T)^{1/2}.
  \end{equation*}
  Therefore the definition of $\delta_{iT}$ using $\epsilon_T$ implies that $2e^{-2T\delta_{iT}^2} = O(T^{-\kappa})$.  Finally, given $c$, for large enough $T$ we have
  \begin{align*}
    \prob{\lambda_T + 1 > T \delta_{iT}(\epsilon_T)} \leq \prob{\lambda_T + 1 > c \kappa^{1/2} T^{1/2} (\log T)^{1/2}},
  \end{align*}
  and by Assumption~\ref{assume:lambda_tail} we may choose $c$ such that the latter sequence of probabilities is $O(T^{-\kappa})$.
\end{proof}

\begin{remark}\label{lambdaT_bound}
Condition~\ref{assume:lambda_tail} is nearly equivalent to making the assumption that $\lambda_T$ behaves like the sum of independent subgaussian random variables. To see this, suppose that with $\mu_T = \ex{\lambda_T}$ and (given $\kappa$) $\sigma_T = \sqrt{T / 2 \kappa}$, we have the Hoeffding bound $\prob{(\lambda_T - \mu_T) \geq t} \leq \exp\{ - t^2 / 2 \sigma_T^2 \}$ for all $t > 0$.  Then $\prob{(\lambda_T - \mu_T) > c T^{1/2} (\log T)^{1/2} } \leq T^{-\kappa}$. If, in addition, $\mu_T = o(T^{1/2} (\log T)^{1/2})$, then this implies our assumption.
\end{remark}

The following lemma shows that the check function satisfies a triangle inequality, and a sort of reverse triangle inequality.  The inequality $|\rho_\tau(u) - \rho_\tau(v)| < |u - v|$ for $\tau \in (0, 1)$ is used often in the quantile regression literature, but for the computational property of the penalized estimator described above in Lemma~\ref{lem:upperbnd}, a sharp inequality is required, which is what is shown in the second part of the following lemma.
\begin{lemma} \label{triangle}
  Let $\rho_\tau(u) = u(\tau - I(u < 0))$ for $\tau \in (0, 1)$ and $u \in \RR$.  Then 
  \begin{enumerate}
    \item $\rho_\tau(u + v) \leq \rho_\tau(u) + \rho_\tau(v)$ 
    \item $|\rho_\tau(u) - \rho_\tau(v)| \leq \max\{\tau, 1 - \tau\} |u - v|$.
  \end{enumerate}
\end{lemma}
\begin{proof}[Proof of Lemma~\ref{triangle}]
  It can be verified that $\rho_\tau(u) = \max\{(\tau - 1) u, \tau u\}$.  This implies both $(\tau - 1)u \leq \rho_\tau(u)$ and $\tau u \leq \rho_\tau(u)$.  Therefore $\tau (u + v) = \tau u + \tau v \leq \rho_\tau(u) + \rho_\tau(v)$ and $(\tau - 1) (u + v) = (\tau - 1) u + (\tau - 1) v \leq \rho_\tau(u) + \rho_\tau(v)$, which together imply
  \begin{equation*}
    \rho_\tau(u + v) = \max\{ (\tau - 1) (u + v), \tau (u + v) \} \leq \rho_\tau(u) + \rho_\tau(v).
  \end{equation*}
  Next, this inequality implies $\rho_\tau(u) \leq \rho_\tau(u - v) + \rho_\tau(v)$ and $\rho_\tau(v) \leq \rho_\tau(v - u) + \rho_\tau(u)$.  Then
  \begin{equation*}
    \rho_\tau(u) - \rho_\tau(v) \leq \rho_\tau(u - v) = \max\{(\tau - 1) (u - v), \tau (u - v)\} \leq \max\{\tau, 1 - \tau\} |u - v|
  \end{equation*}
  and similarly, $\rho_\tau(v) - \rho_\tau(u) \leq \max\{\tau, 1 - \tau\} |u - v|$.  This implies the result.
\end{proof}

\section{On the cross-sectional pairs bootstrap with fixed $N$ and $T$}

In this section, we offer a heuristic illustration of some problems with using a cross-sectional pairs bootstrap for the penalized quantile regression estimator. 

Fix $N$ and $T$ and assume that all $\alpha_{i0} \neq 0$ for simplicity. The assumption on $\alpha_{i0}$ reflects the fact that we make no sparsity assumptions in our analysis (see~\citet{kK00} for analogous expressions with some $\alpha_{i0} = 0$).  Define $\bm{\delta} = \sqrt{NT} (\bm{\beta} - \bm{\beta}_0)$ and $\bm{\eta}$ by $\eta_i = \sqrt{T} (\alpha_i - \alpha_{i0})$ for $i = 1, \ldots N$.  Then let
\begin{equation}
  \VV_T(\bm{\delta}, \bm{\eta}) = \sum_{i=1}^N \sum_{t=1}^T \left\{ \rho_{\tau} \left( u_{it} - \frac{\bm{x}_{it}' \bm{\delta}}{\sqrt{NT}} - \frac{\eta_i}{\sqrt{T}} \right) - \rho_{\tau} (u_{it}) \right\} + \lambda_T \sum_{i=1}^{N} \left\{ \left| \alpha_{i0} + \frac{\eta_i}{\sqrt{T}} \right| - | \alpha_{i0} | \right\},  \label{app:pqr2}
\end{equation}
where $u_{it} = y_{it} - \bm{x}_{it}' \bm{\beta}_0 - \alpha_{i0}$. This objective function is equivalent to~\eqref{pqr} in the main text.  Analysis like that of of \citet{rK04} shows that when $T$ is large, letting $f_i = f_{u_{it} | \bm{x}_{it}}$ and defining $\bm{\gamma}_i = (\bm{\delta}' / \sqrt{N}, \eta_i)'$, and letting $A \approx B$ mean that $A$ is approximately distributed as $B$,
\begin{equation} \label{app:vt_sample}
  \VV_T(\bm{\delta}, \bm{\eta}) \approx - \sum_{i=1}^N \bm{\gamma}_i' \bm{B}_{Ti} + \frac{1}{2} \sum_{i=1}^N \bm{\gamma}_i' \bm{D}_{Ti} \bm{\gamma}_i + \frac{\lambda_T}{\sqrt{T}} \sum_{i=1}^N \eta_i \sign(\alpha_{i0}),
\end{equation}
where
\begin{equation*}
  \bm{B}_{Ti} = \frac{1}{\sqrt{T}} \sum_{t=1}^T \begin{bmatrix} \bm{x}_{it} \\ 1 \end{bmatrix} \psi_\tau(u_{it}), \qquad \bm{D}_{Ti} = \frac{1}{T} \sum_{t=1}^T f_i(0 | \bm{x}_{it}) \begin{bmatrix} \bm{x}_{it} \bm{x}_{it}' & \bm{x}_{it} \\ \bm{x}_{it}' & 1 \end{bmatrix}.
\end{equation*}

To examine the validity of the cross-sectional pairs bootstrap, consider an analog loss function for resampled data. Letting $\bm{y}_i$ and $\bm{X}_i$ denote the vector and matrix of response and covariate observations corresponding to unit $i$, a cross-sectional pairs bootstrap procedure resamples $N$ pairs $(\bm{y}_i, \bm{X}_i)$ for $1 \leq i \leq N$ with replacement. Let $n_i^*$ denote the number of times unit $i$ is redrawn from the original sample. Thus, a bootstrapped estimate, the minimizer of the bootstrap objective function, solves
\begin{equation} \label{app:pairblock}
  \tilde{\bm{\theta}} = \left( \tilde{\bm{\beta}}', \tilde{\bm{\alpha}}' \right)' = \argmin_{ \bm{\theta} \in \bm{\Theta} } \sum_{i=1}^N  n_i^* \sum_{t=1}^T \rho_{\tau} \left( y_{it} - \bm{x}'_{it} \bm{\beta} - \alpha_i \right)  +\lambda_T \sum_{i=1}^N  n_i^*  | \alpha_i |.
\end{equation}
Recenter~\eqref{app:pairblock} employing $\hat{\bm{\theta}}$.  We find a bootstrap analog of the original objective function~\eqref{pqr2}, denoting $\hat{u}_{it} = y_{it} - \hat{\bm{\beta}}' \bm{x}_{it} - \hat{\alpha}_i$:
\begin{equation} 
  \tilde{\VV}_{T}(\bm{\delta}, \bm{\eta}) = \sum_{i=1}^N n_i^* \sum_{t=1}^T \left\{ \rho_{\tau} \left( \hat{u}_{it} - \frac{\bm{x}_{it}' \bm{\delta}}{\sqrt{NT}}  - \frac{\eta_i}{\sqrt{T}} \right) - \rho_{\tau}( \hat{u}_{it} ) \right\} + \lambda_T \sum_{i=1}^N n_i^* \left\{ \left| \hat{\alpha}_i + \frac{\eta_i}{\sqrt{T}} \right| - | \hat{\alpha}_i | \right\}.
\end{equation}
Then
\begin{equation} \label{app:vt_bootsample}
  \tilde{\VV}_T(\bm{\delta}, \bm{\eta}) \approx - \sum_{i=1}^N n_i^* \bm{\gamma}_i' \tilde{\bm{B}}_{Ti} + \frac{1}{2} \sum_{i=1}^N n_i^* \bm{\gamma}_i' \tilde{\bm{D}}_{Ti} \bm{\gamma}_i + \frac{\lambda_T}{\sqrt{T}} \sum_{i=1}^N n_i^* \eta_i \sign(\alpha_{i0})
\end{equation}
where
\begin{align*}
  \tilde{\bm{B}}_{Ti} &= \frac{1}{\sqrt{T}} \sum_{t=1}^T \begin{bmatrix} \bm{x}_{it} \\ 1 \end{bmatrix} \psi_\tau(u_{it} - \bm{x}_{it}'(\hat{\bm{\beta}} - \bm{\beta}_0) - (\hat{\alpha}_i - \alpha_{i0})), \\
    \tilde{\bm{D}}_{Ti} &= \frac{1}{T} \sum_{t=1}^T f_i(\bm{x}_{it}'(\hat{\bm{\beta}} - \bm{\beta}_0) + (\hat{\alpha}_i - \alpha_{i0})) \begin{bmatrix} \bm{x}_{it} \bm{x}_{it}' & \bm{x}_{it} \\ \bm{x}_{it}' & 1 \end{bmatrix}.
\end{align*}

As in Section \ref{subsection:cs}, there are two key differences between expressions~\eqref{app:vt_sample} and~\eqref{app:vt_bootsample}. First, $\tilde{\bm{B}}_{Ti} \neq \bm{B}_{Ti}$ and $\tilde{\bm{D}}_{Ti} \neq \bm{D}_{Ti}$ due to the fact that recentering uses $\hat{\bm{\theta}}$, which is biased since the model implies that $\ex{\psi_\tau(u_{it})} = 0$.

Second, there is a problem with variability in the penalty term.  It is straightforward to calculate that the expected value of the objective function with respect to the bootstrap weights (i.e., conditional on the observations) is minimized at $\hat{\bm{\theta}} = (\hat{\bm{\beta}}, \hat{\bm{\alpha}})$. However, let $\mathcal{A}=\{i : n_i^* > 0\}$ denote the ``active" set of units that are included in the penalty term in \eqref{app:pairblock}. In each bootstrap repetition, $\text{card}(\mathcal{A}) < N$, potentially changing the penalty significantly and leading to solutions $\tilde{\bm{\theta}}$ that are very different than the minimizer $\hat{\bm{\theta}}$. 

\section{Additional Simulation Results} 

\begin{figure}
\begin{center}
\centerline{\includegraphics[width=.6\textwidth]{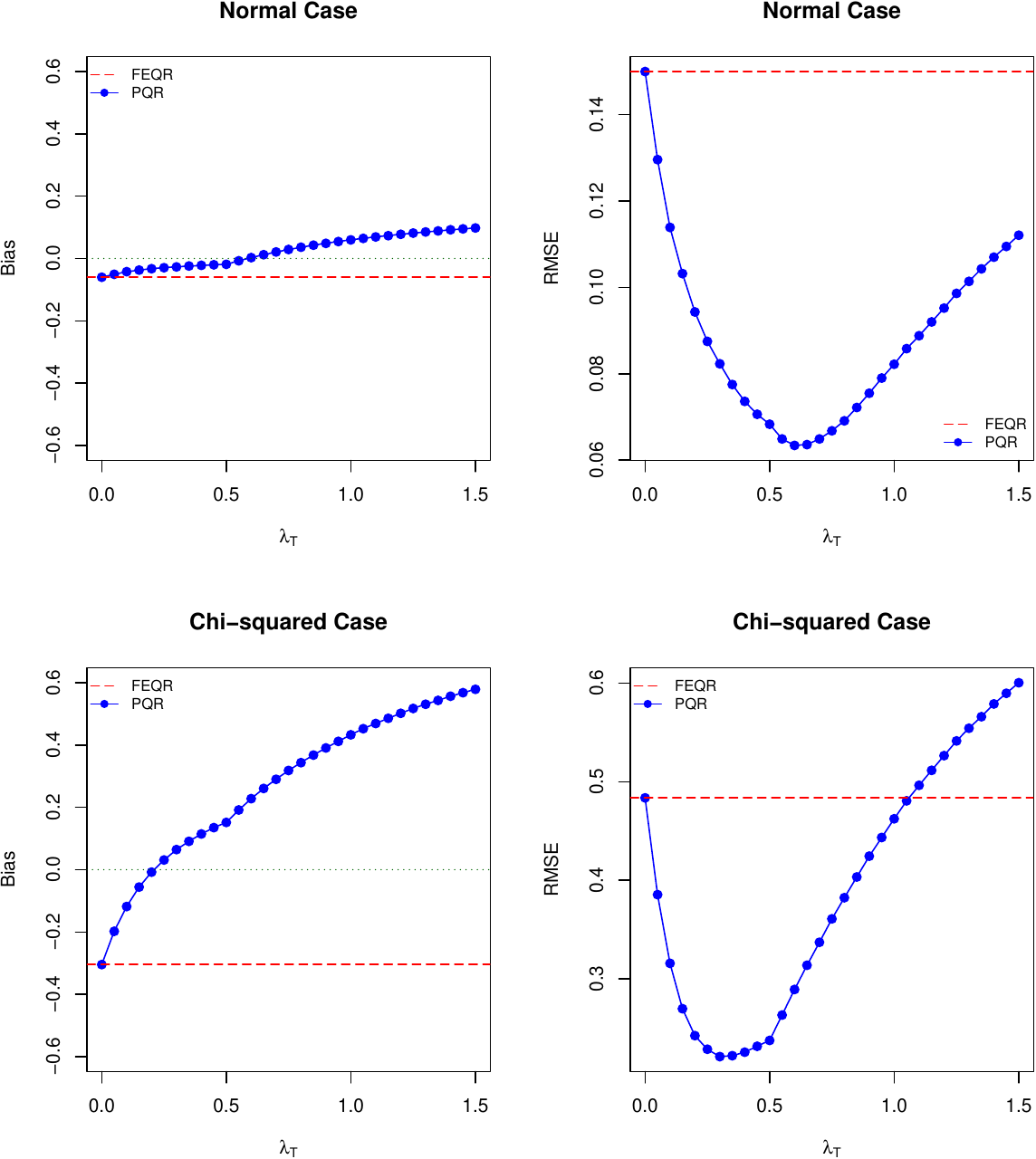}}
\caption{\emph{Small sample performance of the fixed effects (FE) and penalized quantile regression (PQR) in a location-scale shift model.}  \label{supp:fig1}}
\end{center}
\end{figure}

\subsection{Finite Sample Performance of the Penalized Estimator}

Figure \ref{supp:fig1} shows the bias and root mean squared error (RMSE) of the penalized and fixed effects estimator for the slope parameter.  We use the location-scale shift model considered in Section 4 of Kato, Galvao, and Montes-Rojas (2012)\nocite{kK12}. The variables are generated as in their second specification. The parameter of interest is $\beta(\tau) = 1 + 0.5 F_u(\tau)^{-1}$, where $F_u$ is the distribution of the error term. The model is estimated at $\tau=0.75$ considering that the error term, $u_{it}$, is distributed as $\mathcal{N}(0,1)$ or $\chi_3^2$.

The panels in Figure \ref{supp:fig1} show that the fixed effects quantile regression (FEQR) estimator is biased when $N=100$ and $T=5$. The extent of the bias varies with the distribution of the error term. Note in particular that the bias of the fixed effects estimator is -0.28 (or 9\%) when $u_{it} \sim \chi_3^2$, which is consistent with the results in Table 4 in Kato, Galvao, and Montes-Rojas (2012). (See also Koenker, 2004\nocite{rK04} and Harding and Lamarche, 2019\nocite{harding2019}). In contrast, the penalized quantile regression estimator (PQR) reduces the bias and RMSE for small values of $\lambda_T$. The evidence shows that small increases of the tuning parameter lead to  substantial improvements in both the bias profile and the RMSE. 

\subsection{Inference}

\begin{singlespace}
\begin{table}
\begin{center}\footnotesize
\begin{tabular}{c c c c c c c c c c c c c c} \hline
 &  & \multicolumn{6}{c}{Bootstrap Critical Values} & \multicolumn{6}{c}{Bootstrap Standard Errors} \\
 &  &	\multicolumn{3}{c}{Method:} &	\multicolumn{3}{c}{Method:} &	\multicolumn{3}{c}{Method:}	&	\multicolumn{3}{c}{Method:} \\
$N$	&	$T$  &	CS    &	WB1	&	WB2	        &	CS &	WB1	&	WB2  &	CS &	WB1	&	WB2 &	CS &	WB1	&	WB2 \\  \hline

\multicolumn{2}{c}{} & \multicolumn{12}{c}{Location shift model ($\zeta=0$) and $u \sim \mathcal{N}(0,1)$}  \\ 
 &  & \multicolumn{3}{c}{$\alpha_i \sim \mathcal{N}(0,1)$} & \multicolumn{3}{c}{$\alpha_i = i/N$}  & \multicolumn{3}{c}{$\alpha_i \sim \mathcal{N}(0,1)$} & \multicolumn{3}{c}{$\alpha_i = i/N$} \\ \hline

100	&	5	&	0.008	&	0.052	&	0.048	&	0.038	&	0.050	&	0.041	&	0.018	&	0.038	&	0.041	&	0.088	&	0.043	&	0.042	\\
100	&	10	&	0.003	&	0.040	&	0.041	&	0.033	&	0.053	&	0.044	&	0.009	&	0.035	&	0.040	&	0.086	&	0.049	&	0.047	\\
200	&	5	&	0.004	&	0.041	&	0.039	&	0.023	&	0.030	&	0.033	&	0.012	&	0.038	&	0.038	&	0.069	&	0.029	&	0.033	\\
200	&	10	&	0.004	&	0.036	&	0.039	&	0.023	&	0.041	&	0.040	&	0.008	&	0.036	&	0.038	&	0.067	&	0.036	&	0.042	\\   \hline

\multicolumn{2}{c}{} & \multicolumn{12}{c}{Location-scale shift model ($\zeta=0.5$) and $u \sim \mathcal{N}(0,1)$}  \\ 
&  & \multicolumn{3}{c}{$\alpha_i \sim \mathcal{N}(0,1)$} & \multicolumn{3}{c}{$\alpha_i = i/N$}  & \multicolumn{3}{c}{$\alpha_i \sim \mathcal{N}(0,1)$} & \multicolumn{3}{c}{$\alpha_i = i/N$} \\ \hline

100	&	5	&	0.042	&	0.062	&	0.061	&	0.045	&	0.064	&	0.064	&	0.087	&	0.049	&	0.050	&	0.094	&	0.061	&	0.061	\\
100	&	10	&	0.041	&	0.049	&	0.047	&	0.046	&	0.053	&	0.053	&	0.091	&	0.040	&	0.041	&	0.105	&	0.053	&	0.051	\\
200	&	5	&	0.038	&	0.052	&	0.048	&	0.028	&	0.037	&	0.037	&	0.079	&	0.038	&	0.042	&	0.073	&	0.031	&	0.033	\\
200	&	10	&	0.032	&	0.041	&	0.040	&	0.034	&	0.037	&	0.039	&	0.085	&	0.038	&	0.040	&	0.100	&	0.037	&	0.038	\\  \hline

\multicolumn{2}{c}{} & \multicolumn{12}{c}{Location shift model ($\zeta=0$) and $u \sim t_3$}  \\ 
 &  & \multicolumn{3}{c}{$\alpha_i \sim \mathcal{N}(0,1)$} & \multicolumn{3}{c}{$\alpha_i = i/N$}  & \multicolumn{3}{c}{$\alpha_i \sim \mathcal{N}(0,1)$} & \multicolumn{3}{c}{$\alpha_i = i/N$} \\ \hline

100	&	5	&	0.004	&	0.043	&	0.039	&	0.045	&	0.052	&	0.042	&	0.022	&	0.036	&	0.035	&	0.104	&	0.051	&	0.043	\\
100	&	10	&	0.004	&	0.035	&	0.034	&	0.031	&	0.039	&	0.040	&	0.009	&	0.034	&	0.031	&	0.085	&	0.040	&	0.042	\\
200	&	5	&	0.010	&	0.030	&	0.033	&	0.039	&	0.042	&	0.038	&	0.019	&	0.025	&	0.030	&	0.102	&	0.042	&	0.044	\\
200	&	10	&	0.004	&	0.031	&	0.030	&	0.035	&	0.042	&	0.040	&	0.009	&	0.034	&	0.032	&	0.085	&	0.038	&	0.041	\\  \hline

\multicolumn{2}{c}{} & \multicolumn{12}{c}{Location-scale shift model ($\zeta=0.5$) and $u \sim t_3$}  \\
 &  & \multicolumn{3}{c}{$\alpha_i \sim \mathcal{N}(0,1)$} & \multicolumn{3}{c}{$\alpha_i = i/N$}  & \multicolumn{3}{c}{$\alpha_i \sim \mathcal{N}(0,1)$} & \multicolumn{3}{c}{$\alpha_i = i/N$} \\ \hline

100	&	5	&	0.032	&	0.038	&	0.035	&	0.062	&	0.055	&	0.054	&	0.065	&	0.038	&	0.040	&	0.100	&	0.054	&	0.054	\\
100	&	10	&	0.046	&	0.041	&	0.044	&	0.054	&	0.057	&	0.056	&	0.090	&	0.042	&	0.043	&	0.105	&	0.051	&	0.054	\\
200	&	5	&	0.032	&	0.026	&	0.029	&	0.052	&	0.052	&	0.047	&	0.065	&	0.033	&	0.030	&	0.097	&	0.052	&	0.050	\\
200	&	10	&	0.044	&	0.036	&	0.037	&	0.043	&	0.033	&	0.033	&	0.090	&	0.038	&	0.037	&	0.101	&	0.033	&	0.033	\\  \hline

\multicolumn{2}{c}{} & \multicolumn{12}{c}{Location shift model ($\zeta=0$) and $u \sim \chi_3^2$} \\
 &  & \multicolumn{3}{c}{$\alpha_i \sim \mathcal{N}(0,1)$} & \multicolumn{3}{c}{$\alpha_i = i/N$}  & \multicolumn{3}{c}{$\alpha_i \sim \mathcal{N}(0,1)$} & \multicolumn{3}{c}{$\alpha_i = i/N$} \\ \hline

100	&	5	&	0.022	&	0.047	&	0.045	&	0.040	&	0.062	&	0.062	&	0.051	&	0.046	&	0.047	&	0.107	&	0.070	&	0.070	\\
100	&	10	&	0.024	&	0.061	&	0.066	&	0.042	&	0.068	&	0.068	&	0.056	&	0.059	&	0.064	&	0.100	&	0.063	&	0.064	\\
200	&	5	&	0.034	&	0.049	&	0.052	&	0.046	&	0.046	&	0.043	&	0.059	&	0.044	&	0.045	&	0.113	&	0.062	&	0.061	\\
200	&	10	&	0.030	&	0.064	&	0.068	&	0.037	&	0.043	&	0.043	&	0.059	&	0.065	&	0.067	&	0.079	&	0.044	&	0.044	\\  \hline

\multicolumn{2}{c}{} & \multicolumn{12}{c}{Location-scale model ($\zeta=0.5$) and $u \sim \chi_3^2$} \\ 
 &  & \multicolumn{3}{c}{$\alpha_i \sim \mathcal{N}(0,1)$} & \multicolumn{3}{c}{$\alpha_i = i/N$}  & \multicolumn{3}{c}{$\alpha_i \sim \mathcal{N}(0,1)$} & \multicolumn{3}{c}{$\alpha_i = i/N$} \\ \hline

100	&	5	&	0.046	&	0.055	&	0.055	&	0.052	&	0.066	&	0.064	&	0.076	&	0.057	&	0.058	&	0.095	&	0.074	&	0.070	\\
100	&	10	&	0.037	&	0.056	&	0.056	&	0.049	&	0.068	&	0.069	&	0.086	&	0.053	&	0.054	&	0.114	&	0.067	&	0.069	\\
200	&	5	&	0.061	&	0.056	&	0.057	&	0.053	&	0.057	&	0.053	&	0.098	&	0.068	&	0.067	&	0.107	&	0.074	&	0.074	\\
200	&	10	&	0.043	&	0.061	&	0.062	&	0.038	&	0.046	&	0.043	&	0.113	&	0.060	&	0.061	&	0.097	&	0.045	&	0.045	\\ \hline
\end{tabular}
\end{center}
\vspace{3mm}
\caption{\emph{Empirical rejection probabilities of $H_0: \beta_0(0.5) = 1 + \zeta F_u(0.5)^{-1}$. CS denotes cross-sectional pairs bootstrap, WB1 denotes wild bootstrap estimator \eqref{penboot}, and WB2 wild bootstrap estimator \eqref{penbootthr}.}}
\label{supp:table1}
\end{table}
\end{singlespace}

We now turn our attention to the performance of tests using the bootstrap. To this end, Table \ref{supp:table1} reports empirical rejection frequencies for the null hypothesis $H_0: \; \beta_0 = 1 + \zeta F_u(0.5)^{-1}$.  As in Table \ref{mc.table1}, 
we consider different sample sizes $N \in \{100,200\}$ and $T \in \{5,10\}$, different distributions $F_u$, and different assumptions on $\alpha_i$. We report results using two different approaches. The cross-sectional pairs 
bootstrap (CS) samples over $i$ with replacement, keeping the entire block of
time series observations. The
wild bootstrap is implemented as discussed in Section \ref{subsec:wild}. We first obtain residuals $\hat{u}_{it}$ using the penalized quantile regression
estimator. The estimator \eqref{penboot} is labeled `WB1' and the
estimator \eqref{penbootthr} is labeled `WB2'. As in the case of the wild bootstrap estimator proposed by Feng, He, and Hu (2011), 
a finite sample correction is recommended. We adjust the residuals with the influence function and sign function following the Bahadur representation of the estimator derived in Theorem \ref{thm:AN}. Then, we generate $u^\ast_{it} = w_{it} | \hat{u}_{it}
|$, where $w_{it}$ is an i.i.d. random variable distributed as a two-point
distribution with probabilities $\tau$ and $1-\tau$ at $w_{it} = -2 \tau$ and
$w_{it} = 2 (1-\tau)$. Lastly, we generate the dependent variable as
$y_{it}^\ast = \hat{\alpha}_i + \hat{\beta} x_{it} + u^\ast_{it}$. 

The first columns report results based on bootstrap critical values obtained from the distribution of $\sqrt{NT} (\beta^\ast  - \hat{\beta})$, where $\hat{\lambda}_T$ is obtained as in Table \ref{mc.table1}. The last columns report results obtained using bootstrap standard errors, which are denoted by $\mbox{se}(\beta^\ast)$. In this case, the statistic is $| \hat{\beta} - \beta_0 | / \mbox{se}(\beta^\ast)$ and it is compared to $\Phi^{-1}(1 - \alpha/2)$. The theoretical size of the tests is equal to 5\%. As it can be seen in the upper block of Table \ref{supp:table1}, the wild bootstrap procedure tends to produce empirical sizes that are closer to the nominal values. The lower panels of Table \ref{supp:table1} show results for a DGP when the error term is distributed as $t_3$ and $\chi_3^2$ and offer similar conclusions. We do not observe significant differences between probabilities estimated by bootstrap critical values or bootstrap standard errors. 

\small

\bibliographystyle{econometrica}
\bibliography{wild}

\end{document}